\newcommand{\kms}{{\rm km\, s^{-1}}}
\newcommand{\Msun}{{\rm M_\odot}}
\newcommand{\Lsun}{{\rm L_\odot}}
\newcommand{\pc}{{\rm pc}}
\newcommand{\Myr}{{\rm Myr}}
\newcommand{\pcc}{{\rm cm^{-3}}}

\documentclass[preprint2]{pp7}

\usepackage{amsmath,amssymb}
\usepackage{xcolor}

\input{pp7.h}

\begin{document}

\title{\textbf{\LARGE THE LIFE AND TIMES OF GIANT MOLECULAR CLOUDS}}

\author {\textbf{\large M\'elanie Chevance}}
\affil{\small\it Astronomisches Rechen-Institut, Zentrum f\"ur Astronomie der Universit\"at Heidelberg, M\"onchhofstrasse 12-14, D-69120 Heidelberg, Germany}
\author {\textbf{\large Mark R. Krumholz}}
\affil{\small\it Research School of Astronomy and Astrophysics, Australian National University, Canberra ACT 2601, Australia}
\author {\textbf{\large Anna F. McLeod}}
\affil{\small\it Centre for Extragalactic Astronomy/Institute for Computational Cosmology, Department of Physics, Durham University, South Road, Durham DH1 3LE, UK}
\author {\textbf{\large Eve C. Ostriker}}
\affil{\small\it Department of Astrophysical Sciences, Princeton University,
  Princeton, NJ 08544, USA}
\author {\textbf{\large Erik W. Rosolowsky}}
\affil{\small\it Department of Physics, University of Alberta, Edmonton, Alberta T6G 2E1, Canada}
\author {\textbf{\large Amiel Sternberg}}
\affil{\small\it School of Physics \& Astronomy, Tel Aviv University, Ramat Aviv, 69978, Israel; Center for Computational Astrophysics, Flatiron Institute, 162 5th Avenue, New York, NY 10010, USA}

\begin{abstract}
\baselineskip = 11pt
\leftskip = 1.5cm 
\rightskip = 1.5cm
\parindent=1pc
{\small 
Giant molecular clouds (GMCs) are the sites of star formation and stellar feedback in galaxies. Their properties set the initial conditions for star formation and their lifecycles determine how feedback regulates galaxy evolution. In recent years, the advent of high-resolution telescopes has enabled systematic GMC-scale studies of the molecular interstellar medium in nearby galaxies, now covering a wide range of physical conditions and allowing for the first studies of how GMC properties depend on galactic environment. These observational developments have been accompanied by numerical simulations of improving resolution that are increasingly accurately accounting for the effects of the galactic-scale environment on GMCs, while simultaneously improving the treatment of the small-scale processes of star-formation and stellar feedback within them. The combination of these recent developments has greatly improved our understanding of the formation, evolution, and destruction of GMCs. We review the current state of the field, highlight current open questions, and discuss promising avenues for future studies.
 \\~\\~\\~}
\end{abstract}  

\section{\textbf{INTRODUCTION}}
\label{sec:intro}

\subsection{Historical background and modern context}

While the discovery of molecules in the ISM dates back to the first optical observations of the diffuse interstellar bands \citep{Swings37a, McKellar40a}, the discovery of giant molecular clouds (GMCs) proper can be more closely dated to shortly after the first detection of interstellar CO via the $J=1-0$ rotational transition at 2.6 mm
by \citet{Wilson70a}. Studies of individual objects with radio telescopes 
revealed that CO emission could be found in large complexes that came to be known as GMCs \citep{Lada76a, Blair78a, Blitz80b}, and the first Galactic plane surveys in the $J=1-0$ CO line established that much of the emission could be decomposed into dense cloud-like structures with masses of $\sim 10^4 - 10^6$ M$_\odot$, 
mean density $n_{\rm H_2} \sim 100$  $\pcc$, and temperature $T\sim 10$ K \citep[e.g.,][]{Solomon87a, Scoville87a, Dame87a, Dame01a}. These objects made a substantial contribution to the total mass budget of the Galactic ISM, and there was a strong correlation, albeit with a great deal of scatter, between GMCs and sites of ongoing star formation \citep{Mooney88a}. The advent of millimeter interferometer arrays extended these studies to nearby galaxies, first targeting the environments of H~\textsc{ii} regions \citep{Vogel87, Wilson90a}, then surveying the Magellanic Clouds \citep[e.g,][]{Fukui01a} and eventually the full disks of nearby galaxies \citep[e.g.,][]{Engargiola03,Blitz2007}. Today radio observations 
are beginning to probe the cold, dense molecular gas in high-redshift galaxies, including at $z \sim 2$ when galaxy assembly was occurring most rapidly \citep[see the reviews][]{Carilli2013,Combes2018,Tacconi20}. In this article we focus on knowledge gained from Milky Way studies and observations of nearby galaxies that are able to probe GMC cloud scales.

Not surprisingly, given the rate of progress and the centrality of GMCs to star formation, there has been at least one review chapter on GMCs in every Protostars \& Planets volume from the first \citep{1978prpl.conf..153E}. The most recent of these reviews, \citet{Dobbs14a}, was published just before the Atacama Large Millimeter Array (ALMA) came online. 
At that time, most mm-wavelength extragalactic surveys of molecular gas either integrated over whole galaxies \citep[e.g.\ COLD GASS,][]{Saintonge11}, or resolved only relatively large ($\sim 1$ kpc) structures within them (e.g.\ HERACLES, \citealt{Leroy09}; the JCMT NGLS, \citealt{Wilson12}; ATLAS–3D CO, \citealt{Alatalo13}), without separating individual clouds. While variation of cloud properties was observed with environment, for example between galaxy centers and disks \citep{Sandstrom13, Leroy13, Longmore13} or in different regions of M51 \citep{Schinnerer13, Pety13}, systematic cloud-scale studies in external galaxies remained highly challenging, and were possible at all only in the most nearby galaxies of the Local Group \citep{Fukui99, Engargiola03, Rosolowsky07,Bolatto08, Koda09, Wong11, DonovanMeyer13}, which probe a limited range of environments.

Due to the lack of observational constraints, many questions remained unanswered. For example, the molecular gas depletion time -- defined as the ratio between the molecular gas mass and the star formation rate, 
\begin{equation}
t_{\rm dep} \equiv \frac{M_{\rm mol}}{\dot M_*}, \label{eq:tdep}
\end{equation}
has much more scatter on GMC scales \citep[e.g.][]{Schruba10, Onodera10} than on kpc scales \citep{Leroy13}, indicative of the cycling between cloud formation, star formation, and residual gas dispersal \citep[e.g.][]{Kruijssen2014}.   
Despite this, the lack of resolved observations of both molecular emission and star formation signatures in large statistical samples of clouds meant that significant uncertainty remained over individual clouds' lifetimes \citep{Scoville1979, Elmegreen2000, Koda09, BallesterosParedes07, Kawamura2009}.  This situation has significantly improved, and we discuss the important question of empirical constraints on GMC lifetimes in Section \ref{sec:evolution}.

Early work indicated that 
GMCs had comparable kinetic and gravitational energy, and a common assumption was that GMCs are in virial equilibrium \citep[e.g.][]{1993prpl.conf..125B}. 
Similarly, GMCs in the Milky Way 
and in nearby galaxies 
seem to follow the Larson relations \citep{Larson81, Solomon87a,Bolatto08, Fukui08, Heyer09, RomanDuval10, Wong11}, 
in that the observed (nonthermal) line-width increases with cloud size as approximately the $1/2$ power. For constant cloud surface density \citep[as appeared to hold in early observations, e.g.][]{Blitz2007}, this would imply a constant ratio of kinetic-to-gravitational energy.  However, the line width-size relation also extends within clouds \citep{heyer04},
and may simply reflect the power law scaling expected for compressible turbulence \citep[e.g.][]{McKee_Ostriker07}.  Explorations over a wider range of galactic environments show a large range of GMC surface density, but that clouds nonetheless have comparable kinetic and gravitational energies \citep[e.g.][]{Dobbs14a,sun20a}.
An exact accounting of energy is made difficult by the 
uncertainty in tracing mass via CO emission, and some have suggested that GMCs 
are even in free-fall collapse \citep{BallesterosParedes11}, but 
current work discussed in Section \ref{sec:gmc_env} suggests that low-mass structures are unbound while high-mass GMCs are marginally bound.

Recent years have seen immense progress in our empirical understanding of GMCs, driven largely by the synergy between high-resolution Galactic studies and the statistical power that comes from extragalactic surveys, aided by the combination of millimeter observations (from ALMA and other millimeter telescopes) with complementary data in the infrared (\textit{Spitzer} and \textit{Herschel}) and optical (particularly integral field spectrographs such as MUSE). The
systematic analysis of large, multiwavelength data sets
have made it possible for the first time to begin empirically constraining the stages of the GMC life cycle in different environments and quantifying the major factors driving that evolution.

Theoretical models have evolved in tandem with observations. The earliest models of molecular clouds featured either simple geometries coupled to quasi-static evolution controlled by magnetic fields \citep[e.g.,][]{Mouschovias76a, Shu87a, Mouschovias99a}, or pure free-fall collapse and gravity-driven fragmentation \citep[e.g.,][]{Zinnecker84a, Klessen98a, Bastien91a, Bonnell1992, Bonnell97a}. As observations improved and clouds' complex internal structures became resolvable, these approaches were supplemented by models based on the physics of turbulence to interpret observed velocities, the origin of internal structure, and support against gravity (or lack thereof) \citep[e.g.,][]{Vazquez-Semadeni94a,Gammie1996, Bate03a}.
Magnetically-dominated models have been disfavored by observations indicating that magnetic fields are insufficiently strong to prevent collapse \citep[and references therein]{Crutcher12a}, leaving (magneto)hydrodynamic (MHD) turbulence and gravity as the main physical elements in many theories \citep[e.g.][]{Krumholz05a,Hennebelle2008}. 

Development of robust and efficient numerical methods for fluid dynamics and the growth of computational power ushered in an era in which it became possible to study GMCs using large-scale numerical MHD simulations \citep{  Vazquez-Semadeni95a, Stone1998,Klessen00a,Ostriker2001, Mac-Low04a,Bate03a,Padoan07a}. The chapters of \citet{PPIV_2000} and \citet{PPV_2007},  respectively in {\it PPIV} and {\it PPV},  reviewed theoretical modeling of GMCs, drawing largely on numerical MHD simulations that focused on the aspects of kinematics, structure, and evolution that can be explained as a consequence of (M)HD turbulence and gravity, such as the predicted timescales for turbulent dissipation, global evolution, and small-scale collapse; the origin of 
log-normal density PDFs;  and distinctions in structure and 
evolution between strongly- and weakly-magnetized clouds.

At the time of the {\it  PPVI} review \citep{Dobbs14a}, numerical simulations based just on imposed MHD turbulence and gravity continued to provide insights into the development of star formation in GMCs, such as the connection between the power-law portion of density PDFs and the onset of gravitational collapse \citep[e.g.,][]{Kritsuk2011,Collins12a,Federrath2012}.  Simultaneously, serious study of the effects of the energy returned by star formation feedback was also beginning. Initial simulations included one or two feedback processes, for example non-ionizing radiation \citep[e.g.,][]{Krumholz07a, Bate09a,Raskutti2016}, ionizing radiation \citep[e.g.,][]{Dale07a, Peters10a}, protostellar outflows \citep[e.g.,][]{Nakamura07a, Krumholz12a}, stellar winds \citep[e.g.,][]{Dale08a, Rogers13a}, and supernovae \citep[e.g.,][]{Dobbs11a, Rogers13a}. Almost all of these started from isolated clouds rather than including the full galactic context of GMC formation, almost  none  included  more  than  one  or  two feedback mechanisms, and comprehensive parameter studies were lacking.  
The situation now is quite different (see Sections \ref{sec:destruction} and \ref{sec:accomplishment}): a number of groups have published simulations following the full GMC life cycle in the galactic context including feedback, and at the same time controlled parameter studies of individual GMCs have elucidated the relative importance of different feedback mechanisms to cloud destruction and lifetime star formation efficiency, as well as the dependence of evolution on integrated cloud properties (mass, size, magnetization).

\subsection{GMC life-cycle overview and chapter outline}
In this chapter, we will discuss progress in our understanding of GMCs,  both on the theory and observational sides. We shall focus on advances since {\it PPVI}, referring to the review of GMCs in that volume by \citet{Dobbs14a} as needed.  
An overall picture of the GMC life-cycle within the context of the dynamic, multiphase ISM is by 
now well established.  The ISM consists of gas 
at a wide range of temperatures and densities, 
with molecular gas the coldest and densest component.   The gas in the ISM is constantly in flux, with no permanent structures: clouds form where there are converging flows, and are torn 
apart  by diverging flows and by shear.  The 
energy that drives these ISM flows derives from 
stellar feedback (radiation, winds, and supernovae) and from the work done by 
the gravity (as gas accretes
toward the galactic center, interacts with spiral arms, interacts with itself, and possibly as extragalactic gas accretes onto the galaxy).  Specific mechanisms  leading to GMC formation, and dependence on galactic environmental parameters, were reviewed in \citet{Dobbs14a}; recent work (see Section  \ref{sec:formation}) has further investigated these mechanisms with increasingly realistic numerical simulations.  Structures with a large 
range of masses and sizes are formed  in this 
way, with the smaller-column clouds consisting 
of cold atomic gas and the larger, UV-shielded ones largely converting to molecular gas.  Larger structures may form 
in part out of warm diffuse gas that accretes or concentrates and changes phase, and in part by collection of pre-existing smaller cold condensations. While the term ``cloud collision'' has often 
been used in the past for mergers of pre-existing cold condensations, here we avoid 
this terminology in view of the impermanence of 
structures and continuum nature of the ISM. 
In sufficiently massive clouds, there are overdense regions where gravitational 
energy densities exceed the kinetic,
magnetic, and thermal values, leading to local 
collapse and star formation.

When enough massive stars form, the resulting energetic feedback  disperses the remaining gas 
in the cloud.  In this way, feedback determines the lifetime star formation efficiency of a GMC. Clouds that do not host
massive star formation are instead dispersed by processes of 
external origin.  Destruction of GMCs
involves phase changes for some fraction of the mass, with molecules being photodissociated 
and/or photoionized.  However, photoionized 
gas rapidly returns to neutral forms when 
hot stars die, after relatively
brief lifetimes.  Since {\it PPVI} there have been important advances in quantitative assessment 
of the star formation feedback effects described above, both theoretically and observationally, and we review these results here  (see Sections \ref{sec:destruction} and  \ref{sec:accomplishment}). Timescales of each stage of the GMC life-cycle are increasingly well constrained (see Section \ref{sec:evolution}), although further work is needed to obtain more precise 
predictions and measurements as a function of environment. 

The plan of the remaining sections is as follows.  
In Section \ref{sec:properties} we present the 
properties of GMCs revealed by recent observations, with interpretation informed by theory and simulations. We then review the formation physics of GMCs in Section \ref{sec:formation}, covering the transition from atomic to molecular gas and the role of various mechanisms in driving overdensities in the ISM. In Section \ref{sec:evolution} we characterise the evolution of GMCs, from their initial assembly to their dispersal;
we also discuss how clouds' star formation behavior depends on their properties. In Section \ref{sec:destruction} we review the different processes that can destroy GMCs, focusing on theoretical predictions, results of numerical simulations, and constraints from observations on the role of the different feedback mechanisms. We then describe the ``products’' of GMCs during their lifetime in Section \ref{sec:accomplishment}, including some characteristics of the resulting star formation. Finally, in Section \ref{sec:future} we provide an outlook on current and future observations and perspectives for synergies between observations and theoretical modeling.

\section{GMC Properties}
\label{sec:properties}

\begin{figure}[b!]
    \centering
    \includegraphics[width=\columnwidth]{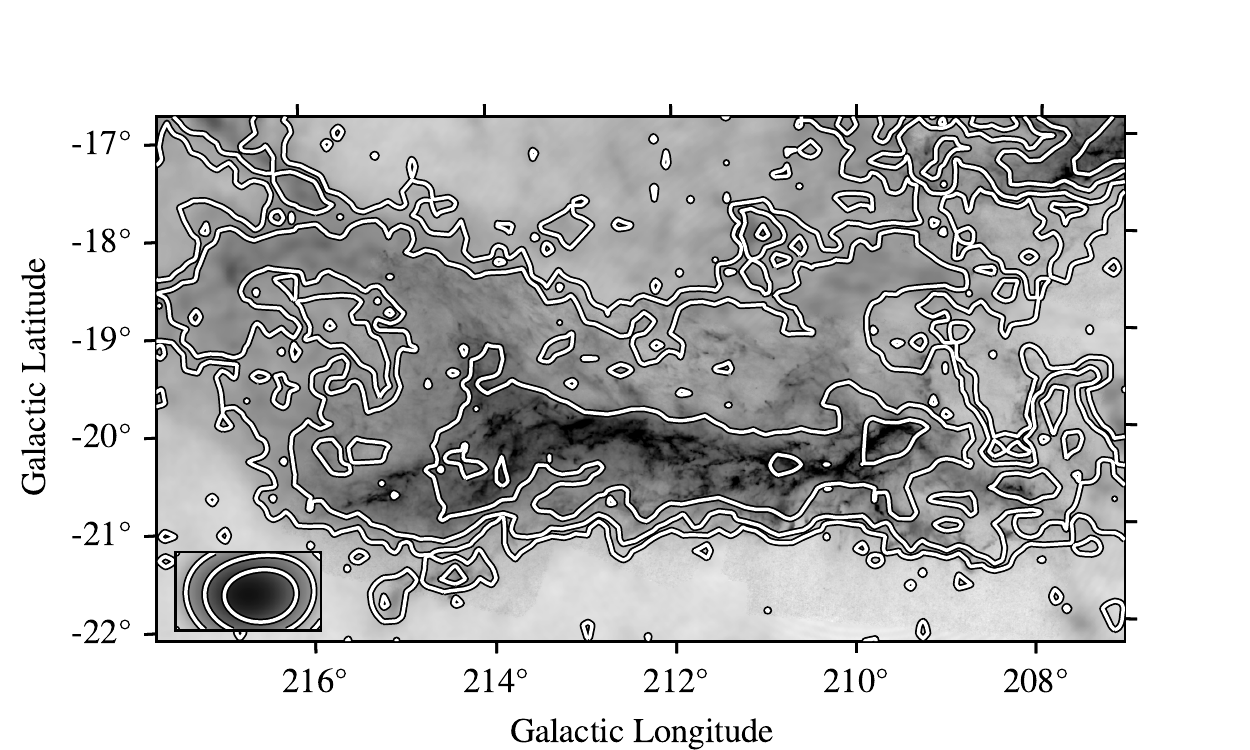}
    \caption{The Orion A GMC as observed in emission from dust \citep[grayscale, logarithmic stretch; ][]{lombardi14} and $^{12}$CO(1-0) \citep[contours; logarithmic spacing;][]{wilson05}. The two tracers both illustrate a typical morphology for a GMCs: moderate aspect ratio, filamentary substructure. The inset figure shows the same maps convolved to 50 pc linear resolution, similar to the resolution that is typically achieved for observations of GMCs in nearby galaxies.}
    \label{fig:oriona}
\end{figure}

\subsection{Is a GMC an entity?}
Given the dynamic nature of GMCs, it is unclear whether there is a physically motivated definition for a GMC.  Even so, the field often treats GMCs as physical entities either implicitly or explicitly.  Here, we review GMC definitions, how those definitions influence our physical interpretations, and how this thinking has developed since {\it PPVI}. 

Figure \ref{fig:oriona} shows two views of Orion A, a prototypical GMC with mass $M\sim 8\times 10^4~M_\odot$, using dust and molecular line emission \citep[][respectively]{lombardi14, wilson05}. The dust map traces dust emission at 353 $\mu$m from {\it Herschel} and {\it Planck} data, taken as a nearly-linear tracer of the combined atomic and molecular gas column density.  The molecular line emission is from $^{12}\mathrm{CO}(1{-}0)$ mapping, which traces the low density molecular regions of the cloud ($A_V > 1$,  $n_\mathrm{H_2}\gtrsim 10^2~\mathrm{cm}^{-3}$) but saturates at high column and volume densities.  These maps show that Orion A has a typical morphology of a GMC: a modest aspect ratio (60 pc $\times$ 20 pc) and filamentary structure throughout the cloud, where the high column density filaments host the massive star formation in this GMC. These two tracers illustrate the challenges in defining GMCs. They offer similar perspectives on cloud structure, though the dust tracers illustrate better that the GMC is part of a large complex of atomic and molecular gas.  In contrast, the CO emission has a well-defined boundary from which the cloud can be distinguished from other clouds in the region. 

These maps illustrate two possible routes for identifying GMCs as discrete objects: using thresholds in molecular line emission or boundaries in dust column density. Typically, molecular line emission is observed in spectral-line data cubes and GMCs are identified as isolated regions in  position-position-velocity space.  Using line emission ensures that the regions studied are indeed in the molecular phase, but CO is not a linear tracer of H$_2$ throughout GMCs \citep[e.g.][]{Pineda2008,Shetty2011,bolatto13}. In contrast, dust emission \citep[e.g.][]{motte10, molinari16, planckpaper35} and extinction \citep[e.g.][]{Lombardi2010,Lombardi2011} is a more linear tracer of the neutral ISM but dust shows no indication of the atomic-to-molecular transition in the ISM.  Dust is usually used as a tracer of column density along the line of sight.  

Recent work has turned to using differential extinction toward stars throughout the ISM to map out the distances to different parts of the dust extinction, which can reveal the three-dimensional structure of the ISM.  When paired with {\it Gaia} parallaxes, differential dust extinction and protostar parallaxes offer a uniquely powerful opportunity to resolve cloud structure in three spatial dimensions \citep[for Orion A see, e.g.,][and a more detailed discussion in the chapter by Zucker et al.]{grossschedel18, rezaeikhan20}.  While the low spatial resolution of the three dimensional molecular cloud maps precludes directly measuring molecular cloud structure in detail, GMCs still stand out as overdensities \citep{chen20}.

Dust-based observations of molecular clouds are restricted to the Solar neighbourhood because of the lack of background sources (extinction) or blending (emission). Extragalactic observations of GMCs in dust emission are restricted to the Local Group because of limited resolution and sensitivity \citep[e.g.,][]{williams18, forbrich20}.  In contrast, molecular line emission can be used to resolve GMCs throughout the Galaxy \citep{umemoto17, pety17, Kong18}, across the Local Group \citep{schruba17, Wong19, kondo21}, and beyond \citep{hirota18, imara19, leroy21a, miura21}.  However, in the inner Milky Way and the Galactic center, kinematic distances are ambiguous and the emission is heavily blended.   Extragalactic observations suffer from poor resolution and blending in molecule-rich regions.  In these cases, molecular cloud identification relies on identifying discrete structures within the blended emission.  Several approaches to this problem are present in the literature including watershed decomposition  \citep[e.g.,][]{williams94, cprops, fellwalker}, identifying discrete three dimensional regions in the spectral-line data cube above a set of brightness temperature contours \citep[e.g.,][]{Solomon87a, shetty12}, or using more sophisticated object identification algorithms \citep{colombo15, MivilleDeschenes17}.  

Object identification in spectral-line data cubes is fraught since the velocity dimension of the cube is not a true spatial dimension. 
\citet{pan15} and \citet{khoperskov16}  compare GMCs extracted from simulations, both in the simulated volume and in simulated observations. These studies find that different approaches to cloud extraction lead to similar scaling relationships between cloud properties and thus similar conclusions about the physical state of the molecular ISM.  However, the extraction of specific individual objects is not robust.  Furthermore, object identification algorithms all have user-defined free parameters that influence the set of objects that are recovered so that different GMC populations can be extracted from a single observation given the tuning parameters of the algorithm \citep{pineda09}.  Given these limitations, especially in extragalactic observations, \citet{leroy16} eschew cloud decomposition, instead proposing to use a fixed spatial scale for analysis of individual lines of sight.  While such an approach does not yield scalings between cloud size and other properties, it provides a robust and well-defined method for comparing observations to each other and to simulations. Extending to multi-scale maps (e.g., at $2\times, 4\times,8\times \cdots$ the resolution) offers an alternative approach to quantifying variations of molecular gas properties with spatial scale.  

In summary, GMCs are observationally defined, in most cases as overdensities or bright, compact features.  This identification is relative to the local medium, and the boundaries between GMCs and their surroundings are not clearly demarcated.  Indeed, theoretical and observational work shows that GMCs are not isolated from their surroundings and evolve over a few crossing times (see Sections \ref{sec:formation}-\ref{sec:destruction} below). {\bf Thus, GMCs should not be regarded as a well-defined set of discrete entities}. Cloud properties must always be interpreted in the context of the object identification strategies that produced them. These properties are illuminating in comparative studies but the translation to any absolute measures requires care. 

\subsection{Internal Structure of GMCs}\label{sec:internal_props}

The internal structure of GMCs is shaped by gravitation, turbulence and propagating shocks, magnetic fields, heating and cooling, and chemistry; many of these effects derive from stellar feedback. Cloud kinematics indicate that the gas motions are turbulent as expected from the high Mach ($\mathcal{M}_s>10$) and Reynolds ($\mathrm{Re}>10^8$) numbers in cold molecular gas \citep{heyerdame15}.  Observational estimates of the velocity structure functions \citep{heyer04, brunt09,  Schneider11a} find the expected power-law scaling between velocity and spatial separations that characterizes turbulent flows.  The size-line width relationships of cloud populations discussed in Section \ref{sec:gmc_env} are measurements of these structure functions on the outer scales of clouds.

Theoretical models of self-gravitating, turbulent gas predict that the volume density distribution should follow a log-normal distribution with a tail that can be described with either a single  \citep[e.g.,][]{BallesterosParedes11, Collins12a, Burkhart17a} or double \citep{Jaupart20a, Khullar21a} power-law.  Observational measurements of the volume density distribution are limited and typically rely on fitting the expected model distributions to data \citep[e.g.,][]{ginsburg13, kainulainen14}. The column density distribution of clouds is better studied and can be described with power-law and log-normal distributions or combinations of the two \citep[e.g.][]{Schneider_2015a,Schneider_2015b}, but edge effects in cloud mapping preclude clearly distinguishing between these cases \citep[e.g.,][]{lombardi15, spilker21}. 

Density and column density distribution functions 
describe the range of internal conditions.  Additionally, the mass in molecular clouds is highly structured 
and is frequently described as {\it filamentary} with such structures being common in self-gravitating turbulent gas \citep{andre14}.  Mass flows along filaments are linked to cluster assembly \citep[e.g.,][]{Kirk13} and define where stars form within GMCs (see chapters by Zucker et al. and Hacar et al.).  Statistically, the mass structure of clouds is strongly spatially correlated as measured through two-point statistics and Fourier analysis \citep{padoan06, pingel13, alvesdeoliveira14}, providing another diagnostic of the turbulence in clouds.  

\subsection{Global properties of GMCs and environmental variation}\label{sec:gmc_env}

Despite the internal complexity of GMCs, there is a long history of interpreting GMCs as a population of discrete entities and using a population-based analysis of single-point descriptions of molecular clouds (mass, radius, etc.).  These simple measures are averages over a wide range of internal conditions, and implicit in this approach is the assumption that the internal physical processes are well summarized by these average properties. Studies of internal cloud structure have not falsified this assumption yet, but it bears constant examination.  Here, we summarize the canonical properites and recent developments.

Both molecular line and dust-based tracers measure cloud mass ($M$) and cloud size (in terms of cloud area $A$ or radius $R$ under a spheroidal approximation).  Molecular line observations can also measure the velocity dispersion of molecular clouds ($\sigma_v$) along the line of sight. The exact definition of how these properties are measured from observations varies across different studies \citep[e.g.,][]{cprops}.  Physical conclusions are rarely changed as a consequence of different definitions, but some caution must be taken in comparing different studies to each other. 

Studies of the Milky Way disk showed that these macroscopic properties are correlated:
\begin{equation}
    \sigma_v = \sigma_0 \left(\frac{R}{1~\mathrm{pc}}\right)^{c_1}\quad \mathrm{and} \quad M = \Sigma_\mathrm{mol} \pi R^{c_2},
    \label{eq:scaling}
\end{equation}
finding $c_1 \approx 0.5$ and $c_2\approx 2$, \citep[see review by ][and references therein]{heyerdame15}.  The size-line width relationship is interpreted as a proxy for the structure function of a turbulent gas flow \citep{Larson81} and the coefficient $\sigma_0$ is the line width of gas measured on a 1 pc scale. In the Milky Way disk, $\sigma_0 \approx 0.7~\mathrm{km~s}^{-1}$ \citep{Solomon87a}.  The empirical mass-radius relationship shows that populations of Milky Way GMCs are well-described with an average surface mass density, $\Sigma_\mathrm{mol} \approx 10^{2}~M_\odot~\mathrm{ pc}^{-2}$ with indications of a decline with galactocentric radius \citep[e.g.,][]{Solomon87a, Heyer09}. 

The correlations between the macroscopic properties of clouds are frequently reframed in terms of the {\it virial parameter} \citep{bertoldi92}:
\begin{equation}
    \alpha_\mathrm{vir} \equiv \frac{5\sigma_v^2 R}{GM} \propto \frac{\sigma_0^2}{\Sigma_\mathrm{mol}}
    \label{eq:virparam}
\end{equation}
where the latter proportionality holds for $c_1=0.5; c_2=2$.  The virial parameter describes the balance between the kinetic energy ($E_k$) and gravitational binding energy ($U_g$) of a GMC, so that $\alpha_\mathrm{vir}=2E_k/U_{g}$ for a uniform density isolated sphere. In the absence of external effects (including surface terms and tidal gravity), magnetic effects, order-unity differences in the coefficient due to global geometry and internal substructure and stratification, and changes over time, 
$\alpha_\mathrm{vir}\lesssim 2$ would indicate self-gravitation, though many of these effects are likely significant \citep{mckee_zweibel1992,dib07}. 

\begin{figure}[h!]
    \centering
    \includegraphics[width=\columnwidth]{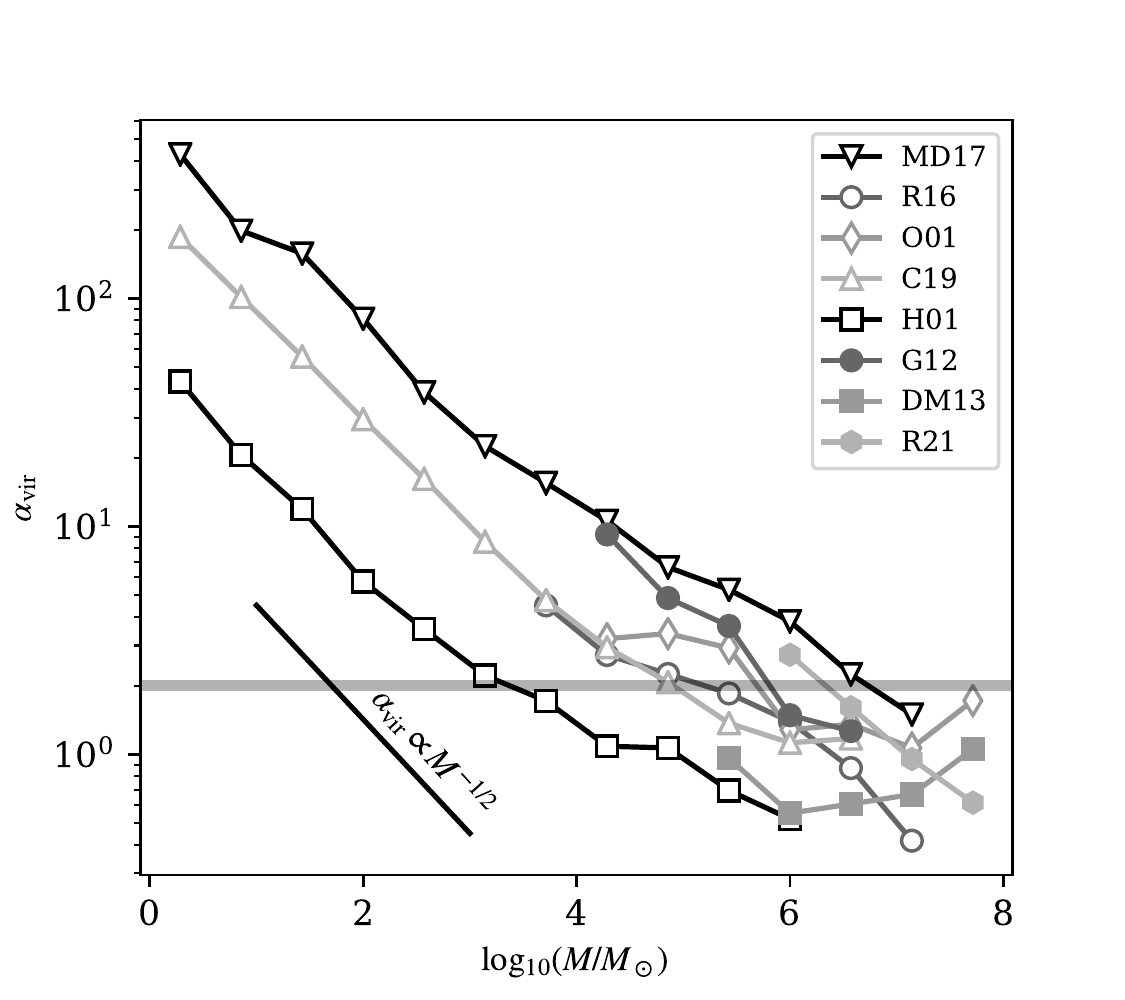}
    \caption{Median virial parameter as a function of CO-derived cloud mass over several cloud catalogues.  Milky Way surveys are indicated with open plotting symbols (MD17: \citealt{MivilleDeschenes17}, R16: \citealt{rice16}, O01: \citealt{oka01}, C19: \citealt{colombo19}, H01: \citealt{heyer01}) and extragalactic surveys are indicated with filled symbols (G12: \citealt{gratier12}, DM13: \citealt{DonovanMeyer13}, R21: \citealt{rosolowsky21}).  The horizontal line indicates the nominal division between bound ($\alpha_\mathrm{vir}<2$) and unbound clouds.  The relation $\alpha_\mathrm{vir}\propto M^{-1/2}$ shows the  scaling expected for observationally censored regions of this parameter space.}
    \label{fig:virparam}
\end{figure}

Figure \ref{fig:virparam} illustrates the average relationship between $\alpha_\mathrm{vir}$ and cloud mass aggregated from several Galactic and extragalactic surveys \citep[see also][]{evans2021}. Cloud mass is measured from the CO emission of a cloud, assuming a constant CO-to-H$_2$ conversion factor of $X_\mathrm{CO} = 2\times 10^{20}\mathrm{cm^{-2}/(K~km~s^{-1})}$ \citep{bolatto13} if unspecified.  Most surveys show $\alpha_\mathrm{vir} \propto M^{-0.5}$ \citep[e.g.,][]{MivilleDeschenes17}.  This scaling can arise from censoring, since this is the same scaling expected for the minimum detectable virial parameter at a fixed surface brightness sensitivity and velocity resolution.  Most CO surveys are not able to isolate low-mass, self-gravitating structures (i.e., small and dense) at large distances.  Such low-mass self-gravitating structures do exist (e.g., prestellar cores) but these are thought to form as substructures within larger, lower density clouds. While massive clouds show  $\alpha_\mathrm{vir}\sim 1$, several studies show $\alpha_\mathrm{vir}\gg 2$ for $M<10^{4}~M_\odot$.  

Finally, the cloud mass distribution function is typically described with a power-law form, 
\begin{equation}
\frac{dN}{dM} \propto M^{c_3}
\label{eq:massdist}
\end{equation}
over a range of GMC masses, where studies find $-2.5 < c_3 < -1.5$ and possible evidence for a truncation at the upper mass end \citep[e.g.,][]{rosolowsky05, RomanDuval10, colombo14, MivilleDeschenes17}.

\subsubsection{Recent Observational Work}

Since \textit{PPVI}, we have seen significant improvements in the quality of measurements of the molecular medium.  In the Milky Way, these improvements have come from new Galactic plane surveys (e.g., FUGIN, \citealt{umemoto17}, SEDIGISM, \citealt{schuller21}, COHRS, \citealt{dempsey13},  MWISP, \citealt{su19}), further development of dust mapping, and more sophisticated analysis approaches.  In extragalactic systems, ALMA and other interferometers have made whole-galaxy surveys of GMCs efficient, and galaxy-to-galaxy variations in the molecular ISM are becoming clear.  Finally, high resolution observations of the Galactic Center study molecular gas and star formation in an environment significantly different than the typical Milky Way environments (see Henshaw et al. chapter). 

Surveying the Milky Way beyond the solar neighborhood requires surveying CO emission, but dust extinction mapping is being extended to progressively larger volumes. \citet{Dame01a} compiled several low-resolution surveys of $^{12}\mathrm{CO}(1{-}0)$ emission into the first full-Galactic Plane map, decomposed into GMCs through dendrogram analysis in \citet{rice16}, and via Gaussian spectral decomposition in \citet{MivilleDeschenes17}.  The catalog of \citet{colombo19} uses spectral clustering \citep{colombo15} to find molecular clouds in the higher resolution data of the JCMT COHRS survey of $^{12}\mathrm{CO}(3{-}2)$ emission \citep[$15''$ vs.~$450''$ of the \citealt{Dame01a} survey;][]{dempsey13}.   Interpreting the emission in terms of molecular mass requires assuming a model for the CO-to-H$_2$ conversion factor.  Dust extinction is unaffected by uncertainties in the conversion factor, but there are (likely smaller) variations in the gas-to-dust ratio.  \citet{chen20} present a uniform catalog of molecular clouds measured from dust extinction and stellar distance measurements. 

Extragalactic studies exchange the excellent linear resolution of Milky Way studies for a vital external perspective, placing GMCs in their local environment. Achieving a sufficiently high linear resolution ($\sim 50~\mathrm{pc}$) to resolve individual molecular clouds at extragalactic distances usually requires interferometers observing CO lines, though studies of Local Group galaxies using single dish telescopes in CO and dust tracers can still reach useful resolutions. Extragalactic studies have been revolutionized by ALMA, which is capable of producing highly resolved maps of nearby galaxies in a few hours of observing.  The largest, uniform study of extragalactic CO in nearby galaxies is PHANGS-ALMA \citep{leroy21a,leroy21b}, which surveyed 90 nearby galaxies in $^{12}\mathrm{CO}(2{-}1)$ emission at 50-150 pc linear resolution.  \citet{DonovanMeyer13} presented observations of 5 nearby galaxies and \citet{rebolledo15} a further 3 targets, with both results using the CARMA interferometer.  Several other studies focus on GMC properties in single, unique targets including dwarf galaxies \citep{schruba17, faesi18, imara19}, elliptical galaxies \citep{utomo15, liu21}, systems hosting AGN \citep{tosaki17, miura21}, as well as starbursts and interacting systems \citep{Whitmore14, leroy15, brunetti21}.

Complementary to high resolution CO observations, mapping dense gas tracer species like HCN, HCO$^{+}$ or N$_2$H$^{+}$ provides an alternative measure of the conditions in molecular ISM, in particular the dense gas thought to be most closely related to star formation. Galactic surveys such as LEGO \citep{Kauffmann17a} and extragalactic studies like EMPIRE \citep{jimenezdonaire2019} map out the rotational lines of molecules with high dipole moments, which are preferentially excited in the densest regions of GMCs. Because the dense-gas emission is significantly fainter than the CO lines, these studies require significantly more observation time to map single targets, usually at lower resolution. On small scales, the relationship between emission from these species and gas conditions is complex, with spatially varying temperatures, densities and chemical abundances all leading to changes in emissivity of the dense gas tracers \citep{Onus18a, barnes20}. Moreover, examining the apparently linear link betweeen dense gas tracer emission and star formation \citep[e.g.,][]{Wu05} has shown to be more nuanced than previously assumed \citep{Longmore13, Usero15a}.

\subsubsection{Summary of Results}\label{sec:summary_properties}

Recent surveys have confirmed some of the early conclusions regarding molecular cloud macroscopic properties, and altered others. This work has shown that GMC properties follow the basic scaling relationships seen in the original Milky Way studies but that these coefficients depend on the environment of the clouds within their host galaxy. 

\paragraph{The surface density of molecular clouds varies throughout galaxies.} In the Galaxy, \citet{rice16}, \citet{MivilleDeschenes17} and \citet{colombo19} all emphasize the variation in cloud properties as seen in CO emission, noting a large range of surface densities of the molecular clouds ($2<\Sigma_\mathrm{mol}/(M_{\odot}~\mathrm{pc}^{-2})<300$) in the mass-radius relationship (Equation \ref{eq:scaling}) as a function of environment.  In particular GMCs in the molecular ring of the Galaxy ($R_\mathrm{gal}\sim 3~\mathrm{kpc}$) have significantly higher surface densities than clouds in the outer Galaxy.  However, \citet{lada20} argue that much of the variation in cloud surface mass density seen in the CO surveys can be explained by a CO-to-H$_2$ conversion factor that varies with galactocentric radius, though they note that such a model still requires a rise in the average cloud average surface density in the molecular ring.  In the Solar neighbourhood, \citet{chen20} find a cloud mass-radius scaling $M\propto R^{1.96}$ implying a near-constant average surface density of $\Sigma_\mathrm{mol}=50~M_\odot~\mathrm{pc}^{-2}$. Clouds in the Galactic center show substantially higher surface densities than the disk \citep{oka01}. Comparing Milky Way observations across the disk of the Galaxy confirm that molecular clouds show variable surface densities, even after accounting for conversion factor changes.

Extragalactic observations amplify the conclusion that $\Sigma_\mathrm{mol}$ is not universal,
particularly for clouds in the central $\sim 1$~kpc of galaxies \citep{oka01, leroy15, sun20a, rosolowsky21, liu21} which show $\Sigma_\mathrm{mol}\sim 10^{2.5}~M_{\odot}~\mathrm{pc}^{-2}$. Starburst galaxies show even higher gas surface densities, frequently approaching $\Sigma_\mathrm{mol}\sim 10^3~M_\odot~\mathrm{pc}^{-2}$ \citep{leroy15, pereira16, miura18}. 
This change in $\Sigma_\mathrm{mol}$ is clearest from extragalactic CO observations, but changes in $X_{\rm CO}$ and the unknown beam filling factors make a conclusive link to a changing physical state less certain.  However, the empirical conclusion is robust: {\it the average CO surface brightness of GMCs varies throughout galaxies when measured on $\sim 10^2$ pc scales and is generally higher where the kpc-scale gas surface density is higher.}.

\paragraph{A single size-line width relationship does not describe all GMCs.} \citet{heyer01}, \citet{Heyer09}, \citet{shetty12}, \citet{MivilleDeschenes17}, and \citet{colombo19} all point out that clouds in the molecule-rich regions of our Galaxy also show higher line widths when measured on a fixed spatial scale.  Extragalactic studies of molecule-bright regions, particularly in the nuclear regions of galaxies, also show elevated line widths for a fixed spatial scale \citep[e.g.,][]{sun20a, rosolowsky21, miura21, liu21} though the measured range of sizes is usually small, preventing an independent assessment of the size-line width relation.  Since the coefficient in the relationship ($\sigma_0$) is changing, steep size-line width relationships \citep[e.g.,][]{liu21} could reflect the standard scaling $c_1=0.5$ with a changing value of $\sigma_0$.   Recalling Equation 1, most studies show reasonable agreement with $\sigma_0^2 \propto \Sigma_\mathrm{mol}$ for GMCs.  Interpreting this in terms of Equation \ref{eq:virparam}, for GMCs ($M>10^{4.5}~\mathrm{M_\odot}$, which account for most of the molecular mass in any given galaxy), the virial parameter is approximately $\alpha_\mathrm{vir}\sim 2$ within a factor of $\sim 3$. This observation holds over a wide range of environments, but low mass clouds (Figure~\ref{fig:virparam}) and barred galaxy centers \citep{sun20a} show significantly higher values. While the detected low-mass clouds show higher line widths for a given surface density, \citet{MivilleDeschenes17} points out that clouds across all mass ranges show a good agreement with the relationship  $\sigma_v \propto (\Sigma R)^{0.43}$. Extragalactic studies also show $\sigma_0^2 \propto \Sigma_\mathrm{mol}$ for high mass GMCs \citep[e.g.,][]{sun20a, rosolowsky21} for GMCs in the disks of galaxies, and \citet{sun20a}
shows a range of $\alpha_\mathrm{vir} \sim 1 {-} 8$, with elevated values of $\sigma_0$ that are primarily associated with the centers of barred galaxies.

\paragraph{The changing physical properties of GMCs are linked to the galactic environments in which they reside.} Observational studies have connected changing GMC properties to locations (e.g.\ outer disks versus central regions), gravitational environment (e.g.\ global galactic potentials, tidal interactions, orbital shearing, epicyclic motions); morphological structures (e.g.\ spiral arms, interarm zones, and stellar bars); and other environmental factors (e.g.\ mid-plane disk pressures, ambient galactic radiation fields, and metallicities) -- see  \citet{Colombo18,meidt18,meidt20,Schruba18,Schruba19,sun20a,rosolowsky21}. GMC properties are also affected by large scale interactions. For example, GMCs in isolated star-forming galaxies differ from those in the compressed molecular zones in colliding merging galaxies \citep{Sun18}. The various environmental effects may vary in different galaxy types in the local Universe, and also at higher redshifts, especially around $z\sim 2$ when galaxy assembly and gas accretion from the intergalactic medium was at its peak \citep{Tacconi13,Tacconi20}.

These studies find the cloud scale surface densities, velocity dispersions, turbulent pressures, and virial parameters, tend to increase toward small galactocentric radii, especially in galaxies containing central stellar bars. This behavior may reflect the larger midplane vertical dynamical pressures in the inner regions \citep{Blitz2006,Sun20b}, in combination with streaming motions and gas accumulations along the bars, with velocity dispersions influenced also by the unbinding effects of the background stellar gravitational potentials \citep{meidt20}. The GMC surface densities are somewhat higher in spiral arms compared to the interarm zones. Finally, galaxies with higher stellar masses and star-formation rates contain GMCs with larger surface densities, higher velocities, and lower velocity dispersions \citep{sun20a}.

\paragraph{The mass (i.e., CO luminosity) distribution of GMCs follows a power-law distribution with index $c_3 \sim -2.0$ in Equation \ref{eq:massdist}}. Two CO-based studies of molecular gas have a good dynamic range in measuring the luminosity distribution ($>2$ orders of magnitude), combined with high linear resolution over a well-defined survey volume \citep{heyer01, colombo19}.  Both studies find $c_3 = -1.8$ within their uncertainties.  Low resolution studies have a limited dynamic range and produce more variation in their derived indices.  Even these high quality studies are subject to biases resulting from the fixed angular resolution projecting to different physical resolutions throughout their survey volume, which tend to result in underestimating the low-mass end of the distribution.  Hence, it is likely that $c_3 < -1.8$.

Extragalactic CO surveys usually recover a limited range in cloud masses and are subject to blending effects that limit the range of recovered cloud masses \citep{rosolowsky21}.  There is good evidence that the mass distribution of the molecular ISM changes with galactic environment, becoming top-heavy with more mass concentrated in high-mass structures in nuclear regions and galactic bars \citep{hughes13, freeman17, rosolowsky21}.  However, the precise functional form of the mass distribution is difficult to constrain and evidence for a high-mass truncation is marginal at best \citep{mok20}.  These studies have found $-2.5 < c_3 < -1.8$ in places where a reasonable estimate for the index can be made. 

\paragraph{Studies of dense-gas tracers imply that the density distribution of molecular gas changes with galactic environment.} The EMPIRE survey  \citet{jimenezdonaire2019} surveyed 9 nearby disk galaxies finding that the implied fraction of relatively dense molecular gas ($n_\mathrm{H_2}>10^4~\mathrm{cm^{-3}}$) changes as a function of galactic environment.  At small galactocentric distances, where the ISM pressures and stellar surface densities are higher, the dense gas emission is brighter relative to the CO emission than it is farther out in the galactic disk.  This suggests that clouds are becoming denser or warmer. However, the star formation efficiency associated with this apparently denser gas is lower than in the outer parts of disks.  This is broadly consistent with the trends seen in the Galactic center \citep[][and chapter by Henshaw et al.]{Longmore13}.

\subsection{Are clouds gravitationally bound or virialized?}\label{sec:bound_vir}

Several observational studies show that high-mass ($M>10^{4.5}~M_\odot$), large-scale structures in the ISM show virial parameters within a factor of 3 around $\alpha_\mathrm{vir} = 2$ (see Figure \ref{fig:virparam} and e.g. \citealt{Heyer09, Bolatto08, faesi18, Sun18}). 
From the pixel-based analysis of galaxies in the PHANGS-ALMA survey, \citet{sun20a} found that $\alpha_{\rm vir}$ covers a range $\sim 1-8$, corresponding to $\sigma \propto \alpha_{\rm vir}^{1/2}\Sigma_{\rm mol}^{1/2}$ for $\Sigma_{\rm mol}\sim 3-300 M_\odot \pc^{-2}$ on 150$\pc$ scale.
However, the observed virial parameter is a coarse measure of a cloud's dynamical state. 
At a minimum, finding $\alpha_\mathrm{vir}\sim 2$ indicates that self-gravity is dynamically important in GMCs, but even when $\alpha_\mathrm{vir}<1$ it does not confirm that clouds are strongly self-gravitating due to the simplifications in defining $\alpha_\mathrm{vir}$. 

The virial parameter is framed in terms of two components of the full virial theorem, and addressing whether the self-gravity of a cloud dominates the dynamics of a GMC requires constraining the additional volume and surface terms.  Theoretical studies have long argued that the surface terms cannot be neglected \citep{dib07}, and detailed studies of the surface terms, magnetic terms, and external effects in numerical simulations \citep[e.g.,][]{Kim_JG2021} confirm that these prevent interpreting gas self-gravity in terms of the simple virial parameter. 

Observational results also highlight the limitations in interpreting the simple virial parameter in terms of self-gravitation.  First, the velocity fields of GMCs are not isotropic but rather show large scale velocity gradients \citep{Bally87, Rosolowsky03a}.  Observations of gradients are ambiguous and can be interpreted as large-scale turbulent modes \citep{heyer04}, rotation \citep{braine18, braine20}, flows that can be the signatures of global collapse or inflow \citep[][and references therein]{Vazquez-Semadeni19} or cloud collisions \citep[e.g.,][]{muraoka20}.  Statistical studies suggest that GMC mass grows over the course of star formation \citep{Kawamura2009, lee16}, which would indicate that fluid flows, changing moments of inertia, and ram pressure could all make significant contributions to cloud energetics.  

Second, GMCs do not exist in isolation but are embedded in a 
gravitational
potential.   In nuclear regions of galaxies, the influence of the external galactic potential is large and can contribute to the significantly larger velocity dispersions, e.g., \citet{liu21} argue that shearing motions contribute significantly to cloud support.  Furthermore, the effect of the external galactic potential may be significantly larger than appreciated in disks as well \citep{meidt18, meidt20}.  When molecular clouds are concentrated in spiral arms (and spurs) in proximity to other clouds, the tidal force of neighboring GMCs can be comparable to or larger than the effects of the galactic potential \citep{Mao_2020}.  

Finally, magnetic fields remain a potentially significant contributor to cloud energetics.  The assessments of the field strength in molecular gas suggests that GMCs are supercritical so that magnetic fields do not dominate the energetics \citep{Crutcher12a}.  However, field geometry in the cloud, and the magnitude and geometry of the external magnetic field, all affect importance of magnetic fields for cloud support (or compression).  Measurements of field geometry have substantially improved since {\it PPVI} thanks to dust polarization maps from {\it Planck} \citep{planckpaper35, soler19} and ground based facilities that map out smaller-scale fields.  These observations show that, on average, magnetic fields are parallel to low-column-density linear structures in maps, but at high column densities this trend reverses and the magnetic field is perpendicular to linear (filamentary) structures.  However, measuring the field orientation does not by itself establish causality: the field could be weak and thus entrained in the fluid flow, or it could be strong and constrain the direction of the fluid flow.  \citet{hu19} interpret the orientation of the dust polarization and the velocity gradients of molecular tracers in the framework of magnetohydrodynamic turbulence to make estimates of the Alfv\'enic Mach numbers,  finding that $M_A\approx 1$ in the five clouds they consider.  \citet{heyer20} also examine the relative orientation of molecular gas features and polarization vectors and conclude that the Alfv\'enic Mach number varies regionally within the Taurus molecular cloud.  These recent observations all point to the magnetic field being significant enough that it cannot be neglected in energetics estimates.

In summary, we must regard the observational consensus that $\alpha_\mathrm{vir}\sim 2$ in GMCs with caution.  These observational results and theoretical expectations all suggest that the simple virial parameter is a crude tool for interpreting cloud dynamics. For example, the simulation work of \citet{Mao_2020} shows that owing to magnetic fields, internal structure/stratification, surface stresses, and external potentials, only a small fraction of cloud mass ($<10\%$) in clouds is actually bound despite clouds showing $\alpha_\mathrm{vir}\lesssim 2$ on large scales.  Despite these challenges, observationally establishing where GMCs are dominated by self-gravitation remains a critical question.  Star formation is gravitationally mediated and theory predicts that the degree of self-gravitation is one of the driving parameters in regulating star formation in GMCs (Section  \ref{sec:evolution}).  Quantifying surface terms, magnetic fields, and external potentials through observations will remain challenging, but future instruments and rigorous comparison to simulated observations provides promising routes forward (Section \ref{sec:future}).

\section{The Formation of GMCs}
\label{sec:formation}

GMC phenomenology provides clues to their lifecycles, including formation, evolution and associated star-formation, and disruption. We first discuss GMC accretion and condensation out of the larger scale and diffuse ISM.

\subsection{Accretion from the larger scale ISM}

GMCs represent massive, $\gtrsim 10^{4.5}$~M$_\odot$,  10-100 pc scale over-densities within the ISM, where conversion of the gas to
molecular form readily occurs and in which gravitational collapse and star-formation becomes possible. 
The appearance and formation of GMCs may proceed via a combination of processes that all lead to compression, cooling, and fragmentation of the more widely dispersed and volume filling
diffuse neutral and ionized hydrogen components of the ISM. Proposed GMC formation mechanisms, as discussed extensively in \citet{Dobbs14a}, include (a) gravitationally-induced compression of Jeans and/or Toomre
unstable regions, (b) converging filamentary gas flows driven by local turbulence, (c) shock wave-induced compression at the boundaries
of expanding supernova-driven shells and bubbles, (d) agglomeration via cloud-cloud collisions, (e) 
compression via galaxy spiral density waves, bars, and shearing motions, and (f) compression in large scale galaxy mergers. These various formation mechanisms are not fully distinct. For example, cloud-cloud collisions are a form of converging flows in which there is already pre-existing structure. Expanding supernova remnants are a large scale element of the overall turbulent flows in the ISM.
The time-scales associated with the various mechanisms differ \citep{Jeffreson18} and these may influence the overall
GMC lifetimes, and resulting star-formation efficiencies and cloud mass-functions across entire galaxies \citep{Inutsuka15,Kobayashi18}. Continued GMC formation offsets dispersal (see Section \ref{sec:destruction}) 
as part of the overall GMC lifecycle \citep{Chevance2020b}. An important advance in numerical modeling since {\it PPVI} is development of self-consistent large-scale ISM simulations (with and without spiral structure) that include both formation and destruction of GMCs, and evolve over hundreds of Myr so that a self-consistent overall ISM state and cloud population has time to develop \citep[e.g.][]{Fujimoto16a, KimOstriker17,Semenov17,Semenov_2021,Grisdale_2018,Grisdale_2019,benincasa20,Kim_WT2020,Smith_R2020,Tress_2020,Jeffreson_2020,Jeffreson_2021b}.

GMCs may retain imprints of the original filamentary and turbulent structures of the atomic gas 
from which they formed \citep{Fujii21}. Hydrodynamical simulations suggest the orientations of  dynamically
significant magnetic fields relative to converging atomic flows may control the GMC formation efficiency \citep{Iwasaki19}.
The various formation channels may or may not lead to gravitationally bound or virialized GMCs which may also be
transient or long-lived.
One reason it is difficult to distinguish among different mechanisms observationally is that large-scale equilibrium in the ISM naturally requires a rough balance between gravity and turbulence so that one is not dominant over another \citep[e.g.][]{KimOstriker17}.

\subsection{From neutral to molecular gas}\label{sec:neutral_to_mol}

A defining characteristic of GMCs is that the hydrogen gas is present mainly in molecular form as H$_2$. Conversion to H$_2$ is a necessary (though not sufficient) condition for the formation of CO molecules --
the primary observational tracers of GMCs via the rotational line emissions. For typical ISM gas densities, even at the high over-densities
of GMCs, H$_2$ is rapidly photodissociated by the ambient background stellar far-ultraviolet radiation
(Lyman-Werner band, 912-1108 \AA). Conversion to (and maintainance of) molecular form requires shielding, and this
is via a combination of dust attenuation of the far-ultraviolet (FUV), and molecular self-shielding as the H$_2$ absorption lines through
which the molecules photodissociate become optically thick \citep{Sternberg14}.  Dust grains are also important as the H$_2$ formation sites, via chemisorption and/or physisorption, with formation
rates that scale as the product of the dust-to-gas ratio and the overall density of hydrogen nucleons \citep{Wakelam17}. In 
regions that are optically thick to photodissociating radiation a substantial conversion to H$_2$ occurs,
although the molecules may continue to be destroyed by ionization and dissociation by penetrating low-energy cosmic rays \citep{Dalgarno06}. 

Conversion of H~\textsc{i} to H$_2$ is the initiating step in the production of CO molecules via the combinations of ion-molecule and neutral-neutral gas phase phase reactions involving the heavy atoms C and O. However, because of the lower abundances of C and O relative to hydrogen, the conversion to CO generally requires higher shielding columns. A substantial fraction of the H$_2$ may therefore be present in outer ionized C$^+$ and neutral C layers \citep{vanDishoeck88,Sternberg95,Hu21a} that are by definition ``CO dark". The relative sizes of the H$_2$ and CO rich zones affect the CO to H$_2$ mass conversion factors, $\alpha_{\rm {CO}}$ and $X_{\rm {CO}}$ \citep[e.g.,][]{bolatto13}.

An important timescale for chemical reactions in molecular clouds is the H~\textsc{i} / H$_2$ equilibration time, which is the time required for the H$_2$ abundance to reach balance between formation and dissociation.
This is the overall rate limiting time-scale for molecular cloud chemistry. For a gas consisting of just H~\textsc{i} and H$_2$ (with negligible H$^+$) the equilibration time is
\begin{equation}
\label{eq: teq}
t_{\rm eq} = \frac{1}{2Rn + D + \zeta}
\end{equation}
where $n=n_{\rm HI}+2n_{\rm H{_2}}$ is the number density of H nucleons (cm$^{-3}$), $R$ is the H$_2$ grain-surface formation rate coefficient (cm$^3$~s$^{-1}$),
$D$ is the {\it local}, possibly shielded photodissociation rate, and $\zeta$ is the cosmic-ray ionization rate.
Characteristically, $R=3\times 10^{-17}Z^\prime_d$~cm$^3$~s$^{-1}$ \citep{Wakelam17} where $Z^\prime_d$ is the dust-to-gas
ratio ($Z^\prime_d=1$ for standard Galactic conditions). The {\it unattenuated} photodissociation rate
$D_0=5.8\times 10^{-11}I_{\rm UV}$~s$^{-1}$ \citep{Sternberg14} where $I_{\rm UV}$ is the ultraviolet field strength ($I_{\rm UV}=1$
in the Galactic solar neighborhood). For Milky Way conditions, $\zeta\approx 10^{-16}$ s$^{-1}$ \citep{Dalgarno06} and H$_2$ removal by cosmic-rays is orders of magnitude less effective than by the unattenuated FUV field. 

In a steady state the balanced atomic to molecular density ratio is
\begin{equation}
\label{eq: eq}
\frac{n_{\rm {HI}}}{n_{\rm{H_2}}} = \frac{D + \zeta}{Rn} \ \ \ .
\end{equation}
For $D\gg Rn$, which is usually the case for unattenuated fields, the gas is primarily atomic in a steady state, and 
$t_{\rm eq}\approx 1/D$ is the photodissociation time $t_{\rm diss}$. The photodissociation time is independent of the gas density and is inversely proportional to the photodissociating field intensity. For unattenuated FUV radiation,
\begin{equation}
    t_{\rm diss} \approx \frac{5\times 10^2}{I_{\rm UV}} \ \ \ {\rm yr}
\end{equation}
In this regime
the equilibration time is short. If starting from a fully molecular state, $t_{\rm eq}$ is the short time required for (almost) complete photodissociation. If starting from a fully atomic gas $t_{\rm eq}$ is the time required to
build up the small steady-state abundance of H$_2$ molecules. For $D\ll Rn$, which is the case in shielded regions,
the gas is primarily molecular in steady state (unless $\zeta/Rn$ is large) and 
$t_{\rm eq}\approx 1/Rn$ is the
H$_2$ formation time $t_{\rm {H_2}}$. The H$_2$ formation time is inversely proportional to the product of the gas density and the dust-to-gas ratio, and is independent of the FUV field strength. In this limit
the equilibration time is long and equal to
\begin{equation}
t_{\rm{H_2}}\approx \frac{1}{Z^\prime_d}\times \frac{10^9 }{n} \ \ \  {\rm yr} \ \ .
\end{equation}
If starting from an atomic state it is the long time required for an
almost complete conversion to molecular form.  If starting from a fully molecular state, $t_{\rm eq}$ is the time required to slowly destroy some of the H$_2$ and produce hydrogen atoms in the shielded regions. Importantly, the H$_2$ formation time is longer for the lower dust-to-gas ratios that occur in low metallicity systems.

For a given gas density, Equations (\ref{eq: teq}) and (\ref{eq: eq}) set lower limits on the required attenuation factors, and hence the time-scales for GMC formation that by definition includes conversion to H$_2$. For example, for $n=100$~cm$^{-3}$, and $I_{\rm UV}=1$, an FUV attenuation factor of $5\times10^{-5}$
is required for conversion to at least 50\% H$_2$, and this takes 10$^7$ yr. For $Z^\prime_d=1$, the required shielding is provided by a combination of dust and molecular self-shielding
behind a total gas column $\sim 10^{21}$~cm$^2$.

In realistic clouds, the local gas densities, dust abundances, radiation fields, and shielding columns
may all be fluctuating in time and space, due to the large scale hydrodynamic filamentary flows, thermal and gravitational instabilities, and magnetically regulated turbulent motions. This
can give rise to time-varying and also out of steady state atomic and molecular abundances. For example, the ratio of the H$_2$ formation time to the free-fall time is
\begin{equation}
    \frac{t_{\rm{H_2}}}{t_{\rm ff}}\approx \frac{2}{Z^\prime_d}\left(\frac{n}{100~{\rm cm}^{-3}}\right)^{-1/2} \ \ \ .
\end{equation}
The time scales are comparable for $n=100$~cm$^{-3}$. Similarly, for a turbulent medium obeying the observed line-width size relation the eddy turnover time $t_{\rm turb}\approx 1\ {\rm Myr} \  ({L/{\rm pc}})^{1/2}$ so that 
\begin{equation}
    \frac{t_{\rm{H_2}}}{t_{\rm turb}}\approx \frac{1}{Z^\prime_d}\left(\frac{n}{10^3 \ {\rm cm}}\right)^{-1/2} \left(\frac{L}{{\rm pc}}\right)^{1/2} \ \ \ ,
\end{equation}
indicating that conversion to H$_2$ may be limited by the turbulent motions that induce density fluctations and cycle gas between shielded and unshielded locations 
\citep[e.g.][]{Bialy2017,Gong_2017}. GMC dynamics and chemistry are coupled, and the
 behavior is captured, at least in part, in recent hydrodynamical simulations \citep{Glover07,Padoan16,Richings16, Seifried20,Gong20,Hu21}.

\subsection{Hydrodynamics and Chemistry}

In recent years several simulation studies have appeared that focus on the incorporation of chemistry for H$_2$ and CO formation and destruction with the hydrodynamics of cloud evolution.  These simulations (hopefully) provide a more realistic picture of the GMC physical states and their coupled chemical compositions.  This includes predictions for time-dependent probability distribution functions for H$_2$ and CO surface densities, variable CO-to-H$_2$ mass conversion factors, CO-dark fractions, and effects of metallicity.

As part of the SILCC-Zoom project, \cite{Seifried20} present 0.1 pc resolution simulation results for molecular cloud formation. These use the AMR code FLASH with a MHD solver, with a simplified chemical network for the H$^+$, H~\textsc{i}, H$_2$, C$^+$, C and CO species \citep{Glover07}. The chemistry is time-dependent and computed on the fly. A background radiation field is included and dust and molecular self-shielding (for both H$_2$ and CO) are included using a multi-ray scheme for each simulation cell (see also \citealt{Safranek-Shrader17}). The resulting H$_2$ and CO distributions differ, with substantial diffuse H$_2$ and CO dark gas extending into low density
($\lesssim 10^3$~cm$^{-3}$) regions, with the CO rich condensations occurring in denser and cold ($\lesssim 50$~K) portions. A mean CO-to-H$_2$ conversion factor, $X_{\rm CO}\equiv N_{\rm{H_2}}/W_{\rm CO}$, equal to $\sim 1.5\times 10^{20}$~cm$^{-2}$ (K km s$^{-1}$)$^{-1}$ is found, consistent with observational estimates, but with a large scatter.  

\cite{Gong_18,Gong20} study the behavior of $X_{\rm CO}$, and the associated CO-dark mass fractions, in a suite of 3D MHD TIGRESS simulations \citep{KimOstriker17}  using the Athena code,  for gas surface densities $\sim 10 - 100\, M_\odot \, {\rm pc}^{-2}$. An evolving three-phase ISM structure (with star formation and resulting supernovae and photoelectric heating)  is computed for a shearing patch of the ISM disk, with H$_2$ and CO obtained for a range of assumed metallicities, background photodissociating and ionizing radiation fields, and cosmic ray ionization rates. The chemistry and radiative transfer are incorporated in post-processing given the MHD cloud structures and column densities. Several spatial resolutions are adopted for a convergence study.  The H$_2$ distributions are hydrodynamically converged since much of the H$_2$ is distributed in medium to low density gas.  CO is confined to the denser portions with CO masses that remain sensitive to the adopted sink particle mass thresholds, although $X_{\rm CO}$ is insensitive to resolution. The resulting $X_{\rm CO}$ factors (for the CO $J=$1-0 and 2-1 transitions) decrease with metallicity, due to increased radiative trapping and excitation temperatures. The conversion factors are also smaller for larger cosmic-ray ionization rates due to the enhanced heating and again larger resulting excitation temperatures. There is much less sensitivity to the FUV field intensity since most of the CO is built up in shielded gas, and with abundances consistent with a steady state since the densities are high.

\cite{Hu21} carry out a coupled hydrodynamical and chemical simulation for a supernova driven self-regulated ISM. The gravitational interactions and hydrodynamics are computed using GIZMO \citep{Hopkins15} together with the HealPix algorithm \citep{Gorski05} for estimating the dust and molecular self-shielding along lines of sight to each gas particle. The H~\textsc{i}/H$_2$ balance is computed time-dependently on the fly, and alternatively assuming a steady state at each point.  For both options, the C$^+$/C/CO chemistry is solved in post-processing, given the H~\textsc{i} and H$_2$ abundances. This is done for overall metallicties, $Z^\prime$, ranging from 0.1 to 3 times solar. As in the other simulations, the H$_2$ is produced over a wider range of gas densities than is CO. The H$_2$ cloud density PDF peaks at $n=20$ to 400 cm$^{-3}$ for $Z^\prime = 3$ down to 0.1, whereas the CO density PDF peak ranges from 300 to $6\times 10^4$~cm$^{-3}$. The long molecular formation timescales, especially at low metallicities, result in lower H$_2$ masses compared to steady state in the diffuse components \citep[see also][]{Krumholz11b, Gong_18}.

\cite{Smith_R2020} (``Cloud Factory") carry out AREPO computations including gravity and supernova feedback to study the separate effects of large scale galactic potentials (including shear) versus smaller scale behavior including cloud collapse and feedback, on conversion to H$_2$ and CO. They follow the growth of filamentary networks emerging from the large scale evolution. Local (clustered) feedback enhances the H$_2$ gas fractions at low densities compared to cloud formation driven only by the large scale potentials.

\cite{Jeffreson_2021a} carry out AREPO simulations of isolated galaxy disks, at spatial resolutions spanning 10 pc to 1 kpc, with time resolutions of 1~Myr.  They are able to track the dynamical evolution of 80,000 individual clouds over a wide range of masses, in environments probing the Toomre~$Q$ parameter, shearing properties, and mid-plane pressure. Most of the low mass clouds interact and merge with other clouds, at rates that are consistent with the crossing time in a supersonic turbulent ISM with a fractal structure. Remarkably, the physical properties of the interacting clouds do not differ from those that evolve in isolation.

A recent example of chemical post processing of hydrodynamical simulations that also includes self consistent radiative transfer and cooling is presented in \cite{Armillotta20}. They carry out GIZMO simulations of the Milky Way Central Molecular Zone (CMZ). The temporal resolution is sufficient to follow the formation of dense gas concentrations on their trajectories within the rings. Much of the dense gas remains unbound, consistent with the relatively low star-formation rates.  Molecular line maps for CO and also the dense gas tracers, NH$_3$ and HCN are computed in post-processing using the DESPOTIC tool \citep{Krumholz14a} for coupled chemistry and NLTE line transfer.

Hydrodynamic simulations are expensive, inspiring efforts to leverage the information they provide. A common resulting feature of the simulations are tight power-law correlations, $N_{\rm eff} \sim n^{0.3-0.4}$ between the
volume gas densities $n$ in each hydro cell (or particle) and the
appropriately angle-averaged gas column density, $N_{\rm eff}$, surrounding each volume element. With this in mind, \cite{Bisbas19} suggest that the chemical mass partitions (H~\textsc{i}/H$_2$, C$^+$/C/CO) of the hydrodynamical simulations may be mimicked in two simple steps. First, assume a volume density PDF, e.g.~a log-normal as expected for a turbulent medium (and with a tail if gravity is included) and then map this to a column density PDF using the correlation between $n$ and $N_{\rm eff}$ found in simulations. Second, extract the associated H$_2$ and CO gas columns from preexisting sets of ``classical" PDR computations for the given $N_{\rm eff}$ and $n$, background UV field strength and cosmic-ray ionization rate. A limitation of this method is that steady-state molecular abundances are assumed.

\subsection{The role of galactic environment}
\label{sec:galactic_environment}

Environment and location affect a wide range of GMC properties. This includes the GMC masses and surface densities, cloud mass spectra, the overall dense gas to stellar mass fractions, turbulent velocity dispersions, viral parameters, internal pressures, depletion times, angular momenta, cloud shapes, and chemical properties including especially the H~\textsc{i}/H$_2$ and CO/H$_2$ partitions; Sections \ref{sec:summary_properties} and \ref{sec:bound_vir} summarize recent findings on environmental dependence of GMC properties.

Because dust shielding against H$_2$ photodissociation depends on the dust-to-gas ratios, and overall metallicities $Z^\prime$, it is expected that H~\textsc{i} to H$_2$ mass ratios should vary inversely with $Z^\prime$ along lines of sight through galaxy disks. Evidence for such behavior is presented in \cite{Schruba18} in a study of H~\textsc{i}, H$_2$, and metallicity gradients across many sightlines in a large sample of local galaxies, at 50~pc to ~kpc scales \citep[see also][]{Wong2013}. The H$_2$ column densities are estimated via observations of the star-formation rates and assuming a constant molecular gas depletion time,
i.e.~$\Sigma_{\rm H_2} = \Sigma_{\rm SFR}/t_{\rm dep}$, rather than via CO data and an adopted CO/H$_2$ conversion factor. For any $Z'$ the measured H~\textsc{i} columns saturate for sufficiently large total (H~\textsc{i}+2H$_2$) gas columns.  The maximal saturation columns do indeed vary inversely with $Z^\prime$ as expected for shielding based theories of atomic to molecular conversions.  This supports the notion that the metallicity dependent H~\textsc{i}-to-H$_2$ chemical conversion time is a lower limit on GMC formation times.

\section{The Evolution of GMCs}
\label{sec:evolution}

In this section, we explore how the gas evolves after the assembly of GMCs, collapsing to form stars and eventually dispersing. 
We can mark the boundary between the formation and evolution phases by the onset of collapse and star formation, but we emphasize that this is far from a sharp line. Because GMCs contain a huge range of densities and thus of dynamical times, the densest structures within them begin to collapse and form stars while accretion of lower-density material is still underway, and it is entirely possible that accretion of new gas occurs throughout a GMCs's lifetime.
Nonetheless, because the main evolution phase is distinguished from the assembly phase by the onset of collapse, the central questions with which we will be concerned in this section are, first, is collapse localized to small parts of GMCs, or is most of the mass in GMCs collapsing (Section \ref{sec:collapse})? How long after molecules first become detectable does it take for star formation to begin, and how long does star formation go on before the accumulated feedback is sufficient to disrupt the cloud (Section \ref{sec:lifetime})? In those regions that are collapsing, how rapidly does star formation proceed (Section \ref{sec:cloud_SFR})? We defer questions regarding the integrated outcome of GMC evolution -- in particular, the time-integrated (rather than instantaneous) efficiency of star formation, 
and the dynamical properties and spatial arrangement of those stars, to Section \ref{sec:accomplishment}.

\subsection{Hierarchical versus global collapse of GMCs}
\label{sec:collapse}

Two main scenarios are commonly presented regarding the collapse of GMCs towards star formation.
In the first one, only a small fraction of the cloud (the densest regions) can collapse, giving birth to stars \citep[e.g.][]{Dobbs11b}.  
The second scenario suggests that clouds experience a global collapse \citep[e.g.][]{Vazquez-Semadeni17, Vazquez-Semadeni19, Elmegreen18}. 

On large scales in galaxies, the complex multi-scale system constituted by GMCs and the ISM appears to be in equilibrium, which may result from statistical averaging \citep[e.g.][]{Ostriker10,Ostriker11, Krumholz18, Sun20b}. On the cloud-scale, extragalactic observations reveal clouds close to virial equilibrium (see Section~\ref{sec:bound_vir} and Figure~\ref{fig:virparam}). However, a difficulty to observationally distinguish between the above two scenarios comes from the fact that the energy signature of virial equilibrium ($\alpha_{\rm vir} \sim 2$) is similar to gravitational collapse \citep[e.g.][]{BallesterosParedes11}, often to within the observational uncertainties. On sub-cloud scales, non-linear instabilities and turbulence-driven structure lead to non-homogeneous GMCs \citep[e.g.][]{BallesterosParedes20}. As a result, GMC structure is filamentary, clumpy, and the small scale structures can collapse and form stars even though the large scale structure is not necessarily bound \citep[][]{Hacar2013, Henshaw14, Henshaw16b, Clarke2017}.
The question is then to understand what fraction of the gas is self-gravitating. 

A central challenge for both the local and global collapse scenarios is explaining why the efficiency of star formation in GMCs is low, both on the basis of fraction of mass converted to stars per free-fall time (as discussed below in Section \ref{sec:cloud_SFR}) and on the basis of the total fraction of cloud mass converted to stars over a cloud lifetime (Section \ref{sec:sfe_obs}).
The local collapse scenario explains the low star formation efficiency of GMCs by the fact that clouds are mostly unbound and only a small fraction of the gas actually satisfies the conditions to form stars \citep[e.g.][]{Dobbs11b, Semenov17, Semenov18, meidt18, meidt20,Mao_2020}.

In the second scenario of global hierarchical collapse, presented by \citet{Vazquez-Semadeni19}, self-gravity is driving motions within clouds and all scales accrete from their parent structures. This process is non-linear and starts very slowly, with no star formation during the first Myr, and accelerates progressively until the bulk of the cloud material is collapsing.
This global (hierarchical) collapse generates non-thermal motions within clouds and provides a possible interpretation for the origin of the large linewidths of GMCs \citep[e.g.][]{Heitsch09, BallesterosParedes11, Zamora-Aviles12a, Traficante18}, which are expected to dissipate quickly if not sustained by external or internal mechanisms \citep[e.g.][]{Dobbs14a}.
In this scenario, only the destruction of the cloud by feedback prevents the star formation efficiency from rising to values that exceed observational constraints.

Observational evidence based on gas and stellar kinematics in favor of one scenario or the other remains mixed.
The global collapse scenario seems to be supported by recent observations of filamentary accretion flows, up to several parsec long, feeding a central clump \citep[e.g.][]{Kirk13, Peretto14, Lu18, Chen19, Shimajiri19}, and signs of a flow directed towards these filaments further away from the central hub. More direct evidence of infall signatures have been observed around clumps \citep{Barnes18} and GMCs \citep{Schneider_2015b}, as revealed by a systematic shift between the $^{12}$CO and $^{13}$CO emission lines.
However, these dense clouds represent only a small part of the hierarchy of scales within GMCs and it remains unclear whether the collapse extends to GMC scales \citep[e.g.][]{Henshaw16b, Henshaw20}.

On the other hand, stellar kinematics for the most part appear to be inconsistent with global collapse. In this scenario almost all stellar velocity vectors should point radially towards or away from a dense, bound cluster that is the collapse center, while in a scenario where collapse is local, much less organisation is expected.
Recent \textit{Gaia} observations have largely failed to detect the radial signature expected for global collapse
\citep[][]{Kounkel18, Ward18, Kuhn19, Dzib21a}.
On the contrary, observations of 109 OB associations shows that they are not the relic of expanding clusters but were formed with low levels of expansion, tracing the fractal structure of their parent GMC \citep{Ward20_gaia}.
There are, however, some counter-examples that do appear to show signs of radial expansion
\citep[e.g.,][]{Lim20a, Swiggum21a}.

Finally, in both scenarios, it is plausible that part of the gas gets in fact disrupted by stellar feedback before it gets a chance to collapse. This will be discussed in Section~\ref{sec:destruction}. In the following subsection, we investigate the duration over which cloud collapse and star formation take place, and explore the possibility that the low star formation efficiency per free fall time is compensated by an extended duration of the star formation.

\subsection{Durations of the inert, star-forming, and dispersing phases of GMCs}
\label{sec:lifetime}

\begin{figure*}[t]
    \centering
    \includegraphics[width=\textwidth, trim=0mm 40mm 0mm 40mm, clip]{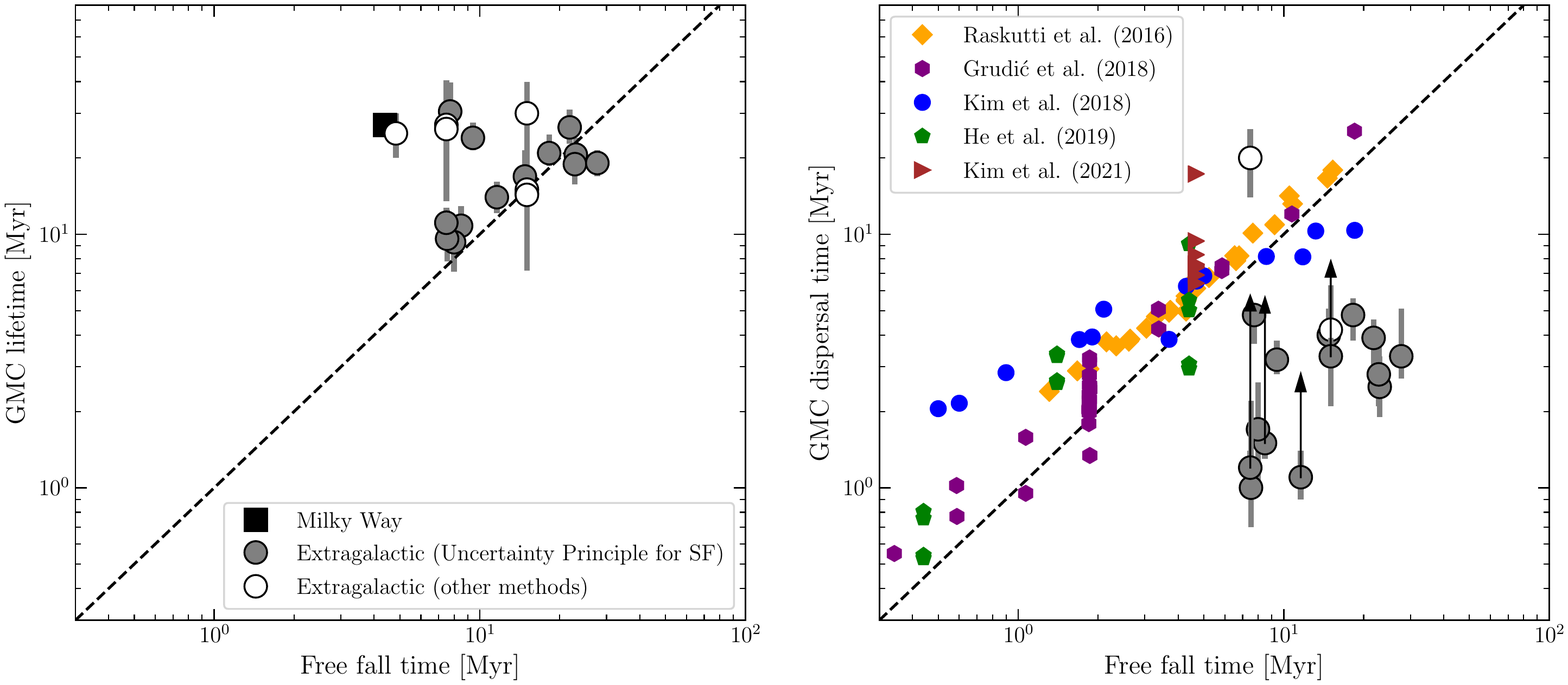}
    \caption{Measured GMC lifetime (\textit{left}) and GMC dispersal time (\textit{right}) in the Milky Way \citep[black square;][]{Murray11} and nearby galaxies, using the uncertainty principle for star formation \citep[grey circles;][]{Kruijssen2019, Chevance2020a, zabel2020, KimChevance21} and other analysis methods \citep[white circles;][]{Engargiola03, Blitz2007, Kawamura2009, miura12, Corbelli17, Meidt15}. The dispersal time is measured from the moment where H$\alpha$ emission from massive stars becomes visible. This neglects a potential phase of dust-obscured massive star formation. In the most nearby galaxies, \cite{KimChevance21} measured the duration of this embedded phase using 24$\mu$m emission. This increases the GMC dispersal time as indicated by the black arrows. In addition, beam dilution might results in longer measured cloud free-fall times. These two effects likely explain the longer duration of the GMC dispersal time found in simulations \citep{Raskutti2016, Grudic18a, Kim_JG2018, He_Ricotti2019, Kim_JG2021}, measured from the formation of the first star.}
    \label{fig:timescales}
\end{figure*}

In a self-gravitating system such as a molecular cloud, the natural reference timescale is the free-fall time,
\begin{equation}
    t_{\rm ff} = \sqrt{\frac{3 \pi}{32 G \rho}},
\end{equation}
where $G$ is the gravitational constant and $\rho$ the density of the gas. This is the minimum timescale required for a spherical object to collapse under the influence of self-gravity only.
The free-fall time therefore represents a lower limit on the GMC collapse time,  which can be lengthened by several causes. For example, the non-spherical geometry of the clouds (likely filamentary) extends the minimum duration of the collapse time \citep[e.g.][]{Toala12, Pon12}. The presence of magnetic fields \citep[e.g.][]{Heitsch01, Padoan11, Federrath2012, Girichidis18, Kim_JG2021} may contribute to a longer contraction time and a reduced star formation efficiency per free-fall time. However, this effect stays relatively limited in the case of magnetically supercritical clouds \citep{Crutcher12a}. Shear induced by galactic differential rotation can also potentially lengthen the duration of the collapse \citep[e.g.][]{Dobbs14a, meidt18, meidt20, Jeffreson18}. Turbulence can also provide support against collapse \citep[][]{Klessen00a, Krumholz05a, Padoan11, Hennebelle11, Dobbs14a, Burkhart18}, but because the turbulent dissipation time is comparable to the crossing time \citep{Stone1998,maclow1999}, collapse still commences on a timescale comparable to the free-fall time in magnetically supercritical clouds \citep[e.g.][]{Ostriker99a, Klessen00a, Kim_JG2021}. Finally, the lifetime of a cloud can be extended due to ongoing accretion of low-density material \citep[e.g.][]{Smilgys2017, Rieder2022}, which if it occurs fast enough can replace the gas that collapses or is ejected by feedback, maintaining a roughly constant or even increasing gas mass despite ongoing collapse \citep[e.g.,][]{Goldbaum11a, Feldmann11a}.

Observationally, indirect methods are required to measure the lifetimes of GMCs. Over the past decade, several statistical methods have been developed, leading to a great variety of results. Some of these approaches are based on object classification \citep[e.g.][]{Kawamura2009, Corbelli17}, others on stellar age spreads (which exclude a potential inert cloud phase; e.g. \citealt{Hartmann01, Grasha18, Grasha19, Hannon19, Messa21}), and yet another group based on following GMC evolution along orbital streamlines \citep[e.g][]{Engargiola03, Meidt15, Kruijssen15, Henshaw16a, Barnes17}. The dissimilarities between these approaches and the lack of homogeneous large data sets has hindered a systematic determination of the GMC evolutionary cycle as a function of the galactic environment.

This is now changing thanks to the advent of large surveys of the molecular ISM enabled by ALMA \citep[e.g. PHANGS,][]{leroy21a}, major progress in numerical simulations of star formation on the cloud scale \citep{Dale15, Walch15, Grudic18a, Haid18, Kim_JG2018,  Semenov18, Kim_JG2021}, and the development of novel analysis frameworks \citep[e.g the `uncertainty principle for star formation';][]{Kruijssen2014, Kruijssen2018}. High resolution observations of the GMC and star-forming region populations in nearby galaxies reveal that tracers of the molecular gas (e.g.~CO) rarely coincide with tracers of star formation (e.g.~H$\alpha$) on the cloud scale ($\sim 100$\,pc; \citealt{Kreckel2018, Kruijssen2019, Schinnerer2019}). This small-scale spatial decorrelation appears to be universal in nearby star-forming galaxies and is interpreted as a sign of an evolutionary cycling between GMCs and young stellar regions \citep[e.g.][]{Feldmann11b, Kruijssen2014, Kruijssen2018, Semenov_2021}. In practice, the observed variation of molecular gas-to-ionizing-radiation flux ratio
as a function of spatial scale when focussing on gas clouds or stellar regions can be fitted by an analytical model \citep{Kruijssen2018} to measure the durations of the successive phases of cloud evolution and star formation (see also Section \ref{sec:obs-sim} and Figure~\ref{fig:simobs}). This decorrelation implies a rapid evolutionary lifecycle, with short observed GMC lifetimes of $\sim 10-30$\,Myr (with variation within and between galaxies) and fast gas dispersal by stellar feedback (within 1-5\,Myr after the onset of massive star formation; \citealt{Kruijssen2019, Chevance2020a, zabel2020}). This is illustrated in Figure~\ref{fig:timescales}.

The mechanisms setting the GMC lifetime can be internal or external and are likely environmentally dependent \citep[e.g.][]{Jeffreson18}. While the GMC lifetime matches the cloud crossing time or free-fall time in low global gas surface density galaxies ($\leq 8$\,M$_{\odot}$\,pc$^{-2}$), the lifetime is instead similar to the free-fall time at the average midplane density (which is comparable to the turbulent crossing time of ISM gas) in denser environments \citep{Chevance2020a, ward2020}, suggesting that external processes might regulate the cloud lifetime there. 
GMCs are observed to have virial parameters of $\alpha_{\rm vir} = 2-10$ in nearby galaxies (see Sections~\ref{sec:gmc_env},\ref{sec:bound_vir}), so they might evolve on a crossing time rather than a free-fall time. But the small difference between these two timescales ($\alpha_{\rm vir}=2-10$ corresponds to $t_{\rm ff}/t_{\rm dyn} = 0.7 - 1.6$) makes it challenging to distinguish observationally (see also \citealt{BallesterosParedes11} for a comparison between free fall velocity and virial velocity).

By contrast, dense star-forming regions within GMCs in the Milky-Way show signs of extended star formation, implying either that collapse takes longer than a free-fall time or that these objects accrete for extended periods, replenishing gas that is lost to collapse. The most prominent example is the Orion Nebula Cluster where most stars are significantly older than a free-fall time \citep{DaRio14, DaRio16, Kounkel18}, and the stellar velocity distribution is close to virialized \citep{Kim19a, Theissen21a}. 
The key question is then how star formation proceeds over the course of the age spread. Is the star formation rate constant, does it decline, or does it accelerate? Several observations indicate that the star formation seems to be gradually accelerating over several million years, being consistent with a $t^2$ power law \citep[e.g.][]{CaldwellChang18}, before a steep decline (as in Upper Scorpius; \citealt{PallaStahler00}). This acceleration has first been interpreted as a result of the contraction of the parent molecular cloud \citep[e.g.][]{PallaStahler99, PallaStahler00}.
However, it is important to note here that the free fall time of an isolated structure gets smaller as this structure collapses and the density increases \citep[e.g.][]{Grudic18a, KrumholzARAA19, Vazquez-Semadeni19, Krumholz20}. Observationally, the instantaneous free fall time might therefore be shorter than the one when the first stars formed, which could explain these seemingly large age spreads. 
\citet{KrumholzARAA19} point out that most simulations of isolated clouds do not reproduce these large stellar age spreads, predicting a typical star formation duration of $\sim 5$\,Myr (i.e. a free fall time; \citealt{Krumholz12a, Kim_JG2018, Grudic18a}). The lack of large scale environment most likely explains this discrepancy \citep[e.g.][]{Jeffreson_2020}. Inflows along filaments (see Section \ref{sec:collapse}) replenish the gas reservoir and maintain the turbulent motions within clouds \citep{Klessen10, Goldbaum11a, Matzner15, LeeHennebelle16a, LeeHennebelle16b}. This is what happens in the `conveyor belt' picture of massive cluster formation \citep[e.g.][and references therein]{Longmore14, Motte18, Krumholz20}, in which originally quiescent, low-density gas is accreted from scales much larger than the resulting cluster. This mechanism is governed by large-scale processes, which can take place over timescales much longer than the free fall time of the central dense clump. If the accretion is faster than the consumption of gas (i.e. if the star formation efficiency is low), the central region can grow in mass, leading to accelerating star formation. The influence of feedback processes on small scales is also a critical element for simulations to take into account (see Section \ref{sec:destruction}), as they can slow down accretion and subsequently increase the duration of star formation.
Observationally, future JWST observations of the dust-obscured star formation in combination with GMC, young H~\textsc{ii} region observations and age-dated star clusters, will enable us to compare the durations and efficiencies of the earlier (embedded) and later (exposed) stages of star formation, and probe a potential acceleration of star formation (see Section \ref{sec:future}).

It remains debated whether all GMCs form stars or if some of them are dispersed (e.g. by galactic dynamical processes) before forming stars. In nearby galaxies, the short duration of the measured CO-visibility timescale \citep[about one dynamical time --][]{Kruijssen2019, Chevance2020a} does not seem to allow for successive episodes of cloud formation and cloud destruction without star formation. This is also observed in simulations \citep[e.g.][]{Jeffreson_2020}.
In addition, in the Milky Way, it is relatively rare to find clouds associated with no star formation at all, although the exact fraction depends on the selection method used to define clouds and the criteria adopted for the presence or absence of star formation; for example, \citet{Wilcock12a} find that $\approx 20\%$ of infrared dark clouds identified with \textit{Spitzer} contain no detectable point sources at 8 or 24 $\mu$m. Clouds are often associated with some level of low-mass star formation, which can happen simultaneously with continuous gas accretion. We note however that feedback from low-mass stars is likely insufficient to disperse these clouds. They might either merge into a higher mass cloud and be dispersed by high-mass stars, or be dispersed by feedback from massive stars in a nearby, cluster-forming cloud (by superbubbles expanding over several hundreds of parsecs).  In extragalactic observations, this low-mass star formation is typically not visible, and a significant fraction of molecular clouds are seen to be unassociated with massive star formation, including embedded star formation as traced by 24 $\mu$m emission \citep{KimChevance21}. This `inert' phase could last for 50 to 80\% of the cloud lifetime. However, these are still early results, and JWST will be crucial to systematically quantify the time over which GMCs are not forming stars, and how the duration of this phase depends on the environment (see Section \ref{sec:future}).

Finally, cloud collapse will end when the energy injected by feedback from massive stars reverses the gas flow, dispersing the remaining gas. Several observational studies, using a variety of methodological approaches, have shown that young stellar regions become unassociated with molecular gas a few Myr after the onset of star formation, both in Galactic and extragalactic observations \citep[e.g.][]{Leisawitz_1989,Whitmore14, Hollyhead15, Grasha18, Grasha19, Hannon19, Chevance2020a, Chevance2020b, Haydon20a, Messa21}. The mechanism(s) through which this happens, and how long this phase lasts, are discussed in Section~\ref{sec:destruction}. We note that the GMC dispersal time is included in the GMC lifetime as measured in Figure~\ref{fig:timescales} and likely represents a small fraction of the total cloud lifetime \citep[$\sim$ 10 to 25\% of the cloud lifetime, e.g.][]{Kruijssen2019, Chevance2020a}.

\subsection{Cloud scale star formation}\label{sec:cloud_SFR}

We next turn to the question of how quickly GMCs convert their gas to stars. Since the free-fall time is the natural timescale for a gravity driven-process like star formation, the most natural parameterization for the star formation rate is the efficiency per free-fall time $\epsilon_{\rm ff}$ \citep{Krumholz05a}; for a cloud of mass $M_{\rm gas}$ and free-fall time $t_{\rm ff}$, which forms stars at a rate $\dot{M}_*$, this quantity is defined as
\begin{equation}
    \epsilon_{\rm ff} = \frac{\dot{M}_*}{M_{\rm gas}/t_{\rm ff}}.
    \label{eq:epsff}
\end{equation}
The value of $\epsilon_{\rm ff}$ plays a major role in determining the evolutionary timescales of clouds and the net, lifetime-integrated efficiencies with which they produce stars and star clusters (Section \ref{sec:accomplishment}).

The value of $\epsilon_{\rm ff}$ can both be determined by observations and estimated from theory and simulations. Theoretical models predict that $\epsilon_{\rm ff}$ should depend strongly on cloud virial parameter $\alpha_{\rm vir}$, and somewhat more weakly on Mach number and strength of magnetization \citep{Clark04a, Krumholz05a, Clark08a, Hennebelle11, Bonnell11a, Federrath2012, Padoan12, Hopkins13b}. The former matters because it parameterizes the competition between gravity causing overdense structures to contract and turbulence leading them to disperse, while the latter matters because it affects the density distribution produced by turbulence. Simulations of isolated GMCs confirm an approximately exponential decrease of $\epsilon_{\rm ff}$ with increasing $\alpha_{\rm vir}^{1/2}$ \citep{Padoan12, Kim_JG2021}.

While there is general agreement on the dependence on the virial parameter, there is less on the absolute value of $\epsilon_{\rm ff}$. Early simulations that included limited or no feedback generally found $\epsilon_{\rm ff} \gtrsim 0.3$, leading to theoretical models where star formation in bound gas is assumed to be efficient \citep[e.g.,][]{Heitsch09, BallesterosParedes11,Hartmann12}. However, as simulation physics has improved, estimates of $\epsilon_{\rm ff}$ have generally come down. Simulations in periodic boxes (intended to represent the interiors of molecular clouds) and that include magnetic fields and feedback in the form of protostellar outflows and radiation tend to produce $\epsilon_{\rm ff} \sim 0.01 - 0.05$ \citep{Myers14a, Federrath15b, Cunningham18a}. Simulations of idealized, isolated GMCs tend to find somewhat higher $\epsilon_{\rm ff}$, in some cases where feedback is ineffective reaching values as high as in the earlier pre-feedback simulations \citep{Grudic18a}, but simulations using initial surface densities, virial parameters, and magnetic field strengths comparable to those of observed GMCs tend to give $\epsilon_{\rm ff} \lesssim 0.1$ \citep{Kim_JG2021}. Simulations of GMCs embedded in full galactic simulations, which capture environmental effects but at the price of lower resolution and heavier reliance on subgrid treatments of feedback, generally give median values of $\epsilon_{\rm ff} \sim 0.01$, albeit with scatters of $\sim 0.5$ dex \citep{Semenov16a, Grisdale_2019, Grisdale21a}.

An observational determination of $\epsilon_{\rm ff}$ requires measurements of the three quantities on the right hand side of Equation \ref{eq:epsff}, each with
its own uncertainties and problems. Measuring $t_{\rm ff}$ 
depends on the volume density, a quantity that is intrinsically difficult to measure from projected, 2D data. Strategies for estimating density and thus $t_{\rm ff}$ depend on the type of data available. For individual clouds, a common approach is to assume that the unseen third dimension is comparable in size to the two visible ones \citep[e.g.,][]{Krumholz12a, Evans14, lee16, Vutisalchavakul16, Heyer16a, Ochsendorf17, Schruba19, Pokhrel21}, though this likely incurs both a systematic error and a scatter of a few tenths of a dex \citep{Hu21a, Hu22a}. For measurements made on galactic scales, it is common to either assume a fixed disk scale height or to estimate the scale height from hydrostatic balance \citep[e.g.,][]{Leroy17, Utomo18}, again leading to a few tenths of a dex difference depending on the exact assumptions made. Yet a third approach  is to measure masses with molecular tracers (most commonly HCN) that, for excitation reasons, select gas in a particular density range \citep[e.g.,][]{Usero15a, Stephens16a, Gallagher18a, Onus18a, jimenezdonaire2019}, though again there is a few tenths of a dex uncertainty on the density that such tracers select \citep[e.g.,][]{Kauffmann17a, Onus18a}. 

Measurement of both $M_{\rm gas}$ and $\dot{M}_*$ has also proven challenging due to evolutionary effects. The most common tracer of the star formation rate available in extragalactic systems -- ionizing luminosity -- only measures the mean star formation rate over the last $\approx 3-5$ Myr \citep[e.g.,][]{Krumholz07b}, comparable to the time inferred for GMCs to disperse (Section \ref{sec:lifetime}). This creates a problem with making consistent measurements of $\dot{M}_*$, $M_{\rm gas}$, and $t_{\rm ff}$: the mass of gas at the time of observation may be much less than was present at the time the stars formed, while the cloud radius may be larger due to feedback, both of which would lead to an overestimate of $\epsilon_{\rm ff}$;  or the instantaneous star formation rate may be much larger than the time-averaged one, leading to an underestimate of $\epsilon_{\rm ff}$ \citep[e.g.,][]{Feldmann11a, Grudic19a, Kim_JG2021}. Averaged over many GMCs many of these errors cancel, but they do not do so cloud-by-cloud. Star formation rates inferred from direct counts of young stellar objects (YSOs), particularly class 0 YSOs, which only have $\approx 0.5$ Myr lifetimes, are much more reliable, but are only available for molecular clouds that are relatively nearby. 

For comparison of observed $\epsilon_{\rm ff}$ to predictions from theory and simulations, a further challenge is that the instantaneous value of $\alpha_{\rm vir}$ in a cloud which has experienced substantial feedback may be much larger than the value of $\alpha_{\rm vir}$ during the main epoch of star formation. This can even lead to a positive correlation between instantaneous observed measures of $\epsilon_{\rm ff}$ and $\alpha_{\rm vir}$ on a cloud-by-cloud basis, the opposite of the theoretically-expected negative correlation \citep{Kim_JG2021}.

Despite these caveats, a number of authors have published measurements of $\epsilon_{\rm ff}$ that paint a relatively consistent picture (see \citet{KrumholzARAA19} for a comprehensive compilation of measurements up to 2018). At the largest and most diffuse scales, $L \approx 100$ pc and $n\sim 10-100$ cm$^{-3}$, studies in nearby galaxies using CO to trace molecular gas and ionization or IR emission to trace star formation consistently find $\epsilon_{\rm ff} \sim 0.01$ \citep{Leroy17, Utomo18, Schruba19}. Studies of individual, spatially resolved GMCs ($L\sim 1-10$ pc, $n\sim 100 - 10^4$ cm$^{-3}$), using dust or CO to trace gas and YSO counts to measure star formation give similar values of $\epsilon_{\rm ff}$ \citep{Evans14, lee16, Vutisalchavakul16, Ochsendorf17, Pokhrel21}. Finally, at the scales of dense clumps traced by cold dust or HCN emission ($L \lesssim 1$ pc, $n\sim 10^4 - 10^5$ cm$^{-3}$), using IR emission or YSO counts as a star formation tracer, again observations find a median value $\epsilon_{\rm ff} \sim 0.01$ \citep{Heyer16a, Onus18a, Gallagher18a}. \textbf{Thus a robust conclusion from the observations is that the median value of $\epsilon_{\rm ff} \approx 0.01$, within $\approx 0.5$ dex, with no evidence for systematic variations over $\sim 4$ dex in density scale}, over scales ranging from large swathes of galactic disks to the dense clumps that are plausibly the progenitors of individual star clusters.

However, \textbf{there is substantial disagreement as to the amount of scatter about this median}, with different analysis methods leading to systematically different answers. Methods that rely on ionizing luminosity to estimate star formation rates on sub-galactic scales generally produce large dispersions of $0.5 - 0.8$ dex \citep{lee16, Vutisalchavakul16, Ochsendorf17}, while those that use direct counts of young stellar objects give much smaller dispersions of $0.2 - 0.3$ dex \citep{Evans14, Heyer16a}. In a particularly deep data set combining \textit{Herschel}-derived gas masses with class 0 YSO counts -- likely the best methods currently available -- \citet{Pokhrel21} and \citet{Hu22a} find only a $0.18$ dex cloud-to-cloud dispersion in the mean value of $\epsilon_{\rm ff}$ over 12 Milky Way GMCs, and only a $\approx 0.3$ dex dispersion within any individual cloud over a $>1$ decade range in surface density. Given the methodological problems with ionizing luminosity discussed above, the smaller dispersions are likely more reliable. However, it is possible that at least some of the difference in dispersion estimates is a result of the studies based on YSO counting being forced to sample a more limited range of environment than the ionization-based ones, since H~\textsc{ii} regions are visible considerably further than individual young stars.

A third point, that builds on the previous two, is that \textbf{there is thus far only very weak evidence for systematic variation of $\epsilon_{\rm ff}$ with GMC properties or environment}. Measured on galactic scales, \citet{Leroy17} report a weak anti-correlation between $\epsilon_{\rm ff}$ and gas surface density within M51, \citet{Utomo18} report a similar anti-correlation with total galactic mass, and \citet{Schruba19} report lower $\epsilon_{\rm ff}$ at higher $\alpha_{\rm vir}$. However, at least part of these effects are plausibly explained as arising from systematic variations in the conversion factor between mass and CO emission, and no similar correlations have been found on sub-galactic scales. This may simply be a result of the various biases and uncertainties discussed above, or it may indicate a real lack of variation in $\epsilon_{\rm ff}$.  A potential way forward would employ a statistical approach based on regional averages of pre-star formation cloud properties and inferred star formation efficiency.  

\section{The Destruction of GMCs}
\label{sec:destruction}
We have seen that observations favor GMC lifetimes of $\sim 10-30$ Myr. These are much shorter than their depletion times, implying that GMCs do not reach the end of their lives by slowly and completely transforming all of their gas into stars, but by some other mechanism(s). In this section we discuss cloud destruction. Since {\it PPVI}, a great deal of effort in the community has been devoted to modeling and measuring how star formation feedback leads to destruction of GMCs, and we review this work here. Section \ref{sec:feedback_mechanisms} summarizes the candidate feedback processes and provides quantitative intercomparisons of their importance, contrasting simple theoretical estimates with measurements from numerical simulations that represent realistic ISM conditions. Section~\ref{sec:obs_fb} then reviews observed constraints on feedback. Finally, Section~\ref{sec:destr_env} discusses investigations aimed at understanding the environmental differences in cloud destruction. 

\subsection{Destruction by external stresses}

Molecular clouds, like other concentrations of gas, can be sheared apart by
the same turbulent flows in the ISM that created them, or by the increased
background galactic shear as gas emerges from spiral arms into the interarm
region.  The increase in shear from arms to interarms is due to the tendency for potential vorticity
to be conserved, which leads to
$2A/\Omega\equiv -d\ln\Omega/d\ln R = 2-\Sigma/\Sigma_0 $
for a tightly-wrapped spiral pattern in a galaxy with a flat rotation curve
\citep{Hunter_1964,Kim_Ostriker2002}, where  $A$ is Oort's $A$ parameter and
$\Sigma/\Sigma_0$ the ratio between
the local and azimuthally-averaged gas surface density. The change in shear
from arms to interarms can be observed with careful kinematic analysis
\citep{shetty_2007}, and indeed the presence of
large giant molecular associations (GMAs) in arms but not interarms is anticorrelated with the shear
parameter, which is higher in interarm regions
\citep{Miyamoto_2014}; see also \citet{rosolowsky21}.   However,
differences in cloud mass spectra between arms and interarm regions
may also owe to other environmental differences, including the duration
of the temporal interval for migration from one arm to another
\citep{Pettit2020}.  

Destruction
of GMCs or GMAs by ambient-ISM turbulence or shear is possible only if
the cloud's self-gravity is relatively small.  This
amounts to the requirement that the timescale for external stresses to
shear the cloud apart is shorter than the gravitational timescale,
\begin{equation}
t_{\rm ff}(\rho_{\rm cloud}) \gtrsim
\frac{\rho_{\rm cloud} \delta v_{\rm cloud} R}{\rho_{\rm amb} [\delta v_{\rm amb}(R)]^2},
\end{equation}
where $\rho_{\rm cloud}$ and $\rho_{\rm amb}$ are cloud and ambient densities,
$R$ is the size of the cloud, $\delta v_{\rm cloud}$ is its internal
velocity dispersion, and $\delta v_{\rm amb}(R)$ is the  velocity
acting on the cloud from the surface Reynolds stress.
In an average sense, $\delta v_{\rm amb}(R)$ will at least equal the velocity amplitude $v(k)$ in the turbulent 
power spectrum of ambient gas at scale $R\sim \pi/k$.
However, high velocity motions with  correlation scales larger
than $\sim R$ (such as expanding shells driven by superbubbles) also
lead to destructive
instabilities when they overrun clouds, in which case
$\delta v_{\rm amb}(R)$ would be the nonlinear amplitude of the instability
at the cloud scale. 

The continual energy injection in-/dispersal of- GMCs with relatively
low overdensity  is important in
limiting their rate of conversion to more strongly self-gravitating states
as their internal turbulence dissipates.
Limiting the fraction of strongly self-gravitating molecular clouds
in turn 
limits the overall molecular depletion time, because the SFR 
per unit molecular mass is expected to decline steeply with increasing virial
parameter -- see Section \ref{sec:cloud_SFR}.
Since both observations and large-scale numerical simulations of
the star forming ISM indicate  that the most 
GMCs are not strongly bound (see Section \ref{sec:bound_vir}), 
a large fraction of GMCs must be dispersed by ``ambient''
stresses before they evolve to become strongly bound. 
However, the GMCs that {\it do} become bound are the 
most interesting from the point of view of star formation.  
In the remainder of
this section, we will concentrate on the destruction of GMCs that are
sufficiently gravitationally bound that they become actively star-forming,
and are subject to energetic stellar feedback.

\subsection{Stellar feedback mechanisms}
\label{sec:feedback_mechanisms}

Stellar feedback mechanisms include (1) jets and wide-angle outflows,
which originate as winds at a range of velocities from circumstellar
disks and disk-magnetosphere interfaces through a magnetocentrifugal
mechanism; (2) ionizing EUV radiation, which photoevaporates dense
material, producing high-pressure ionized gas that is accelerated away
from the source to $\sim 10-30\; \kms$ via internal
pressure gradients, and also helps to eject neutral gas from clouds via
thermal and radiation pressure forces;
(3) non-ionizing FUV/optical radiation, which
deposits momentum when absorbed by or scattered off dust, with the
radiation pressure force subsequently transferred to the gas by
dust-gas collisions; (4) IR radiation produced by dust reprocessing of
UV/optical --  in very high column clouds the IR may have multiple
interactions with dust, applying radiation forces;  (5) line-driven stellar
winds, which shock to create very hot, high-pressure bubbles that then
interact with the surrounding photoionized or neutral gas, (6)
supernovae, which send powerful blast waves into their surroundings,
accelerating the gas impulsively and leaving behind an expanding hot
remnant.

Although low-mass stars are the main product of star formation, they
contribute relatively little to feedback: of the above mechanisms,
only jets/outflows  predominantly arise from low-mass stars, while these
stars also contribute at a low level to the total bolometric luminosity.
The other feedback effects are associated with high-mass, hot, luminous but
short-lived massive O and B stars.

Jets and outflows are important in dispersing the natal cores of
individual stars  \citep[e.g.][]{Matzner_McKee2000},
and also aid in driving turbulence in
the larger-scale cloud \citep[e.g.][]{Nakamura07a}, especially in
locations where low-mass stars
are clustered together.  The driving mechanisms and observed properties
and statistics of jets and outflows have been extensively reviewed
in previous volumes of {\it Protostars and Planets}, including five chapters
in {\it PPV}
and one in {\it PPVI} \citep{Frank_2014}.
Given the comprehensive previous coverage in reviews
\citep[see also][]{Bally_2016} and the fact that jets/outflows
do not have sufficient power to disperse a whole GMC, for the remainder
of this section we will focus on the five other feedback processes that
are associated with high-mass stars.  
In Sections \ref{sec_EUV} - \ref{sec_SNe}, we discuss these five feedback processes in turn.

Before turning to individual feedback 
mechanisms, an important general point 
is that cloud destruction driven by
feedback is inevitably subject to stochasticity in sampling from
the IMF, since the high mass stars that produce the lion's share of
the energy input are rare
(stochastic sampling similarly affects star formation efficiency -- see e.g.
\citealt{Geen_2018,Grudic19b}).
When sampling from the IMF, low mass star clusters
will have a very large variance in the ionizing and non-ionizing photon
input as well as the wind and supernova power. As the total mass of a
cluster increases, the variance from stochastic sampling decreases,
and the median feedback input increases to approach
that from a fully-sampled IMF.  
For example, using SLUG \citep{SLUG_2012,SLUG_2015} to sample from
a \citet{Chabrier_2003} IMF and apply spectral synthesis from
Starburst99 \citep{Leitherer_1999}, Equations 33 and 34 of
\citet{Kim_JG2016} provide fits as a function of cluster mass $M_*$ 
to the median value of $\Xi \equiv Q_i/M_*$
and $\Psi \equiv {\cal L}/M_*$, respectively providing the ionizing photon rate
and bolometric luminosity output.  Above $M_* \sim 10^3 \Msun$, the
median radiation outputs 
approach constant values, $\Xi \sim 2.5 \times 10^{13}\, {\rm s^{-1} \ g^{-1}}$
and $\Psi \sim 1800\, {\rm erg\, s^{-1}\, g^{-1}}$.

\subsubsection{Photoionized gas}\label{sec_EUV}

Photoionization is the form of feedback that has been studied the longest,
since it creates H~\textsc{ii} regions that are observable via traditional optical
nebular diagnostics.   For an H~\textsc{ii} region of radius $r$ powered by a source with
ionizing photon rate $Q_i$, the
density of ions (or electrons) in the ionized gas is $n_i = [3Q_i f_{\rm ion} /(4\pi\alpha_B r^3)]^{1/2}$,
where $\alpha_B$ is the case-B recombination rate coefficient, and
$f_{\rm ion}<1$ accounts for losses to dust absorption and
escape of radiation outside the ionized region (``Str\"omgren sphere''); the above assumes singly ionized H and He.
The ionized gas exerts
a thermal pressure force per unit
area $2 n_i kT$ on neutral structures, which can be augmented by the
``rocket effect'' \citep{Oort_Spitzer1955}  back-reaction force if ionized gas freely expands
from surfaces where it is photoevaporated.  

For an idealized H~\textsc{ii} region of radius $r$ produced by a star cluster of mass $M_*$,
the total thermal pressure force
on its surface would be $8 \pi n_i kT r^2$, which implies a rate
(per unit stellar mass) of momentum injection
\begin{multline}
\label{eq:pdotion}
\frac{\dot{p}_{\rm ion,th}}{M_*} = 190\, \kms\,\Myr^{-1}
\left(\frac{f_{\rm ion} \Xi}{2.5 \times 10^{13}}\right)^{1/2}\\
\times \left(\frac{M_*}{10^3\Msun}\right)^{-1/2}
\left(\frac{r}{10\pc}\right)^{1/2}.
\end{multline}
For fiducial parameters, this exceeds the direct radiation pressure force
(see Section \ref{sec_FUV}) if $r/\pc > f_{\rm ion}^{-1} M_*/10^4\Msun$
\citep{Krumholz_Matzner2009}.  
For all but the most massive clusters and earliest evolutionary stages, this inequality is satisfied; photoionization is 
therefore believed to be the most important feedback mechanism over the lifetime of GMCs with properties similar to those in the Milky Way. However, for a very massive (luminous)
cluster at early evolutionary stages,
the radiation pressure exceeds the thermal gas pressure, compressing the ionized gas towards its boundary
\citep{Draine_2011}.

In realistic inhomogeneous clouds, the photoionized gas will tend to expand
into low-density regions to fill much of the original volume 
\citep[e.g.][]{Walch2012,Geen2015},
while dense structures will remain neutral and be accelerated more slowly.   
The density within the ionized gas is still expected to be comparable to the
Str\"omgren estimate above
when the H~\textsc{ii} region has a given size.  
In this more realistic case of an  ``unconfined'' H~\textsc{ii} 
region, the   classical \citet{Spitzer1978}
expansion solution (which assumes a surrounding 
swept-up neutral shell of gas) does not apply.
Instead, provided the gravitational potential well is not too deep \citep[requiring escape speed $\lesssim 10\, \kms$ -- see][]{Dale_2012,Dale_2013a},  
 the ionized gas
accelerates under its own pressure gradients
and escapes \citep[e.g.][]{Whitworth79a,Matzner02a}. The characteristic
momentum flux of escaping ionized gas is
equal to $1+(v_i/c_i)^2$ times 
the thermal momentum injection rate of Equation \ref{eq:pdotion},
where the
outflow velocity of ionized gas $v_i$ is typically two or three times the sound speed $c_i \sim 10\, \kms$
\citep[e.g.][]{Tenorio-Tagle1979,Yorke1986,Ali2018,Kim_JG2018}.
The corresponding escaping mass
flux in ionized gas $\dot M_{\rm ion}$ is of order the momentum flux divided by $v_i$.
Thus, if the solid angle
filled by expanding ionized gas 
exceeds the solid angle filled by neutral structures, loss of
momentum and mass in ionized gas would dominate over loss of neutral
gas accelerated by the pressure of photoionized gas. That is, 
direct photoevaporation would be more important 
than the ``rocket effect.'' Neutral structures
are driven out more slowly than the ionized gas because they are overdense
(the lower-density regions are preferentially photoionized) and accelerate
more slowly for a given momentum input rate per unit area;
from simulations
the difference is a factor $\sim 2-3$ in velocity \citep{Kim_JG2018}.  

The scenario of GMC destruction
driven by photoevaporating H~\textsc{ii} regions has been demonstrated in numerical simulations.
Using adaptive
ray tracing to obtain an accurate solution of the ionizing radiation
field in turbulent, star-forming clouds
over a wide range of parameters, the numerical radiation
hydrodynamic (RHD) simulations  of \citet{Kim_JG2018} 
show that the mean photoevaporation rate
is $\dot N_{\rm ion}\equiv\dot{M}_{\rm ion}/(1.4 m_p) \approx c_i (Q_{\rm i,max} R_0/\alpha_B)^{1/2}$
for $Q_{\rm i,max}$ the maximum
ionizing photon input rate and $R_0$ the initial cloud radius.
In addition, the momentum
injection rate in these simulations from thermal pressure forces
is very close to $\dot M_{\rm ion} c_i$.
While the functional dependence on parameters 
is the same as theoretically predicted, this measured momentum
injection rate is a factor $\sim 5-10$ below the simple spherical model (Equation \ref{eq:pdotion})
with $f_{\rm ion}=1$.
Part of the reason for this reduction is that a significant
fraction of the ionizing photons either escape the cloud or are
absorbed by dust \citep[see also][]{Kim_JG2019}.
In addition, distributed star formation (as opposed to a single central cluster) leads to force cancellation.

\subsubsection{Direct radiation pressure}\label{sec_FUV}

For a star cluster of luminosity ${\cal L}$, the rate (per unit mass) of momentum
input from 
direct radiation  is
\begin{equation}\label{eq:pdotrad}
\frac{\dot{p}_{rad}}{M_*} =\frac{{\cal L}}{c M_*}= 19\, \kms\,\Myr^{-1}
\frac{\Psi}{1800 \, {\rm erg\, s^{-1}\, g^{-1}}},
\end{equation}
where the fiducial value assumes full sampling
of the IMF. Stellar FUV 
radiation interacts with the material of the surrounding cloud, which 
consists of an interconnected network of clumpy
filaments.  If we consider an individual structure of surface density
$\Sigma$ with total optical depth $\tau$ (averaged over the radiation
spectrum), the  radiation force per unit area
for a flux $F$ incident to its surface is $(F/c)(1-e^{-\tau})$.
With a dust crossection per hydrogen of
$\sigma \sim 10^{-21} {\rm cm}^2$  in the FUV (and slightly higher in  EUV),
essentially all overdense structures within
molecular clouds will be optically thick to direct radiation,
$e^{-\tau} \ll 1$.  
For radiation directly incident from a star cluster of mass $M_*$
at distance $r$,  $F=\Psi M_* /(4 \pi r^2)$, and the
gravitational force per unit area from the same star cluster on the structure is
$GM_*\Sigma/r^2$.  The Eddington
ratio between the radiation force and gravitational force of
the star cluster on the structure is then $f_{\rm Edd,*} = \Sigma_{\rm Edd,*}/\Sigma$,
where
$\Sigma_{\rm Edd,*} = \Psi/(4 \pi cG)= 340\, \Msun\, \pc^{-2} (\Psi/1800 \, {\rm erg\, s^{-1}\, g^{-1}})$.  $\Sigma_{\rm Edd,*}$
is the maximum surface density a structure could have that  could
be ejected by direct radiation pressure.

In idealized models of cloud destruction by radiation pressure,
star formation produces a central star cluster, and a uniform spherical
shell of (dusty) gas is ejected when the SFE becomes large enough for the
radiation force to exceed the total gravity (stars plus the gas)  
\citep[e.g.][]{Fall10a,Kim_JG2016,Raskutti2016,Rahner_2017}.
Defining $\Sigma_{\rm cloud,0} \equiv M/(\pi R^2) $ for $M$ and $R$ the cloud 
mass and radius, this leads to
a predicted SFE $\sim\Sigma_{\rm cloud,0}/\Sigma_{\rm Edd,*}$ when 
$\Sigma_{\rm cloud,0} \ll \Sigma_{\rm Edd,*}$,
while the SFE
approaches unity when $\Sigma_{\rm cloud,0}\sim \Sigma_{\rm Edd,*}$. This
kind of idealized spherical
model can be generalized from direct radiation pressure 
to other sources of momentum injection that scale linearly in $M_*$
\citep[e.g.][]{Li_Vogelsberger2019} -- see Section \ref{sec:SFE_theory}.

Although spherical models are attractive in their conceptual
simplicity, in reality both observations and theory indicate that the
distribution of surface
densities in molecular clouds follows a log-normal functional form (see Section \ref{sec:internal_props}), in which most of the mass is in structures with surface
densities well above $\Sigma_{\rm cloud,0}$. This makes cloud destruction by
radiation pressure more difficult, and 
requires analysis beyond simple spherical models.
In  particular, predictions for the SFE and cloud destruction timescale can be obtained 
by taking into account the log-normal distribution of surface density,
and assuming that only the fraction of mass in structures with sub-Eddington surface density ($f_{\rm Edd}>1$
including gas gravity)
is ejected. These
models predict much higher star formation efficiency than idealized
spherical models (for bound clouds), and agree well  with RHD simultations
in which radiation pressure is the only feedback effect included
\citep{Thompson_2016,Raskutti2016,Kim_JG2018}.  The log-normal PDF model
also yields an analytic prediction for the velocity distribution in
outflowing gas  which agrees with numerical simulations \citep{Krumholz17e,Raskutti17a}. 

As noted in Section \ref{sec_EUV}, the characteristic
momentum input rate from direct radiation pressure (Equation \ref{eq:pdotrad})
is below the characteristic value from photoionized gas pressure
(Equation \ref{eq:pdotion})
unless an
H~\textsc{ii} region is powered by a very luminous source and/or is very compact
\citep{Krumholz_Matzner2009}.  
Numerical RHD simulations indeed show that the net radial force
from ionized gas pressure gradients exceeds the net radial force from radiation
pressure unless the surface density of a cloud is very high
\citep{Kim_JG2018}. 
However, the
transition from gas-pressure to radiation pressure dominance occurs at higher cloud surface density ($\sim 10^3 M_\odot \pc^{-2}$)
than would be predicted by simple spherical models
\citep[e.g.][]{Krumholz_Matzner2009,Kim_JG2016}.  

Similar to the situation for ionized gas pressure forces, the actual
net radiation pressure force applied to gas from stars formed within a GMC
is likely much lower than it would be in an idealized situation
of single star cluster in a uniform, optically-thick cloud. 
For turbulent clouds, the numerical RHD simulations of
\citet{Kim_JG2018} show that the net radial force on cloud material due to
direct radiation pressure forces is a factor $\sim 5-10$ below the maximal
rate ${\cal L}/c$.  Several effects contribute to this reduction: much of the radiation may escape
from the cloud \citep{Kim_JG2019}, radiation forces from distributed
sources (as opposed to a single central cluster) partially cancel, and radiation momentum deposited within gravitationally bound regions is advected inward with the collapse \citep{Krumholz18c}.

\subsubsection{Reprocessed radiation pressure}\label{sec_IR}

For a spherical cloud with a central radiation source ${\cal L}$,
the rate (per unit mass)
of momentum input from reprocessed IR radiation is $\dot{p}_{\rm IR}/M_* = {\cal L}\tau_{\rm IR}/(c M_*)$, where $\tau_{\rm IR}$ is the center-to-edge optical depth.
Reprocessed IR becomes more important
than direct radiation if this (dust)
optical depth exceeds unity, which requires very high column density.  For
this reason, reprocessed radiation is of most interest in cluster-forming
clumps and in extremely compact GMCs where super-star clusters form
\cite[e.g.][]{Murray_2010}.   
Assuming spherical symmmetry, for any given fluid element 
the Eddington ratio between the radiation force (per unit volume)
$\rho \kappa_{\rm IR} {\cal L}/(4 \pi r^2c)$
and the gravity of the central star cluster $\rho G M_*/r^2$
is $f_{\rm Edd,*} = \kappa_{\rm IR} \Psi/(4 \pi c G)
= 0.7 \kappa_{\rm IR}/10\, {\rm cm^2 \, g^{-1}}$, where we have
adopted  $\Psi \equiv {\cal L}/M_* = 1800\ {\rm erg\, s^{-1}\, g^{-1}}$ assuming a normal IMF (but see further discussion below). 
For radiation temperature $T<10^3$ K and Solar neighborhood dust-to-gas ratio,
the Planck-weighted mean opacity for normal interstellar dust
\citep[][Chapter 23]{Draine_book2011} is below the value
$\kappa_{\rm IR, crit} \equiv 4 \pi c G/\Psi \approx 14\, {\rm cm^2 \, g^{-1}}$
required to attain $f_{\rm Edd,*}>1$.

In reality, the radiation field is not spherically symmetric,
and gas as well as (distributed) stars contributes to gravity.  Nevertheless, 
based on RHD simulations of self-gravitating, turbulent,
star-forming clouds with 
a range of (spatially-uniform) $\kappa_{\rm IR}$, \citet{Skinner_Ostriker2015} indeed
found that only in models with $\kappa_{\rm IR} > \kappa_{\rm IR, crit} $ does
IR radiation expel significant mass from a cluster-forming cloud
\citep[see also][]{Tsang_Milos2018}.   However,
there is not a simple inverse relationship between star formation
efficiency and $\kappa_{\rm IR}$, as would apply in spherical symmetry.  
Because a standard IMF and dust-to-gas ratio
will not yield $\kappa_{\rm IR} > \kappa_{\rm IR, crit}$ unless the temperature is very high,
\citet{Skinner_Ostriker2015} argued that 
either a top-heavy IMF or enhanced dust abundance is required for IR 
radiation to become dynamically important.  

Since the radiation energy density and temperature generally increase
toward the interior
of a star-forming cloud, and the Planck-weighted
$\kappa_{\rm IR}$ increases with radiation temperature, interior regions of clouds
are more susceptible to IR radiation forces than outer regions,
and this can also lead to radiation Rayleigh-Taylor instabilities
\citep{Krumholz_Thompson2013,Davis_2014}.
For a normal IMF and dust-to-gas ratio, the surface density of a cloud
has to be very  large before the opacity even in the center exceeds $\kappa_{\rm IR, crit}$.  Adopting $\kappa_{\rm IR} \propto T^2$, \citet{Crocker_2018} found that 
$\Sigma \sim 10^5$ $\Msun$ $\pc^{-2}$ would be required
for the central region of a cloud to become super-Eddington at a star
formation efficiency of 50\%.  Taking into account the realistic flattening of
the temperature-dependent
opacity curve between $T\sim 10^2-10^3$ K (associated with the observed decline
in the extinction shortward of $10 \mu {\rm m}$), the required surface
density to reach $\kappa_{\rm IR,crit}$
would be an  order of magnitude larger. For dust that has experienced
agglomeration, the Planck-mean or Rosseland-mean opacity can even decline
in the range $T=10^2-10^3$ K \citep{Semenov_2003}, never
reaching $\kappa_{\rm IR,crit}$ for a normal IMF or dust abundance.

Taken together, the current evidence suggests that reprocessed radiation would
only be able to play a role in cloud destruction in conditions where the
total surface density is extraordinarily high ($>10^6$ $\Msun$ $\pc^{-2}$)
and there is also an enhancement in the dust abundance and/or a top-heavy
IMF.  The IMF may be either globally top-heavy or have higher-mass stars concentrated in the center of a forming cluster -- there is evidence of both, especially in extreme environments 
\citep[e.g.][]{McCrady05,Lu13,Hosek19}.  If high-mass stars are segregated in the core of a cluster, reprocessed radiation may be important there even if it is not on larger scales.

\subsubsection{Stellar winds}\label{sec_winds}

Stellar winds are driven at $1000\; \kms$ or more from the surface of
O stars, and the combined shocked winds from clustered stars
produce a hot, high-pressure bubble
that expands into the surrounding cloud.   
For a star cluster of mass $M_*$ that fully samples the Kroupa IMF,
the specific energy and momentum input rate averaged over 1 Myr
are respectively
${\cal L}_w/M_* \approx 2.5\; \Lsun/\Msun$ and
$\dot{p}_w/M_* \approx 8.6\; \kms/\Myr$, using
Starburst99 \citep{Leitherer_1999}.

Based on the classical \citet{Weaver_1977} solution,
in which the leading shock radiates away 35\%  of the input energy and the remaining is shared between the hot bubble and the cool shell it accelerates,
the momentum input rate (per unit mass) from a wind-driven bubble would be \citep{Lancaster_2021a}
\begin{multline}\label{eq:pdotwind}
\frac{\dot{p}_w}{M_*}  = 880 \frac{\kms}{\Myr}
\left(\frac{t}{\Myr}\right)^{2/5}
\left(\frac{M_*}{10^3 \Msun}\right)^{-1/5}\\
\times 
\left(\frac{n_{\rm H}}{10^2 \pcc}\right)^{1/5} 
\end{multline}
for $n_{\rm H}$ the ambient gas density. 
While momentum injection at this level
would in principle have significant impact on the
parent cloud (e.g.~unbinding an
Orion-like cloud within a Myr), in practice observations
 show that the energy
of the wind bubble is only $\sim 1-10\%$ of what would be expected in
the classical solution \citep{Rosen_2014}.

Some of the hot bubble's
energy, especially at late times,
may be lost through ``leakage'' from the cloud \citep{Harper-Clark_2009}.
However, recent theory and numerical simulations
\citep{Lancaster_2021a,Lancaster_2021b,Lancaster_2021c} indicate that
the majority of the energy is likely lost through turbulent mixing and
subsequent rapid cooling at the interface between the hot bubble
and surrounding gas, with the cooling significantly
enhanced by the fractal structure of the interface.
With most of the energy lost
to cooling, the transfer of momentum to the cold gas of the
surrounding cloud by the bubble amounts to
only a few times the initially-injected value,
i.e. $\dot{p}_w/M_* \sim 10-30\; \kms/\Myr$, with the larger value
applying for less-massive clusters \citep{Lancaster_2021b}.  This is
comparable to the momentum input rate from direct radiation pressure
(Equation~\ref{eq:pdotrad}).

\subsubsection{Supernovae}\label{sec_SNe}

Supernovae are the largest source of momentum injection to the ISM from
stellar feedback.
The momentum injection from a single SN can be estimated analytically based on the
properties of the Sedov-Taylor solution at the time
$t_{\rm SF}$ that expanding blast wave first becomes radiative and 
shell formation occurs
\citep[][section 39.1.1]{Draine_book2011}; the result for a uniform ambient medium  with density 
$n_{\rm H}$ is
$p_{\rm SF} \approx  2 \times 10^5\; \Msun\, \kms (n_{\rm H}/ 1\pcc)^{-0.1}$.  Numerical
simulations in a uniform ambient medium
show that the momentum subsequently increases by $50\%$, with
the momentum-conserving stage reached by
$3\, t_{\rm SF}$ \citep{Kim_Ostriker2015a}, and that the presence of
inhomogeneous structure in the ambient medium does not significantly
affect the total momentum injection \citep{Iffrig_Hennebelle2015,Kim_Ostriker2015a,Martizzi_2015,Walch_Naab2015}.  However, the 
spatial {\it distribution} of injected momentum is strongly affected by  inhomogeneity, which in GMCs is extremely high, given the very low effective volume filling factor resulting from highly supersonic turbulence. In an inhomogeneous medium, shocks  preferentially propagate along low-density directions, so that even if a SN occurs within a GMC, much of the energy it produces is likely to emerge into (and be radiated by) the surrounding diffuse ISM \citep[see also][]{Geen_2016,Ohlin_2019,Zhang_Chevalier2019,lucas20}.  Fully following the effects of supernovae numerically (to obtain both the correct terminal momentum and hot gas mass produced) requires sufficiently high resolution. In particular, the mass resolution must be $\ll 10^3 M_\odot$, so that  evolution will be well resolved up to the point when the leading shock becomes radiative \citep{Kim_Ostriker2015a}; the resolution requirements in simulations including multiple SNe may be even more severe \citep{Kim_Ostriker_Raileanu2017,Gentry_2019}. If the Sedov stage of evolution cannot be resolved (as is typically true for global galaxy simulations), supernova feedback is often treated via momentum feedback.

For the perfectly spherical case,
the space-time correlation of supernovae due to stellar clustering creates
an overpressured superbubble analogous to that from a stellar wind, which
in principle could enhance the momentum injection 
\citep[e.g.][]{McCray_Kafatos1987,Gentry_2017}.  For the idealized spherical problem, the multiple-SN solution approaches the limit of continuous energy input 
\citep[cf][]{Weaver_1977} when individual blast waves become subsonic before reaching the cooled shell.   This occurs after $\sim 10/(1-\theta)$ SN events for $\theta$ the fraction of energy lost to cooling via mixing at the interface; after this point the momentum per SN scales as $\hat p \propto (1-\theta)^{4/5} \Delta t^{1/5} t^{2/5}$ for $\Delta t$ the SN interval \citep{El-Badry_2019}.  

For a realistic inhomogenous, turbulent medium, the superbubble shell is nonspherical, and instabilities at the interface between the hot interior and cool swept-up shell lead to mixing and cooling, so that $1-\theta$ (which varies inversely with the turbulent diffusivity, enhanced by fractal structure of the interface)  becomes small \citep{El-Badry_2019, Fielding_2020,Lancaster_2021a}. When $1-\theta$ is small, individual 
blasts  remain supersonic, and indeed  \citet{Kim_Ostriker_Raileanu2017} found  that for an inhomogeneous medium the momentum
injection per event is quite similar to that from the
arrival of a single supernova
shock.  From simulations of multiple SNe exploded in a uniform ambient medium, \citet{Gentry_2019} 
found that momentum per event  increases with clustering, but were only able to set a lower limit on the enhancement because even at extreme resolution heat transfer was dominated by numerical rather than physical conduction; based on an analytic extrapolation to the expected physical conduction rate, they estimate the momentum increase might be an order of magnitude. 

Even without a correlation boost, the momentum injection
from supernovae per unit cluster mass
$p_{\rm SN}/M_* \sim 1-3 \times 10^3 \; \kms$
is large compared to other feedback mechanisms.  
The advent of supernovae at the end of massive stars' lives
(with
a roughly constant momentum injection rate from $\sim 3-40$ Myr after a
massive cluster 
forms -- \citealt{Agertz_2013}), may, however, be too late to play
much role in parent cloud destruction, based on current estimates of
GMC dispersal times (see Section~\ref{sec:lifetime}).  Instead,
due both to the late advent of SNe and the escape of SN energy through low-density channels,
much or most of
the feedback power from supernovae
will be applied not to the natal GMC where a massive star
is born, but to the larger-scale ISM \citep[e.g.][]{Smith_MC2021}.  As a result, expansion of
superbubbles can be important to destruction of GMCs proximate to the driving cluster, as noted above.

\subsection{Observational constraints on feedback}\label{sec:obs_fb}

We have thus far discussed feedback from a theoretical perspective. However, thanks to multiwavelength observations of GMCs in the Milky Way and the Magellanic Clouds, it has now become possible to estimate the strength of various feedback mechanisms directly from observations. 
As ongoing observations probe a wider range of environments (e.g., metallicity, dust content, properties of host galaxy) in the near future, we can expect further 
interesting constraints on the relative importance of different feedback mechanisms at varying stages of evolution and for different local conditions (e.g. outer disk, spiral arm, galactic center, starburst).

Many of the observational estimates take the form of a pressure. A caveat to be kept in mind is that, for the purposes of understanding contributions from different feedback mechanisms to global destruction of GMCs, the quantity that matters is not the pressure but the net outward force. For gas pressure, the force is obtained as an integral over volume of the radial pressure gradient, which becomes a pressure multiplied by an area if the pressure gradient is dominated by  a local discontinuity across a surface.  For radiation pressure, the radial force is an integral over volume of the radiation flux multiplied by $\kappa \rho/c$, which in the diffusion limit (for reprocessed radiation) is an integral of the radial gradient of radiation pressure.  Pressure (either gas or radiation) may be high  locally (e.g. close to a photoevaporating surface or radiation source) but may have a  limited effect  on global dynamics because the high pressure region is confined to small scales with  low global filling factor.

\subsubsection{\bf Photoionized gas pressure and direct radiation pressure}

The pressure of photoionized gas is empirically estimated in several different ways.  From maps of either extinction-corrected H$\alpha$ or free-free radio emission, a total ionizing photon rate $f_{\rm ion}Q_i$ may be measured, and then for an assumed volume based on the size of a spatially-resolved H~\textsc{ii} region the usual Str\"omgren estimate is applied to obtain the mean electron density (as in Section~\ref{sec_EUV}).  This approach, using free-free emission, was adopted by  \citet{lopez14} for a sample of 32 H~\textsc{ii} regions in the LMC and SMC.   
\citet{olivier21} also used free-free radio emission to obtain ionized gas pressure for a sample of 106 young, small ($R<\pc$) H~\textsc{ii} regions.
\citet{mcleod19} studied some of the same HII regions as \citet{lopez14} but at higher spatial resolution with {\it MUSE}, using line ratios to obtain estimates of the electron density;    \citet{mcleod2020} applied the same methodology  to five H~\textsc{ii} regions in the nearby dwarf galaxy NGC 300.  Since the ionized gas is nonuniform and the gas contributing to emission is preferentially at higher density, the line ratio probes higher density gas and this method will generally yield higher pressure than the method based on integrated emission measure and  constant-$n_e$ ionization equilibrium (the ratio of the two density estimates is the square root of the effective volume filling factor for ionized gas).
For a sample of 5 H~\textsc{ii} region complexes in the much more extreme environment of the Milky Way Galactic Center, \citet{barnes20} employed radio observations with a similar approach to \citet{lopez14} to estimate the photoionized gas pressure. Their data set collects both single-dish and interferometric data at a range of wavelengths from the literature; individual pointings at different wavelength and from different studies may therefore represent the same HII region observed at different beam scales and/or with a different instrument.   We note that to make a proper comparison among different studies of ionized gas pressure, it is necessary to multiply the estimated ionized gas pressure by an effective area to obtain the outward force on the cloud gas.  This area is comparable to the cloud size (or beam size) for the H$\alpha$ or free-free method, but smaller than this for the line ratio method.   

Empirical estimates of the direct radiation pressure are obtained by dividing a bolometric luminosity by $c$ and an effective area.   This area is $4 \pi R^2$ for the radiation pressure that would be applied to the opaque shell of radius $R$ bounding an optically-thin H~\textsc{ii} region, or a factor 3 smaller for the nominal volumetric mean radiation pressure (setting radiation pressure to the volume-weighted average of $L/(4 \pi r^2 c)$ for a sphere).  
For their LMC and SMC sample, \citet{lopez14} scale the extinction-corrected H$\alpha$ to obtain a bolometric luminosity and use an effective area $4\pi R^2/3$.  \citet{olivier21} adopt bolometric luminosities obtained from YSO SED fits in the literature.  In their studies, \citet{mcleod19} and \citet{mcleod2020} sum over luminosities of individual O stars derived from spectral types, with the first study using an effective area $4\pi R^2$ and the second using $4\pi R^2/3$.  
For their Galactic Center study, \citet{barnes20} scale either the radio emission or IR emission to obtain bolometric luminosities associated with  individual structures (radio beams or ``leaves'' from dendogram analysis of IR maps), and adopt area $4\pi R^2/3$.  
Just as for the ionized gas, it is necessary to multiply the estimated direct radiation pressure by an effective area (cloud or beam size) to quantify the environmental impact. For a centrally  concentrated source, the net direct radiation force on a cloud is $(1-f_{\rm esc})L/c$, independent of area. 

\citet{lopez14} found that the ionized gas pressure exceeds the direct radiation pressure by an order of magnitude or more.  Given the relatively large size ($\sim 10-100$ pc) and evolved state of their sample, this is consistent with expectations.  The  higher-resolution observations of \citet{mcleod19} and \citet{mcleod2020} mostly target smaller H~\textsc{ii} regions, and therefore one might expect the radiation pressure to be closer to the ionized gas pressure.  However, these sources are still much larger than the characteristic radius expected for radiation pressure to dominate (see Sections \ref{sec_EUV} and \ref{sec_FUV}).  In addition, the line-ratio method returns higher ionized gas density than using integrated fluxes.  As a consequence, \citet{mcleod19} and \citet{mcleod2020} found that the ionized gas pressure far exceeds the radiation pressure. However, as noted above, a proper comparison requires applying an effective area factor; for a low volume filling factor of ionized gas, the impact would be reduced.   The Galactic Center H~\textsc{ii} region complexes investigated by \citet{barnes20} have outer scales $\sim 10$pc, with  IR maps at $\sim \pc$ resolution and interferometric radio data extending down to much smaller sizes ($\sim 10^{-3}$ pc for Sgr B2).  The measured ionized gas pressure estimates increase with decreasing effective beam size roughly $ \propto r^{-1}$.  Since the direct radiation pressure is assumed to scale $\propto r^{-2}$, this exceeds the gas pressure at sufficiently small size, with a crossover at $\sim 0.1$ pc.  For their sample of individual 
very young H~\textsc{ii} regions, \citet{olivier21} found comparable radiation pressure and ionized gas pressure at scales $\sim 0.1$ pc, with the former  shifting slightly higher than the latter at smaller scales.   

While the above results are quite interesting, we caution that care must be taken in interpreting them.  As discussed in Sections \ref{sec_EUV} and \ref{sec_FUV}, actual measurements of the time-averaged gas pressure force and radiation pressure force in numerical simulations with ray-tracing radiative transfer show that they are both a factor of 5-10 below the values  ($2 k T_i [12 \pi Q_i R/\alpha_B]^{1/2}$ for gas pressure or $L_{\rm bol}/c$ for radiation pressure) that would be obtained using the above empirical approaches.

\subsubsection{Reprocessed radiation pressure}

Estimates of reprocessed radiation pressure require multiwavelength IR-to-submm observations and a dust model in order to obtain a dust temperature $T_d$, with the radiation temperature assumed to be equal to $T_d$ and the reprocessed radiation pressure $a_{\rm rad} T_d^4/3$.  From multiwavelength IR observations fit to dust models, \cite{lopez14} obtained estimates for the volume-weighted mean  reprocessed radiation pressure that are below the ionized gas pressure but above the direct radiation pressure.  The median ratio of reprocessed to direct radiation pressure is $\sim 10$, implying a large IR optical depth.
For their Galactic Center sources, \citet{barnes20} also used multiwavelength IR observations fit to dust models, and found that reprocessed radiation pressure is typically a factor of a few below the direct radiation pressure at $\sim $pc scales.
\citet{olivier21} fit SEDs to source models including disk and envelope to obtain the dust temperature profile, and then computed a mean radiation pressure as a simple volumetric average over the source of $a T_d^4/3$.  This leads to reprocessed radiation pressure estimates that exceed both ionized gas pressure and direct radiation pressure for sources with sizes $<1$ pc.  
Since, however, the reprocessed radiation pressure estimate is based on a model fit at scales below the resolution of the observations,  and since reprocessed radiation pressure increases towards smaller scale as $\sim \kappa \rho L/(4 \pi cr)$, the high reprocessed radiation pressure may correspond to an effective spatial scale that is smaller than the scale applicable for the ionized gas and direct radiation.

\subsubsection{Stellar winds}
For their sample of H~\textsc{ii} regions in the LMC and SMC, \citet{lopez14} used {\it ROSAT} and {\it Chandra} spectral data to obtain temperatures and densities which yields a hot gas pressure (for the LMC) or an upper limit (for the SMC).  In almost all cases, the hot-gas pressure was lower than the pressure of photoionized gas.  These results on low hot gas pressure are consistent with theoretical predictions and numerical simulations showing that much of the energy of hot shocked winds is lost at early stages to mixing and cooling, and at late stages to leakage (see Section \ref{sec_winds}). \citet{Lancaster_2021a} also show that the estimates of hot-gas pressure in the Orion nebula using the {\it XMM-Newton} data obtained by \citet{Gudel08} are consistent with expectations for a wind with highly efficient cooling.  
Finally, we note that the C~\textsc{ii} 158 $\mu$ observations of  \citet{Pabst_2019,Pabst_2020} in Orion indicate a shell expansion rate of 13 km s$^{-1}$.  While they advocate for this expansion being due to the stellar wind, it is likely that direct radiation pressure and the pressure of photoionized gas also contribute.

\subsubsection{Supernovae}
To obtain empirical estimates of the net momentum injected by supernovae, it is necessary to select remnants that are old enough to have reached radiative stages, that are reliably associated with a single supernova (the number contributing cannot be known in the case of multiple successive sources), and for which the surrounding shell is either complete or at least clearly defined, with information for both mass and velocity.  From a set of 7 supernova remnants with 21 cm observations (in cases where interaction with a molecular cloud is involved, this is augmented by CO and HCO$^+$ observations) and estimated ages $\sim 10^4 -10^5$~yrs, \citet{Koo_BC2020} measured the total momentum of the expanding shells.  The results, in the range $1-5 \times 10^5 \, \Msun  \, \kms$, are consistent with expectations from theory and numerical simulations for $E_{\rm SN}\sim 10^{51}$~erg.

\subsection{Importance of different destruction mechanisms for
varying cloud and environmental properties}
\label{sec:destr_env}

The most direct constraints on the relative importance of different physical
mechanisms  to cloud destruction as a function  of cloud properties come
from numerical simulations of individual clouds.  With this approach,
it is possible to explicitly follow effects of feedback in inhomogeneous
clouds, treating radiation-gas interactions
with RHD techniques, resolving the interactions of shocked winds and
supernova blasts with surrounding gas, and controlling whether a given
mechanism is turned off or on.  However, simulations of this kind generally
adopt idealized initial conditions for turbulent clouds, and
do not capture important environmental effects.  To  follow GMC dispersal
driven by feedback from earlier star formation in  neighboring clouds,
cloud mergers or destruction from cloud-cloud collisions, or 
environmental biases (e.g. spiral-arm vs. interarm)
in the differential importance of destruction mechanisms,
larger-scale simulations are crucial.  These larger-scale simulations,
however, have much lower spatial and mass resolution and \textit{per force}
adopt subgrid treatments to model feedback.  Below we review
current results from these two different types of simulations.  For
the future, an exciting prospect is the marriage of the two approaches,
in which high-resolution simulations of individual clouds are used
to calibrate subgrid models which are then deployed in realistic large-scale
galactic ISM simulations. From resolved cloud-scale simulations it is clear that  a large fraction of the radiation, wind, and supernova energy injected in a cloud is either radiated away or escapes through low-density channels to larger scales. To account for radiative losses, subgrid models adopted in global galactic simulations typically use momentum rather than energy injection.  
The level of momentum injection adopted is generally 
based on analytic spherical solutions that do not allow for energy escape from GMCs, however, implying that the effect on cloud dispersal is overestimated.  This can be improved by employing subgrid models in which feedback momentum injection is instead calibrated from high-resolution GMC simulations that allow for escape of radiation, wind, and supernova energy.

\subsubsection{Individual GMC simulations}
\label{ssec:individual_gmc_sims}

Cloud-scale simulations generally find that the timescale
$t_{\rm dest}$, defined as the interval between the onset
of star formation and a cloud's destruction, is comparable to the initial
free-fall time of the cloud (see Figure \ref{fig:timescales}).  
\cite{Raskutti2016}, who considered only direct radiation pressure
with an M1 RHD solver, found $t_{\rm dest}/t_{\rm ff} \sim 0.8-1.2$, slightly increasing over
$\Sigma_{\rm cloud} = 25 - 250 \, \Msun \pc^{-2}$.
\citet{Kim_JG2018}, who used ray-tracing RHD to model both photoionization
and direct radiation pressure, found
$t_{\rm dest} /t_{\rm ff} \sim 0.6-4 $, increasing over $\Sigma_{\rm cloud} = 10 - 10^3 \, \Msun \pc^{-2}$; this corresponds to a decrease of 
$t_{\rm dest}$ from $\sim 10 \, \Myr$ to  2 $\Myr$ with 
increasing $\Sigma$. 
In simulations with momentum injection modeling stellar winds (with $\dot p/M_*$ set to values ranging up to $\sim 100\, \kms\, \Myr^{-1}$) 
\citet{Li_Vogelsberger2019} found that $t_{\rm dest}$ increases roughly
linearly  with the initial free-fall time of the cloud.
\citet{Grudic19a}, based on simulations with a variety of feedback treatments, found that star formation was complete within $\sim 1~t_{\rm ff}$ of the first collapse, corresponding to a range $\sim 3-15$ Myr for their cloud models with surface density $\Sigma =64 \Msun\, \pc^{-2}$ and a range of initial masses and radii.  The simulations of \citet{Grudic18a}, which had the same feedback treatment but a wider range of gas surface density and rapidly-rotating initial conditions, similarly found a star formation duration $\sim 1~t_{\rm ff}$.    
\citet{He_Ricotti2019}, based on simulations treating photoionization
feedback with an M1 RHD solver, found that $t_{\rm dest}$ scales roughly linearly with the initial cloud size, which
for their setup corresponds to a ratio $t_{\rm dest}/t_{\rm ff}$ that increases
weakly with $\Sigma_{\rm cloud}$.
In simulations with both photoionization and radiation pressure, 
varying the initial virial parameter has very little effect on cloud lifetime. A very high magnetization increases $t_{\rm dest}$, but only because the SFE and therefore feedback
is reduced \citep{Kim_JG2021}. The ray-tracing RHD simulations of
\citet{Fukushima_2020} 
for $\Sigma_{\rm cloud} \sim 10 -300 \,  \Msun\, \pc^{-2}$ and metallicity
$Z/Z_\odot=10^{-2} - 1$ 
with both photoionization and radiation pressure (scaled $\propto Z$)
found  total cloud lifetimes (from the initiation of the simulation until 90\% of star formation is complete) only slightly longer than $t_{\rm ff}$, increasing
in a relative sense at larger $\Sigma_{\rm cloud}$.  The insensitivity
to metallicity indicates that photoionization of hydrogen is the main
cloud destruction mechanism. 
Destruction timescales in the above simulations with radiation
feedback are generally similar to observed estimates (see Section \ref{sec:lifetime}).  

Consistent with general theoretical expectations, numerical simulations show
that cloud destruction by feedback is ``easier'' (producing
a higher fraction of unbound gas after a given evolutionary time, or 
requiring a lower lifetime SFE to destroy the cloud)
in clouds with lower surface density or escape speed.  
Although they only ran their photoionization simulations 
up to 3 Myr, \citet{Dale_2013a} found the fraction of unbound gas
at that point was a strongly decreasing function of increasing
cloud escape speed, with only a few percent of the gas unbound when
$v_{\rm esc}$ is comparable to the sound speed of ionized gas.
The simulations of \citet{Raskutti2016} with radiation pressure,
and those of \citet{Kim_JG2021} with both photoionization and radiation
pressure, 
found that that lower SFE was needed to destroy clouds in models
with higher initial virial parameter (i.e. lower escape speed relative to the turbulent velocity).  
\citet{Dale17a} found, in simulations with combined
photoionization and wind feedback, that the gas that becomes unbound
due to feedback  (correcting for mass-loss in control models without feedback) 
does so at a higher rate in models with a larger initial virial parameter.
Simulations with feedback have consistently found a trend of increased SFE at higher surface density (see Section \ref{sec:SFE_theory}). 

From Sections \ref{sec_EUV} and \ref{sec_FUV}, the general
theoretical expectation is that photoionization should be more effective
in destroying clouds than radiation pressure except for very massive
or compact clouds.  For a range of cloud properties
(with $\Sigma\sim 10-10^3 \Msun\, \pc^{-2}$), \citet{Kim_JG2018} compared 
RHD simulations with
radiation pressure only, photoionization only, and combined effects,
and showed that the net momentum injection induced by photoionization
exceeds that from radiation pressure except in massive ($M\ge 10^5\Msun$)
clouds at $\Sigma \gtrsim 500 \Msun\, \pc^{-2}$.
Simulations presented in \citet{ali21}, considering a turbulent
cloud with initial mass $10^5~\Msun$ and radius $10~\pc$ and
employing a Monte Carlo method
for radiation transfer, 
also showed 
that except at sub-pc scales near radiation sources, the radiation
pressure is lower than the gas pressure created by photoionization.  
In all the models of
\citet{Kim_JG2018} with both photoionization and radiation pressure,
the mass loss of ionized gas exceeds the mass loss of neutral gas.  This
means that \textit{photoevaporation -- i.e. photoionization followed by
acceleration of the ionized medium under its internal
pressure gradients  -- is the main
mechanism for cloud destruction}, exceeding mass loss driven by 
ionized gas pressure (plus radiation pressure) acting to accelerate neutral gas out of the cloud.
In the \citet{Kim_JG2018} simulations, the
fraction of the original cloud mass lost as  
ionized gas ranged from $>90\%$ to $30\%$, decreasing as a function
of cloud surface density.  Gas photoevaporated from GMCs would recombine
when radiation is no longer able to reach it, either because of intervening
extinction or because ionizing radiation sources are no longer available. 

Also consistent with theoretical expectations, in Milky Way-like GMCs
the dynamical effects of stellar winds are relatively
unimportant compared to that of photoionization.  
The simulations of \citet{Dale_2013b} comparing effects of stellar winds 
(using just the initial momentum injection) with photoionization
feedback showed that the latter is considerably more effective in
producing unbound gas, while their simulations with combined
feedback \citep{Dale_2014} were only slightly more effective than
photoionization feedback acting along.
\citet{Geen_2021}, in turbulent GMC simulations that investigated
(using controlled studies with an M1 RHD solver) relative effects from
photoionization, winds, and radiation pressure from individual very massive
stars, similarly found at most a $\sim 10\%$ enhancement in the outflowing
momentum for combined models compared to photoionization-only models.

\subsubsection{Galactic-scale ISM simulations}
\label{ssec:galactic_scale_sims}

There have been many galactic-scale ISM simulations (cosmological
zoom, global isolated galaxy, and kpc-scale disk patch) with star
formation and a variety of treatments of feedback, and
several have characterized the statistical properties of structures
similar to molecular clouds for comparison to observations (see Section \ref{sec:properties}).  A subset of these have further made efforts to assess cloud
lifetimes and/or probe the effects of feedback on
cloud destruction.  For example, in their high-resolution local-patch
resimulations of a spiral-arm global model without and with ionizing
radiation feedback, \citet{Bending2020} found that ionizing feedback
is effective in breaking up massive clouds into lower mass clouds,
with clouds defined by a density threshold $n_{\rm H}=300~\pcc$ most
affected. \citet{Haid_2019} zoomed in on two individual molecular
clouds in kpc-scale patch simulations to compare evolution with and
without radiation feedback (computed with a tree method), and found
that even with very similar initial masses, the two clouds evolve
differently 
because one has much higher extinction.

Although their simulations did not include ionizing radiation
feedback, \citet{Grisdale_2018} found masses and surface densities of
clouds much lower in global galaxy simulations with supernova feedback
than without feedback, concluding that feedback is essential for
preventing clouds from becoming too massive and dense compared to
observations. Interestingly, the tracking analysis of clouds in the
same simulations, as presented in \citet{Grisdale_2019}, shows that
most live only 3-4 Myr, suggesting that ionizing radiation feedback
may not be necessary for short cloud lifetimes. \citet{benincasa20}
also applied tracking analysis to clouds at masses $> 10^5\Msun$ in
their cosmological zoom simulations, 
and found that a comparable number die by losing
mass and by merging with other clouds, with cloud lifetimes
independent of the stellar content formed in a cloud, but longer for
clouds with low virial parameter at a reference time.
\citet{Smith_R2020}, comparing
simulations with clustered SN/random/no SN feedback,
zoomed in on individual cloud complexes to follow their evolution and
destruction and found that  masses and sizes of filaments were smaller
with clustered feedback.

Using the ``tuning fork'' diagram (see Figure \ref{fig:simobs}) to investigate
the scale-dependent (de)correlation of dense gas and star formation in
simulations with a range of prescriptions for star formation and feedback,
\citet{Semenov_2021} found that only with their ``full feedback'' model
(including ionizing radiation, early momentum injection to
model radiation pressure and winds, and supernovae) is it possible
to recover the observed level of decorrelation between dense gas and stars
in the age range $2-5\, \Myr$, especially at scales $\lesssim 100~\pc$.  
\citet{Fujimoto19a} were unable to recover the observed 
decorrelation at scales $\lesssim 100~\pc$, which they attributed to ineffective
early feedback.

\citet{Jeffreson_2021b} implemented a subgrid model in moving-mesh simulations
to make momentum injection from early feedback more effective even  where H~\textsc{ii} regions cannot be resolved (mass
resolution $\gtrsim 10^3~\Msun$ and softening length $\gtrsim 20~\pc$).
This treatment is based on an analytic model of H~\textsc{ii} region expansion,
which allows for merging of H~\textsc{ii} regions and directional momentum input
for blister-type regions, but does not allow for the 
factor of 5-10 reduction in momentum injection of both radiation
pressure and ionized gas pressure associated with distributed star
formation and escape of radiation in inhomogeneous clouds (see Sections 
\ref{sec_EUV} and \ref{sec_FUV}).  With this
implementation, the small-scale decorrelation of star formation and
molecular gas is in better agreement with observations.  Cloud-tracking
techniques applied to the Milky-Way-like isolated galaxy simulations
of \citet{Jeffreson_2020} show cloud lifetimes of $\sim 10-20\, \Myr$,
flat above and increasing below a scale of $100\, \pc$
\citep{Jeffreson_2021a}.  In similar simulations for a more massive galaxy
\citep{Jeffreson_2021b},
there is a peak in the lifetime at $M \sim 10^5-10^6~\Msun$.   Also, the
lifetimes of the lowest mass clouds are reduced when the subgrid H~\textsc{ii}
region model is applied, while the lifetimes of high-mass clouds
($\gtrsim 10^6 \Msun$) are insensitive to inclusion of a subgrid
treatment of H~\textsc{ii} regions.

\section{Lifetime Accomplishment}
\label{sec:accomplishment}

We have seen above that GMCs are generally destroyed on timescales of at most a few free-fall times, and that observed GMCs also convert their mass into stars at a rate of only a few percent per free-fall time. Together, these observations suggest that, at least in bulk, GMCs convert relatively little of their mass to stars before being dispersed. However, with new observations and models we can now investigate these questions more precisely: how efficiently do GMCs convert mass to stars, and does this vary with GMC properties or environment? Under what circumstances does this conversion leave behind bound star clusters, rather than unbound field stars? How does the spatial structure of the stellar populations produced -- bound or unbound -- relate to that of the parent GMC?

\subsection{Net star formation efficiency}

\subsubsection{Observational constraints}\label{sec:sfe_obs}

The net star formation efficiency of a GMC, which we denote $\epsilon_*$, is the fraction of the total cloud mass (including all gas accreted over its lifetime) that is converted to stars. It is important to distinguish this from both the star formation efficiency per free-fall time $\epsilon_{\rm ff}$, and the instantaneous ratio of stellar mass to gas mass $M_*/M_{\rm gas}$, which is sometimes also referred to as an efficiency \citep[e.g.,][]{Myers86a, lee16}; while $M_*/M_{\rm gas}$ necessarily goes to infinity as a GMC is dispersed, $\epsilon_*$ is, by definition, bounded between 0 and 1. The net star formation efficiency is fundamentally harder to measure than either $M_*/M_{\rm gas}$ or $\epsilon_{\rm ff}$ (Section~\ref{sec:cloud_SFR}). The difference is that $M_*/M_{\rm gas}$ and $\epsilon_{\rm ff}$ are both instantaneous quantities, which can in principle be determined from observations of a single cloud made at a single time. By contrast, directly measuring $\epsilon_*$ for a single cloud would require observations at (at least) two different times: the gas mass early in the evolution of the cloud, and the stellar mass once it has been fully dispersed, $\sim 10-30$ Myr later \citep{Feldmann11a}. Consequently, measurements of $\epsilon_*$ are generally derived from statistical arguments applied to cloud populations, rather than observations of single GMCs. In the simplest cases the distribution of $\epsilon_*$ follows quite straightforwardly from the observed distribution of $M_*/M_{\rm gas}$. For example, if the star formation rates in individual clouds are constant and most gas is acquired before and dispersed after the star-forming epoch, the mean value of $\epsilon_*$ is simply $2\langle M_*/M_{\rm gas}\rangle$; such a scenario is potentially applicable to $\lesssim 0.1$ pc-scale protostellar cores that form individual stars, and is often implicitly assumed in analyses of such systems \citep[e.g.,][]{Konyves15a}. However, more sophisticated statistical analyses are needed if the processes of accretion, star formation, and gas dispersal overlap in time, or if star formation rates are not constant, as is likely to be the case at GMC scales. 

\citet{McKee97a} and \citet{Williams97a} made one of the earliest attempts to measure $\epsilon_*$, by comparing the distributions of GMC mass and OB association luminosity in the Milky Way. They concluded that on average the net star formation efficiency for Galactic GMCs is $\epsilon_* \approx 5\%$; while the exact figure is dependent on strong assumptions (in particular, the number of ``generations'' of OB associations formed per molecular cloud), the basic result that $\epsilon_* \ll 1$ follows just from the simple observation that the largest GMCs in the Milky Way have masses $\gtrsim 10^6$ M$_\odot$, while the largest OB associations have ionising luminosities that correspond to stellar masses $\approx 10^5$ M$_\odot$.

More precise measurements of $\epsilon_*$ require making use not just of luminosity distributions, but also spatial information, which allows one to correlate GMCs with sites of star formation. \citet{Kruijssen2018} point out that, once the typical GMC lifetime $t_{\rm GMC}$ is deduced from the decorrelation of star formation and gas (Section \ref{sec:lifetime}), the mean efficiency is simply $\epsilon_* = t_{\rm GMC}/t_{\rm dep}$, where $t_{\rm dep}$ is the mean depletion time (see Equation~\ref{eq:tdep}), which can be measured directly from the ratio of total gas mass to star formation rate. Application of this method to the nearby galaxies surveyed by PHANGS-ALMA yields values of $\epsilon_* = 0.02-0.10$ \citep{Kruijssen2019, Chevance2020a}, consistent with \citeauthor{Williams97a}'s Milky Way value. They find no obvious correlations between $\epsilon_*$ and large-scale galaxy properties, though the sample is still quite small.

Unfortunately there has been limited observational exploration to date of the extent to which $\epsilon_*$ varies with either GMC properties or galactic environment. Such a measurement would be challenging, since it would require a large enough sample that one could subdivide it into bins by, for example, galactocentric radius or arm versus interarm, yet still keep the subsamples large enough to allow statistical inferences. There are also claims in the literature that there is a large cloud-to-cloud variance in star formation efficiency, with some clouds forming stars very efficiently \citep[e.g.,][]{Murray11, lee16, Ochsendorf17}; however, we caution that these claims are based on variations in $M_* / M_{\rm gas}$, which as noted above do not constrain $\epsilon_*$, since variations in $M_* / M_{\rm gas}$ could simply reflect the cloud life cycle, with some clouds being nearly-dispersed ($M_*/M_{\rm gas} \gtrsim 1$) while others are just starting to form stars ($M_*/M_{\rm gas} \ll 1$; \citealt{Feldmann11a}). Thus while there are observational constraints on the cloud-to-cloud variation of $\epsilon_{\rm ff}$ (as discussed above), there are at present no published estimates that constrain the cloud-to-cloud variation of $\epsilon_*$.

\subsubsection{Theoretical models}\label{sec:SFE_theory}

Theoretical models for the low mean value of $\epsilon_*$ generally appeal to stellar feedback. Early authors attempted to estimate $\epsilon_*$ by assessing how much mass various feedback mechanisms -- particularly photoionization -- would remove from the molecular phase per unit mass converted to stars \citep{Whitworth79a, McKee97a, Matzner02a, Zamora-Aviles12a, Zamora-Aviles14a}. These calculations generally yield values of $\epsilon_* \sim 5-10\%$, which can be understood at the order of magnitude level from simple scaling arguments. Consider a GMC that forms a stellar population of mass $M_*$ that produces ionizing photons at a rate $Q_i = \Xi M_*$ (c.f.~Section~\ref{sec_EUV}). These photons produce an ionized region of ion number density $n_i$, which from ionization balance must have a characteristic size $L\sim (Q_i / \alpha_{\rm B} n_i^2)^{1/3}$. If the ionized gas is not confined, either by the gravity of the parent GMC or surrounding dense gas, it will escape with a characteristic speed equal to its sound speed $c_i$, leading to a mass loss rate of order $\dot{M} \sim n_i m_{\rm H} L^2 c_i$. If this persists for the lifetime $t_i$ of the ionizing stars, the resulting ratio of the mass photoevaporated to the mass of the star cluster is
\begin{eqnarray}
    \frac{M_{\rm phot}}{M_*} & \sim & c_i m_{\rm H} t_i \left(\frac{\Xi^2}{M_* \alpha_{\rm B}^2 n_i}\right)^{1/3}
    \\ \nonumber
    & = & 11 \left(\frac{n_i}{30\mbox{ cm}^{-3}}\right)^{-1/3} \left(\frac{M_*}{1000\mbox{ M}_\odot}\right)^{-1/3},
\end{eqnarray}
where the numerical value uses $t_i = 3.7$ Myr, $\Xi = 2.5\times 10^{13}$ s$^{-1}$  g$^{-1}$, and $\alpha_{\rm B} = 2.59 \times 10^{-13}$ cm$^3$ s$^{-1}$, and the density to which we have scaled is typical of observed H~\textsc{ii} region densities. Thus we see that a moderate mass cluster, $M_* = 10^3$ $M_\odot$, can evaporate $\sim 10\times$ its own mass, leading to a final star formation efficiency $\epsilon_* \sim 10\%$.

While this result is numerically consistent with observations, it implicitly assumes that all of the GMC mass is either converted to stars or photoionized, and thereby neglects the possibility that some mass might be removed dynamically by the momentum provided by stellar feedback, without the need to change the phase of the gas being removed from molecular to ionized. Which mass loss process dominates -- phase change or dynamical ejection -- likely depends on the properties of the clouds and their environments. In semi-analytic models, \citet[also see \citealt{Krumholz06a}]{Goldbaum11a} find that dynamical disruption dominates for clouds in low-density environments that accrete slowly, while photoevaporation dominates for massive clouds in dense environments. In simulations of isolated clouds, \citet{Kim_JG2018, Kim_JG2021} find that mass loss by photoevaporation generally dominates for massive ($M\ge 10^5$ $M_\odot$) clouds, but that neutral mass loss becomes increasingly important for clouds of increasing gas surface density, virial parameter, and magnetization. \citet{Kim_JG2018} propose a model, calibrated by simulations, for $\epsilon_*$ as a function of cloud properties for cases where photoevaporation dominates.

For the case where dynamical disruption is dominant, \citet{Fall10a} derived a basic result (which has subsequently been expanded by others, e.g., \citealt{Thompson_2016, Raskutti2016, Li_Vogelsberger2019}): the net star formation efficiency that a cloud can reach before being disrupted depends primarily on its surface density $\Sigma$, its virial ratio $\alpha_{\rm vir}$, and the momentum per unit stellar mass $\dot{p}/M_*$ provided by feedback:
\begin{eqnarray}
    \label{eq:epsstar_fall}
    \epsilon_* & \approx & \left(1 + \frac{\Sigma_{\rm crit}}{\Sigma}\right)^{-1} \\
    \Sigma_{\rm crit} & \approx & \frac{\dot{p}/M_*}{\alpha_{\rm vir} G} 
    \\
    \nonumber
    & = & 2200 \alpha_{\rm vir}^{-1}\left(\frac{\dot{p}/M_*}{10\;\mbox{km s}^{-1}\mbox{Myr}^{-1}}\right) \; \mathrm{M}_\odot\mbox{ pc}^{-2}.
\end{eqnarray}
Simple evaluation of this equation with $\Sigma=100$ M$_\odot$ pc$^{-2}$, $\dot{p}/M_* = 10$ km s$^{-1}$ Myr$^{-1}$, and $\alpha_{\rm vir}=1$, values suggested by a combination of observations and simulations as discussed above, gives $\epsilon_* = 0.04$, in reasonable agreement with observations. Theoretical predictions for $\dot{p}/M_*$ more broadly are discussed in Section~\ref{sec:feedback_mechanisms}, where estimates range from $\sim 20$ km s$^{-1}$ Myr$^{-1}$ for direct stellar radiation to $\sim 200$ km s$^{-1}$ Myr$^{-1}$ for photoionized gas pressure.

\begin{figure}[ht]
\includegraphics[width=\columnwidth]{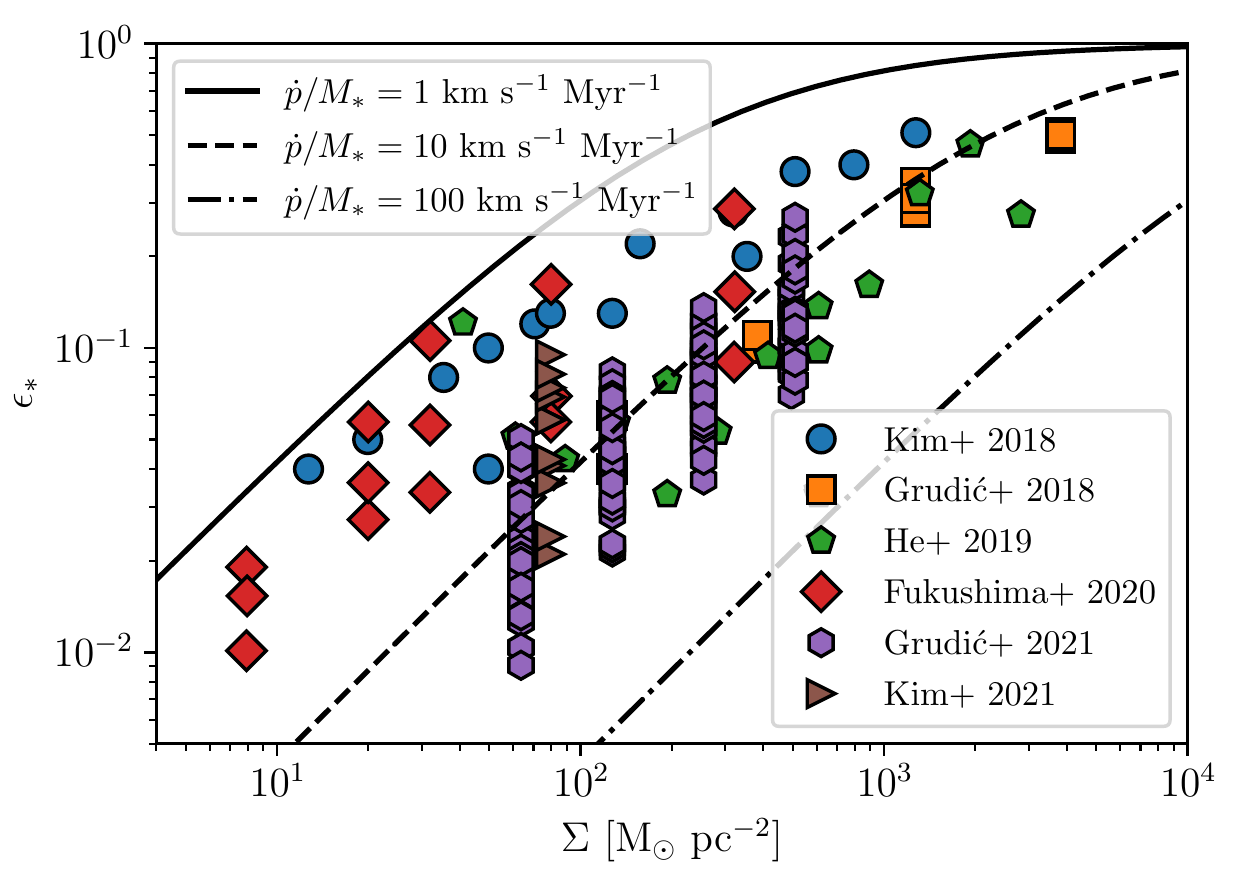}
\caption{
A compilation of integrated star formation efficiencies $\epsilon_*$ obtained in simulations of isolated GMCs with photoionization and other feedback processes; the data shown are from \citet{Kim_JG2018}, \citet{Grudic18a}, \citet{He_Ricotti2019}, \citet{Fukushima_2020}, \citet{Grudic21a}, and \citet{Kim_JG2021} as indicated. Lines show Equation~\ref{eq:epsstar_fall} evaluated with $\alpha_{\rm vir}=1$ and $\dot{p}/M_*$ as indicated in the legend.
\label{fig:sigma_eps}
}
\end{figure}

Full simulations of isolated GMCs with stellar feedback by a number of authors indeed show behaviour similar to the predicted trends of $\epsilon_*$ with $\Sigma$, but also raise some doubts. We show a compilation of simulation results for the dependence of $\epsilon_*$ on $\Sigma$ in Figure \ref{fig:sigma_eps}; we include only cases where authors have simulated a range of column densities, and where the simulations include photoionization feedback, thought to be the most important mechanism as discussed in Section \ref{ssec:individual_gmc_sims}. As the plot shows, the simulations are generally consistent with the functional trend predicted by Equation \ref{eq:epsstar_fall}; experiments by \citet{Li_Vogelsberger2019} in which they independently varied $\dot{p}/M_*$ while leaving other parameters fixed, and by \citet{Kim_JG2021}, who do the same for $\alpha_{\rm vir}$, also yield dependences qualitatively consistent with Equation~\ref{eq:epsstar_fall}. However, it is notable that the quantitative value of $\epsilon_*$ found in simulations for a given $\Sigma$ and $\dot{p}/M_*$ is generally an order of magnitude higher than the numerical estimate provided by Equation \ref{eq:epsstar_fall} \citep[e.g.,][]{Raskutti2016}. The reason for this discrepancy is that Equation \ref{eq:epsstar_fall} implicitly assumes a uniform density distribution, while in a turbulent medium much of the mass is contained in structures with surface densities higher than the mean, and thus more resistant to expulsion. Conversely, turbulence creates underdense regions that can be blown out by feedback even when material at the mean density is retained by gravity. More recent models for $\epsilon_*$ account of these effects, and while they are somewhat more complex, they retain the same basic dependence on $\Sigma_{\rm crit}$ as Equation \ref{eq:epsstar_fall} but numerical coefficients in better agreement with simulations \citep{Thompson_2016, Raskutti2016}.

Figure~\ref{fig:sigma_eps} shows that, for surface densities comparable to those sampled thus far by observations ($\Sigma\sim 100$ M$_\odot$ pc$^{-2}$), simulations of individual clouds with feedback yield $\epsilon_* \sim 0.01 - 0.1$, consistent with the observational constraints. Moreover, Figure \ref{fig:sigma_eps} represents a clear prediction that should be testable by future surveys: integrated star formation efficiencies should be systematically higher in regions where GMC surface densities are higher, for example galactic centers or starburst regions. However, the figure also shows that simulations by different authors often differ at almost the order of magnitude level in their predicted value of $\epsilon_*$ at fixed $\Sigma$. Some of this is due to variations in the choice of initial condition (which affects $\alpha_{\rm vir}$ and the amount of magnetic support in the cloud), but much of the variation appears to be due to differences in the simulation methods. For example, the efficiencies found by \citet{Grudic21a} are systematically a factor of $\sim 5$ smaller than those obtained by \citet{Kim_JG2018} or \citet{Fukushima_2020}, despite the fact that their initial conditions are quite similar. The exact origin of the difference is difficult to pinpoint given the many ways in which the simulations vary; while all three include both photoionization and direct radiation pressure, they differ in resolution, hydrodynamic method (Lagrangian versus fixed grid versus static mesh refinement grid), radiative transfer method (ray tracing versus a variant of the Str\"omgren volume method), and whether they include other forms of feedback beyond photons. The effectiveness of feedback seems to be quite sensitive to these details, along with others such as how the IMF is sampled \citep{Grudic19b}, how radiative momentum is deposited on the computational grid \citep{Hopkins18a}, and how well the dusty absorption regions around individual sources are resolved \citep{Krumholz18c}. This sensitivity means that the precise predictions for $\epsilon_*$ must still be regarded as substantially uncertain.

Thus far we have discussed only models of isolated clouds, because there have been far fewer efforts to study $\epsilon_*$ in the context of simulations that include whole galaxies. This is in part due to purely technical difficulties: doing so would require both tracking individual clouds (as already implemented for example by \citealt{Grisdale_2019}, \citealt{benincasa20}, and \citealt{Jeffreson_2021a}), and tracking which stars formed in which clouds. No published work thus far has attempted such a direct measurement. Indirectly, however, efforts to reproduce the ``tuning fork'' diagram (Section~\ref{ssec:galactic_scale_sims}), necessarily offer some constraint on $\epsilon_*$, since the shape of the lower branch of the tuning fork is sensitive to this value (c.f.~Figure 11 of \citealt{Kruijssen2018}). To the extent that a particular simulation matches observed turning forks \citep[e.g.,][]{Semenov_2021}, its value of $\epsilon_*$ must be in reasonable agreement with observations.

\subsection{Formation of bound clusters}
\label{sec:bound_clusters}

The central problem of bound cluster formation as it relates to GMCs is to understand why only $\sim 10\%$ of stars in the Milky Way and similar galaxies form in clusters that survive more than $\sim 10$ Myr after star formation. What sets the number, and what distinguishes the stars that do end up in bound clusters from those that are destined to be field stars? In principle the answer to this question should be closely related to the question of star formation efficiency, since more efficient star formation should leave behind more bound stellar systems \citep[e.g.][]{Kruijssen2012, Adamo2020, Grudic21a}, and to the question of whether GMCs are collapsing locally or globally (Section \ref{sec:collapse}). However, the exact connection between efficiency, collapse morphology, and boundedness is tied up the larger question of how bound clusters form in the first place. Since this topic was reviewed extensively in \citet{KrumholzARAA19}, we will only summarize parts of this review, and limit further discussion to more recent developments.

As discussed in Section~\ref{sec:collapse} and \ref{sec:lifetime}, star-forming regions appear to form hubs at the centers of networks of filaments, with the filaments acting as conveyor belts to feed the hubs. These hubs are a natural candidate to be the progenitors of star clusters, and within them star formation appears to accelerate, but also to take place over several local free-fall times. As a result of this extended period of star formation, stars have the opportunity to relax and virialize independently from the gas; virial or close-to-virial velocities are observed in several nearby young clusters \citep{Kim19a, Lim20a, Theissen21a}. 
The origin and fate of these hubs and the virialized stars they contain is closely linked to the question of whether GMC collapse is local or global: if the hubs represent the focii of a global collapse with most stars forming at the collapse center, then we arrive at a scenario where most stars are formed in a region that is bound and virialized, but that subsequently becomes unbound when stellar feedback removes the gas, allowing $\sim 90\%$ of the stars to escape \citep[e.g.,][]{Vazquez-Semadeni19}. On the contrary, if collapse is mostly local, feedback can be less violent, because there is no need for it to unbind the virialized stars in hubs. Instead, the explanation for why 90\% of stars do not form in bound clusters is that 90\% of stars do not form in hubs at all; they form in a more distributed arrangment around the hubs, one that never virializes, and that is very easily unbound by even relatively mild feedback \citep[e.g.,][]{Krumholz20}.

As discussed in Section \ref{sec:collapse}, measurements of gas and stellar kinematics have not cleanly distinguished between the two scenarios. Instead, the strongest evidence thus far
comes from a budget argument pointed out by \citet{Krumholz20}. The overall star formation rate of the Milky Way implies that dense star-forming clumps such as those observed by ATLASGAL \citep{Urquhart18a} cannot convert the majority of their mass to stars, i.e., they must have $\epsilon_* \ll 1$. However, observations show that $\epsilon_{\rm ff} \sim 0.01$ is essentially constant in all observed star-forming systems, including the dense clumps (Section~\ref{sec:cloud_SFR}), so for a collapse scenario to produce a star formation history that accelerates, as we observe in dense clusters, the density must increase (and thus the free-fall time decrease) faster than the mass is dispersed by feedback. However, this in turn implies $\epsilon_* \sim 1$, i.e., efficient star formation. There is no way in a global collapse scenario for star formation to simultaneously produce low integrated efficiency $\epsilon_*$, low efficiency per free-fall time $\epsilon_{\rm ff}$, and still produce star formation histories comparable to what we observe. Thus the cluster formation scenario that we tentatively favor is that one where collapse is primarily local rather than global. However, because this is a statistical argument, it can only tell us about the average mechanism of star formation in the Milky Way; it does not preclude the possibility that individual regions might form in a global collapse.

\subsection{How does the ISM hierarchy translate into the young stellar population?}

Bound star clusters are the inner part of a hierarchy of stellar structures that extends over scales of kpc, just as the dense clumps from which clusters form are the inner parts of a much larger hierarchical structure defined by GMCs, and, on even larger scales, associations and cloud complexes. Numerous statistial tools exist in the literature for characterizing these structures: for example the $\Delta$-variance \citep{Schneider11a, Dib20a} and multifractal models \citep{Elia18a, Robitaille19a, Yahia21a} for the gas, and Bayesian mixture \citep{Kuhn14a, Kounkel18} and density clustering descriptors \citep{Joncour18a, Gonzalez21a} for the stars.
As with star clusters, this topic was recently reviewed by \citet{Gouliermis18a}, so we focus on what is new since this work.

While more complex statistics exist, most star-gas comparisons have focused on a simpler descriptor: the two point correlation function, or, equivalently, the fractal dimension. These two are nearly equivalent, since a fractal seen in projection with projected (2D) dimensionality $D_2$ gives rise to a two-point angular correlation function that is a powerlaw of slope $\alpha_2 = D_2 - 2$ \citep{Calzetti89a}. Measurements of either $\alpha_2$ or $D_2$ in cold gas are surprisingly scarce in the literature. \citet{Falgarone91a} find $D_2 \sim 1.3$ for the CO emission from Milky Way molecular clouds, while \citet{Westpfahl99a} find $D_2 \sim 1.2 - 1.5$ for the H~\textsc{i} for several galaxies in the M81 group.
On the other hand, \citet{Grasha19}, find no significant fractal structure in the positions of M51 GMCs with masses $<3\times 10^6$ M$_\odot$; only the most massive clouds are fractally-distributed, with $D_2\sim 1.7$.

The two-point correlation function of young stars has been studied much more extensively. For individual young stars, \citet{Hennekemper08a}, \citet{Kraus08a}, and \citet{Sun17a} all report that the distribution of stellar surface densities (which is equivalent to the correlation function) is well-described by a powerlaw on scales as small as $\sim 0.01$ pc and as large as the sizes of the regions surveyed, $\gtrsim 1$ pc, with a slope corresponding to a fractal dimension $D_2 \sim 1 - 1.5$. Young stellar clusters, which can be seen to much larger distances, also show powerlaw two-point correlation function with similar slopes, $D_2 \sim 1.5$, over scales ranging up to kpc \citep[e.g.,][]{Scheepmaker09a, Gouliermis15a, Gouliermis17a, Grasha17a, Grasha18, Grasha19}. \citet{Menon21a} analyze the correlation function of star clusters in 12 galaxies observed by \textit{HST} as part of the LEGUS survey \citep{Calzetti15a, Adamo17a}. They find that young clusters in all galaxies show fractal structure, but that $D_2$ varies from $0.5 - 1.9$, with no obvious correlation with galaxy properties such as mass or star formation rate. 

While the similarity of the fractal dimensions for gas and stars might seem to suggest that the spatial statistics of young stars are simple reflections of those found in the gas from which they form, direct comparisons between gas and stars do not suggest a simple, linear relationship; instead, the stellar surface density rises superlinearly with the gas surface density, indicating that young stars are more strongly clustered than the gas from which they form \citep{Lada17a, Pokhrel20a, Retter21a}, but that clustering diminishes with age on the crossing timescale of the stellar system:
tens of Myr for star clusters \citep{Grasha17a, Grasha18, Grasha19, Menon21a}, and $\sim 1$ Myr for individual
class I and class II YSOs \citep{Pokhrel20a}. 
This suggests a somewhat more complex evolutionary story: stars are born more clustered than the gas from which they form, because denser gas forms stars more rapidly; indeed, analytic models suggest typical correlation functions at birth should have slopes corresponding to $D_2 \sim 1$ \citep{Hopkins13a, Guszejnov17a, Guszejnov18b}.
However, since this structure is erased on a crossing timescale, the actual spatial structure we observe for a population of small but non-zero age is somewhat less clustered than this theoretical birth limit.

\section{Future prospects}
\label{sec:future}

As the preceding sections have shown, perhaps the greatest observational advance since {\it PPVI}
is that large surveys now provide statistically meaningful samples of GMCs and young stars across a wide range of environments, while simulations now include much more realistic treatments of both stellar feedback and the large-scale galactic environment.
While there have been major advances, current work also has limitations.  Here, we will review these with an eye toward the future, also discussing efforts towards spatially-matched comparisons between theory and observations.

\subsection{Insights from recent studies}

Since {\it PPVI}, it has become qualitatively clear that the environment 
plays a major role in forming, evolving, and destroying GMCs. 
Neverthless, more work on both observational and theoretical sides is needed for fully quantitative characterization of environmental dependencies. 

{\bf The environmental dependence of cloud properties and scaling relations.} 
With ALMA, GMCs are resolved in the Local Group and beyond, revealing variations of cloud surface densities and other properties with environment (see Sections~\ref{sec:gmc_env} and \ref{sec:galactic_environment}). However, complementary dust emission observations (e.g.~with {\it JWST}, see below), as well as observations of the atomic and molecular gas at higher spatial resolution or in different tracers (e.g., HCN to trace the dense gas as demonstrated with the EMPIRE survey in \citealt{jimenezdonaire2019}), are needed to more robustly characterize GMCs in external galaxies and to dissect the dependence of scaling relations on environment.

While the PHANGS-ALMA survey 
has been particularly informative 
owing to the large number of observed galaxies and the relatively good spatial resolution ($\sim$ 50--120 pc scales across about 90 galaxies), 
the PHANGS 
sample covers a limited range of environmental properties. 
A wider variety of galactic environments, from extremely low-metallicity dwarfs to central regions of interacting systems, both in the local and the high redshift Universe \citep[e.g., high-z analogs as probed by the DYNAMO survey  of][]{fisher2017}, will provide additional leverage. These will be crucial to further investigate how the environment affects, or perhaps, regulates e.g., H~\textsc{i}-to-H$_{2}$ conversions, cloud surface densities, velocity dispersions, and virial parameters.        

{\bf The environmental dependence of star formation within GMCs.} While the star formation efficiency per free-fall time, $\epsilon_{\rm ff}$, has been measured across a large number of nearby galaxies, there have been far fewer observations aimed at studying variations of the net star formation efficiency, $\epsilon_*$. Recent studies using a combination of molecular gas and star formation tracers have demonstrated that measuring the instantaneous star formation efficiency from observations is possible, but turning this into $\epsilon_*$ requires statistical methods (see Section~\ref{sec:sfe_obs}). Furthermore, the small number of systems analyzed thus far precludes drawing any conclusions about how $\epsilon_*$ depends on environment or GMC properties. Further investigation requires expanding the number of sampled galaxies, including a larger diversity of environments (e.g., dwarf starburst galaxies, extremely low-metallicity conditions), as well as connecting star formation in the local Universe to star formation at cosmic noon \citep[e.g.,][]{zanella2015,dessauges2019}. These can then be linked to results from theoretical models, which, for example, predict that $\epsilon_*$ increases with GMC surface density (see Section~\ref{sec:SFE_theory}). 

{\bf GMC destruction and lifetime as a function of environment.} As discussed in Section~\ref{sec:destr_env}, the relative importance of the range of possible GMC destruction mechanisms as a function of cloud properties and environment requires further investigation. On the numerical side, this can be achieved by exploiting high-resolution simulations of individual clouds to inform subgrid models used in large-scale simulations. To match this, observations with large scale coverage and high spatial resolution are needed  to capture GMC destruction in a statistical sample across different environments.
In addition to sampling different environmental circumstances (e.g., metallicity, host galaxy properties, arm vs.\ inter-arm, location within the galaxy), multi-wavelength observations are crucial to measure the stellar content and signatures of different feedback mechanisms across the electromagnetic spectrum: SN blasts and shocked stellar winds in the X-Ray regime; direct radiation pressure and ionised gas in H~\textsc{ii} regions in the optical; and dust processed radiation pressure in the IR.

\subsection{Upcoming surveys, telescopes, and instruments}

\subsubsection{Ground-based optical and near-infrared facilities}
A multitude of new instruments and facilities, 
both ground- and space-based, will see first light in the next decade. Among the first will be the Local Volume Mapper (LVM; \citealt{kollmeier17}), an optical integral field unit (IFU) survey that is part of SDSS-V. 
In addition to surveying the Milky Way at $\sim$ 1 pc resolution, LVM will cover the Small and the Large Magellanic Clouds, 
yielding an unprecedented contiguous spectroscopic map at $\sim$ 10 pc resolution, 
which will provide the means to 
study the interplay between the stars and the ISM from from cloud to galaxy-wide scales in these two galaxies. 

Further, the next generation of high spatial and spectral resolution optical and near-infrared IFU instruments is being built for existing facilities. Examples of upcoming AO-supported instruments include VLT/MAVIS (MCAO Assisted Visible Imager and Spectrograph; \citealt{mcdermid20}), 
and GIRMOS (Gemini Infrared Multi-Object Spectrograph, planned for 2023; \citealt{sivanandam18}). Efforts in constraining feedback-driving massive stellar populations will also be facilitated by future IFU instruments covering bluer (near-UV) wavelengths, e.g., BlueMUSE \citep{richard19}, the successor of the highly successful MUSE instrument that has been used to extensively map the spatially resolved ionized ISM in nearby galaxies in the optical \citep[e.g.,][]{kreckel2019,mcleod19,mcleod2020,dellabruna2020,James2020}.

The near-UV, optical, and near-infrared will also see significant advances thanks to the next generation large, 30m-class, ground-based facilities (e.g., the Extremely Large Telescope, ELT; the Giant Magellan Telescope, GMT; and the Thirty Meter Telescope, TMT). HARMONI on the ELT will revolutionize the optical and near-infrared ground-based IFU landscape by providing adaptive optics-supported coverage in the near-IR, allowing the characterization of the resolved stellar and gaseous components in nearby galaxies (e.g., \citealt{gonzalez18}), interfacing with space-based missions like {\it JWST}. Enticing prospects of instruments like HARMONI on ELTs include extending highly resolved stellar feedback studies that directly link individual massive stars to the feedback-driven gas, currently only achievable in the Magellanic Clouds \citep[e.g.,][]{mcleod19}, to extreme systems at Mpc distances such as the starburst galaxy NGC~253.

\subsubsection{The long wavelength and long baseline regimes}
With its very large field-of-view and sub-30" resolution, CCAT-prime \citep[][]{stacey2018} will resolve the early stages of molecular cloud formation (see Section~\ref{sec:neutral_to_mol}) by probing CO-dark H$_{2}$ gas traced by fine structure lines of atomic carbon ([C~\textsc{i}]) and mid-$J$ CO lines \citep{simon2019astro2020}. Programs like GEco \citep[Galactic Ecology of the dynamic ISM --][]{GEcoDecadal} will spectrally map the Milky Way and nearby galaxies,
providing the statistical sample of observations needed to connect the nearby and the high-redshift Universe, where spatially-resolved [C~\textsc{i}] observations are beyond the reach of even ALMA.

Sub-kpc scale mapping of the cold atomic and molecular gas across entire galaxies is a key science goal of the next generation VLA (ngVLA; \citealt{mckinnon19}), 
which will improve on existing and upcoming radio observatories by an order of magnitude in sensitivity and spatial resolution. ngVLA will push GMC lifetime studies \citep[e.g.,][]{Kruijssen2019} on $<$ 100 pc scales to more distant galaxies, therefore covering a much larger dynamic range in galaxy properties, and probing the relative importance of early stellar and SN feedback \citep[see below; e.g.,][]{Semenov_2021}. Moreover, the ngVLA will be able to efficiently mosaic entire nearby galaxies, reaching sensitivities needed to reliably measure cloud luminosities, kinematics, and CO surface brightness \citep[e.g.][]{leroy2018}. For the nearest clouds this means vastly improved turbulence studies (e.g.,\citealt{heyer04}, \citealt{burkhart2015}) and column density distribution function analyses \citep[e.g.,][]{kainulainen09}.

Also on the radio horizon is the Square Kilometre Array (SKA), which will bridge the gap between LOFAR and ALMA by covering the frequency ranges of 350 MHz to 15.3 GHz (SKA1-mid) and 50 to 350 MHz (SKA1-low) at high angular resolution, high sensitivity, and large field of view. One of the main aims of this observatory will be to map the Southern sky in the HI 21-cm line, as well as in hydrogen and carbon radio recombination lines. With the capability of mapping galaxies out to $z \sim 1$ at about 1", as well as reaching sensitivities needed to perform HI studies in star-forming Local Group dwarf galaxies, observations from the full SKA will be perfectly complementary to the high spatial resolution data sets of nearby galaxies obtained throughout the electromagnetic spectrum and mentioned above (i.e.~from optical IFU surveys, {\it JWST}, VLA, and ALMA), providing a full picture of the multi-phase ISM that makes up the evolutionary stages of GMCs as they cool, form stars, and disperse \citep[e.g.,][]{beswick2015}.

\subsubsection{Upcoming space-based missions}
The successful launch of {\it JWST} has opened a new high resolution window in the near- to mid-IR regime. The list of approved Cycle 1 proposals highlights the unique capabilities of {\it JWST} to probe star formation, stellar feedback in galaxies at Mpc distances at the same spatial scales currently achieved only in Milky Way and nearest systems. Of particular interest will be programs that use {\it JWST} in conjunction with co-spatial ancillary data from {\it HST}, optical IFUs, and ALMA.
{\it JWST} will identify and characterize individual (massive) young stellar objects at much larger distances circumventing the necessity of (and uncertainties from) inferring the star-forming activity within young, embedded star-forming regions from integrated properties or unreliable tracers for these evolutionary stages (e.g., H$\alpha$). Crucially, this will trace star formation in nearby galaxies between the earlier and later stages as resolved by ALMA CO and optical H$\alpha$ maps (respectively),
enabling more accurate timescale measurements of the different stages of star formation, and therefore providing vastly improved GMC lifetimes \citep[see e.g.,][]{KimChevance21}. Beyond Mpc distances at redshifts of $\sim$ 1, {\it JWST} will also enable the study of commonly-used optical tracers of the feedback-driven ISM such as H$\alpha$ emission at GMC scales. 

The high-energy regime will see significant advancements with X-Ray missions like Athena \citep{nandra13} and Lynx \citep{gaskin19} planned for the 2030s. 
These missions will be crucial to constrain the contribution of winds from massive stars and supernova events in regulating the evolution and disruption of GMCs (see e.g., \citealt{sciortino13,decourchelle13}). For example, it remains unclear how and what fraction of the energy from stellar winds is lost \citep[e.g.,][]{Rosen_2014,Lancaster_2021a}. This can be addressed with X-Ray coverages reaching $\sim$ 0.2 keV, as they give access to strong cooling lines such as O~\textsc{vi} and O~\textsc{vii} stemming from the hot, stellar wind-driven gas ($\sim 10^{5}$ K $<$ T $<$ 10$^{7}$ K), enabling a quantification of the energy budget from stellar winds.  

\subsection{Connecting observations and simulations}
\label{sec:obs-sim}

With the large statistical samples provided by recent observational and numerical campaigns, it is becoming possible to address 
two broad classes of questions: (i) how accurate are the analyses commonly applied to 
observations of GMCs when used on synthetic data from simulations? (ii) in what ways do the simulations agree and disagree with the observations, and hence, how can the simulations be improved upon? 

High resolution simulations of isolated disk galaxies that include physically-motivated prescriptions for stellar feedback are now detailed enough for  direct comparisons with observations. 
For example, the models presented in \citet{Jeffreson_2021b} and discussed in previous sections of this review, produce molecular clouds which follow the $\sigma - \Sigma$ relation (velocity dispersion - surface density) and that have lifetimes in excellent agreement with observations \citep{Sun18}. Their results also underpin that early stellar feedback (in particular the thermally-driven expansion of HII regions, in agreement with observations, e.g., \citealt{lopez14,mcleod19, chevance2022}) significantly affects the GMC properties while also reducing supernova clustering \citep[see also][]{Smith_MC2021} and therefore reducing the importance of SNe on individual GMC scales. This latter finding adds to a growing numerical evidence that pre-supernova feedback in GMCs produces lower densities and enhanced channels for supernova energy to escape, which is now also target of quantitative observational studies \citep[e.g.,][]{mcleod21}. 

\begin{figure}[h]
    \centering
    \includegraphics[scale=0.43]{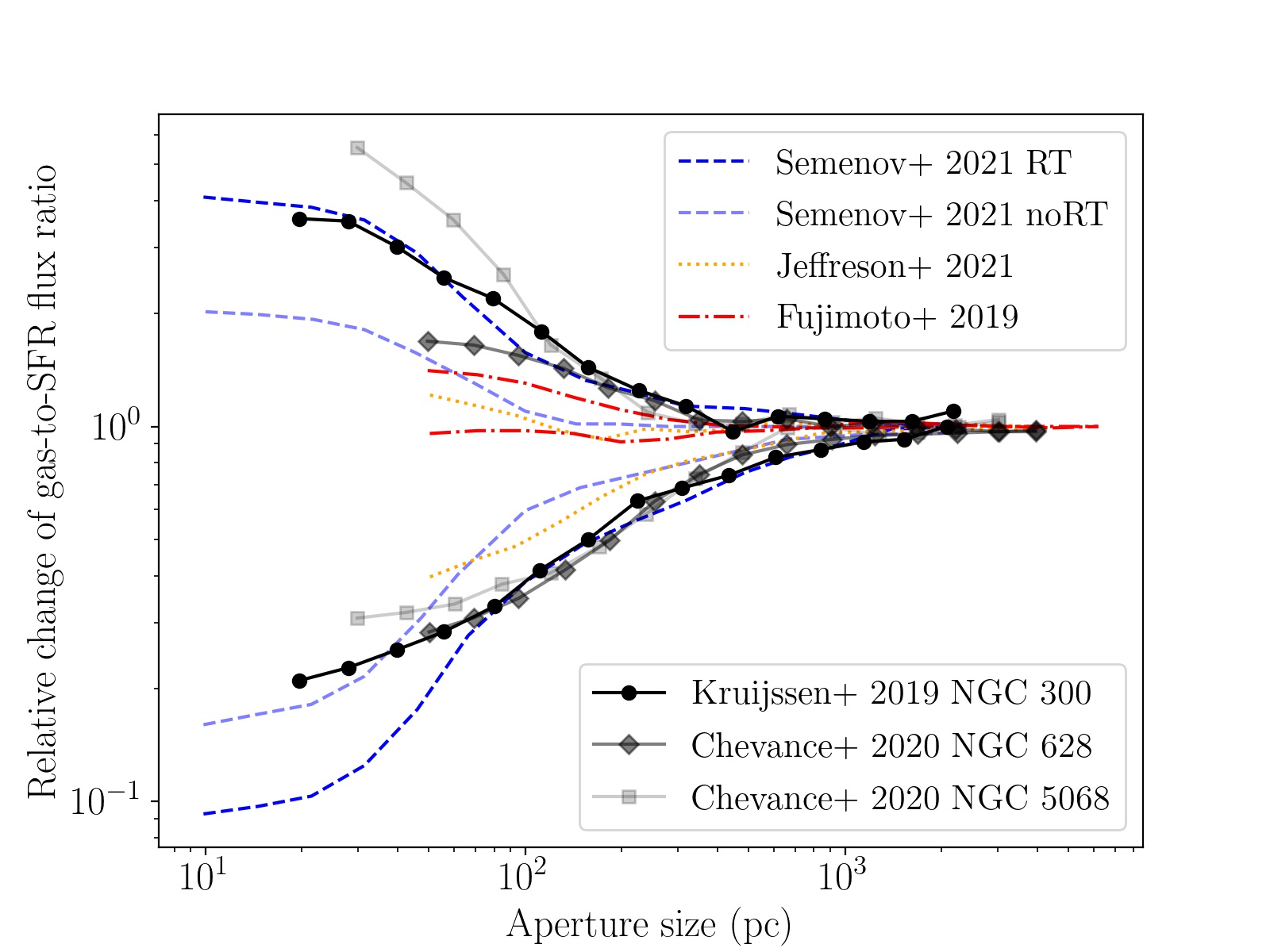}
    \caption{The so-called ``tuning fork diagram" illustrating the spatial decorrelation between gas and stars, as traced by the gas-to-SFR flux ratio relative to the galactic average value as a function of aperture size, obtained from both observations (of NGC~300, \citealt{Kruijssen2019}; of NGC~628 and NGC~5068, \citealt{Chevance2020a}) and simulations (\citealt{Fujimoto19a}; \citealt{Semenov_2021}  with and without explicit radiative transfer, RT, modeling; and \citealt{Jeffreson_2021b}). The upper and lower branches are the result of focusing apertures on molecular gas and stellar peaks, respectively. The lines correspond to best-fit models of \cite{Kruijssen2018} to the synthetic and observed peaks, which yield molecular cloud lifetimes and feedback timescales. The comparison of simulated tuning fork diagrams to observations tests the adopted stellar feedback prescriptions, indicating the importance of early (pre-supernova) feedback in removing gas around emerging clusters,  crucial to reproduce the observed decorrelation at small aperture sizes.
    }
    \label{fig:simobs}
\end{figure}

Of particular interest are analysis methods which can be readily applied to both observations and simulations, thus providing the means for a direct comparison between the two. One of these is a novel method presented in \citet[see also \citealt{Schruba10}]{Kruijssen2018} , which translates the statistical behavior of the relation between molecular gas and star formation to empirical constraints on baryon cycle in the ISM down to GMC scales given tracers for the molecular gas (e.g., CO) and the stars (e.g., H$\alpha$), providing direct measurements of molecular cloud lifetimes and feedback timescales. This method has been applied to a number of nearby galaxies \citep[e.g.,][]{Kruijssen2019,Chevance2020a, ward2020,zabel2020,Kim_JG2021, chevance2022} as well as numerical models \citep[e.g.,][]{Kruijssen2018,Fujimoto19a,Haydon20b,Semenov_2021}.
This is shown in Figure \ref{fig:simobs}, which compares the spatial decorrelation between stars and gas obtained from different simulations and observations. Motivated by the results of \citet{Kruijssen2019}, in their recent study \citet{Semenov_2021} use this tool to analyze how the spatial decorrelation between gas and stars changes as a function of different numerical treatments of stellar feedback and ISM conditions in a suite of simulations tailored to match the nearby galaxy NGC~300. 
They find that on scales $\lesssim$ 100 pc, radiative and mechanical feedback are the main drivers affecting the gas/star decorrelation, with supernovae taking over at scales above $\sim$ 100 pc. Moreover, \citeauthor{Semenov_2021} demonstrate that the distribution of cloud lifetimes is connected to the degree of spatial (de-)correlation. The empirical constraints obtained using this statistical method can be directly turned into subgid recipes for galaxy simulations \citep{Keller2022}, and comparisons between observed or simulated GMC lifetimes using the spatial decorrelation method and those obtained directly from simulations (e.g., \citealt{benincasa20}) will yield further insight into dependencies on e.g., environment or stellar feedback prescriptions.

Another statistical approach to analyzing GMC properties is encoded in {\sc turbustat} \citep{koch19}, a {\sc Python} package specifically designed to extract a variety of different turbulence statistic from spectral line (e.g., CO) data cubes. This tool can be applied to observational and synthetic data cubes alike, thus offering the means of analyzing observations and simulations of turbulent GMCs in the same manner. For example, \citet{boyden18} apply this tool to synthetic CO observations of simulated turbulent GMCs with the aim of assessing the impact of different post-processing assumptions on a variety of turbulence statistics, finding that implementing realistic chemical modeling is crucial to statistically compare models and observations. 
As simulations become progressively more sophisticated, these simulated observations can be used to benchmark commonly used observational diagnostics \citep[e.g.,][]{Mao_2020} and ideally to inspire new approaches for characterizing GMCs.

Overall, both observations and simulations are transitioning towards resolving a wider dynamic range in terms of both spatial scales and ISM phases. 
While pc and sub-pc scale observations are becoming available for a growing number of 
galaxies, numerical efforts will go towards building observationally-informed suites of GMC simulations ranging from individual star-forming molecular clouds to high resolution cosmological zoom-ins, as well as developing meaningful ways of connecting them to observations.

\bigskip
\textbf{Acknowledgments} 
The authors thank Ian Bonnell and an anonymous referee for helpful suggestions on the manuscript.
MC gratefully acknowledges funding from the Deutsche Forschungsgemeinschaft (German Research Foundation) through an Emmy Noether Research Group (grant number KR4801/1-1) and from the European Research Council (ERC) under the European Union's Horizon 2020 research and innovation programme via the ERC Starting Grant MUSTANG (grant agreement number 714907). 
MRK acknowledges support from a Humboldt Research Award from the Alexander von Humboldt Foundation, from the Australia-Germany Joint Research Cooperation Scheme (UA-DAAD), and from the Australian Research Council through awards DP190101258 and FT180100375.  
AFM was partially supported by NASA through the NASA Hubble Fellowship grant No.~HST-HF2-51442.001-A awarded by the Space Telescope Science Institute, which is operated by the Association of Universities for Research in Astronomy, Incorporated, under NASA contract NAS5-26555. 
The work of ECO was partially supported by NASA ATP through grant No.~NNX17AG26G and by the NSF through AARG grant No.~AST-1713949.
EWR acknowledges the support of the Natural Sciences and Engineering Research Council of Canada (NSERC), funding reference number RGPIN-2017-03987. AS acknowledges support by the German Science Foundation via DFG/DIP grant STE 1869/2 GE 625/17-1, and by grants from the Center for Computational Astrophysics (CCA) of the Flatiron Institute, and the Mathematics and Physical Sciences (MPS) division of the Simons Foundation.

\bigskip

\bibliographystyle{ppvi_lim1}
\bibliography{ppvii.bib}

\begin{thebibliography}{501}
\parskip=0pt \itemsep=0pt \small \baselineskip=11pt
\providecommand{\natexlab}[1]{#1}

\bibitem[\protect\astroncite{\emph{{Adamo} et~al.}}{2017}]{Adamo17a}
{Adamo} A. et~al. (2017) \emph{\apj}, \emph{841}, 131.

\bibitem[\protect\astroncite{\emph{{Adamo} et~al.}}{2020}]{Adamo2020}
{Adamo} A. et~al. (2020) \emph{\ssr}, \emph{216}, 4, 69.

\bibitem[\protect\astroncite{\emph{{Agertz} et~al.}}{2013}]{Agertz_2013}
{Agertz} O. et~al. (2013) \emph{\apj}, \emph{770}, 1, 25.

\bibitem[\protect\astroncite{\emph{{Alatalo} et~al.}}{2013}]{Alatalo13}
{Alatalo} K. et~al. (2013) \emph{\mnras}, \emph{432}, 3, 1796.

\bibitem[\protect\astroncite{\emph{{Ali} et~al.}}{2018}]{Ali2018}
{Ali} A. et~al. (2018) \emph{\mnras}, \emph{477}, 4, 5422.

\bibitem[\protect\astroncite{\emph{{Ali}}}{2021}]{ali21}
{Ali} A.~A. (2021) \emph{\mnras}, \emph{501}, 3, 4136.

\bibitem[\protect\astroncite{\emph{{Alves de Oliveira}
  et~al.}}{2014}]{alvesdeoliveira14}
{Alves de Oliveira} C. et~al. (2014) \emph{\aap}, \emph{568}, A98.

\bibitem[\protect\astroncite{\emph{{Andr{\'e}} et~al.}}{2014}]{andre14}
{Andr{\'e}} P. et~al. (2014) in: \emph{Protostars and Planets VI}, (edited by
  H.~{Beuther}, R.~S. {Klessen}, C.~P. {Dullemond}, and T.~{Henning}), p.~27.

\bibitem[\protect\astroncite{\emph{{Armillotta} et~al.}}{2020}]{Armillotta20}
{Armillotta} L. et~al. (2020) \emph{\mnras}, \emph{493}, 4, 5273.

\bibitem[\protect\astroncite{\emph{{Ballesteros-Paredes} and
  {Hartmann}}}{2007}]{BallesterosParedes07}
{Ballesteros-Paredes} J. and {Hartmann} L. (2007) \emph{\rmxaa}, \emph{43},
  123.

\bibitem[\protect\astroncite{\emph{{Ballesteros-Paredes}
  et~al.}}{2007}]{PPV_2007}
{Ballesteros-Paredes} J. et~al. (2007) in: \emph{Protostars and Planets V},
  (edited by B.~{Reipurth}, D.~{Jewitt}, and K.~{Keil}), p.~63.

\bibitem[\protect\astroncite{\emph{{Ballesteros-Paredes}
  et~al.}}{2011}]{BallesterosParedes11}
{Ballesteros-Paredes} J. et~al. (2011) \emph{\mnras}, \emph{411}, 1, 65.

\bibitem[\protect\astroncite{\emph{{Ballesteros-Paredes}
  et~al.}}{2020}]{BallesterosParedes20}
{Ballesteros-Paredes} J. et~al. (2020) \emph{\ssr}, \emph{216}, 5, 76.

\bibitem[\protect\astroncite{\emph{{Bally}}}{2016}]{Bally_2016}
{Bally} J. (2016) \emph{\araa}, \emph{54}, 491.

\bibitem[\protect\astroncite{\emph{{Bally} et~al.}}{1987}]{Bally87}
{Bally} J. et~al. (1987) \emph{\apjl}, \emph{312}, L45.

\bibitem[\protect\astroncite{\emph{{Barnes} et~al.}}{2017}]{Barnes17}
{Barnes} A.~T. et~al. (2017) \emph{\mnras}, \emph{469}, 2, 2263.

\bibitem[\protect\astroncite{\emph{{Barnes} et~al.}}{2018}]{Barnes18}
{Barnes} A.~T. et~al. (2018) \emph{\mnras}, \emph{475}, 4, 5268.

\bibitem[\protect\astroncite{\emph{{Barnes} et~al.}}{2020}]{barnes20}
{Barnes} A.~T. et~al. (2020) \emph{\mnras}, \emph{498}, 4, 4906.

\bibitem[\protect\astroncite{\emph{{Bastien} et~al.}}{1991}]{Bastien91a}
{Bastien} P. et~al. (1991) \emph{\apj}, \emph{378}, 255.

\bibitem[\protect\astroncite{\emph{{Bate}}}{2009}]{Bate09a}
{Bate} M.~R. (2009) \emph{\mnras}, \emph{392}, 1363.

\bibitem[\protect\astroncite{\emph{{Bate} et~al.}}{2003}]{Bate03a}
{Bate} M.~R. et~al. (2003) \emph{\mnras}, \emph{339}, 577.

\bibitem[\protect\astroncite{\emph{{Bending} et~al.}}{2020}]{Bending2020}
{Bending} T. J.~R. et~al. (2020) \emph{\mnras}, \emph{495}, 2, 1672.

\bibitem[\protect\astroncite{\emph{{Benincasa} et~al.}}{2020}]{benincasa20}
{Benincasa} S.~M. et~al. (2020) \emph{\mnras}, \emph{497}, 3, 3993.

\bibitem[\protect\astroncite{\emph{{Berry}}}{2015}]{fellwalker}
{Berry} D.~S. (2015) \emph{Astronomy and Computing}, \emph{10}, 22.

\bibitem[\protect\astroncite{\emph{{Bertoldi} and {McKee}}}{1992}]{bertoldi92}
{Bertoldi} F. and {McKee} C.~F. (1992) \emph{\apj}, \emph{395}, 140.

\bibitem[\protect\astroncite{\emph{{Beswick} et~al.}}{2015}]{beswick2015}
{Beswick} R. et~al. (2015) in: \emph{Advancing Astrophysics with the Square
  Kilometre Array (AASKA14)}, p.~70.

\bibitem[\protect\astroncite{\emph{{Bialy} et~al.}}{2017}]{Bialy2017}
{Bialy} S. et~al. (2017) \emph{\apj}, \emph{843}, 2, 92.

\bibitem[\protect\astroncite{\emph{{Bisbas} et~al.}}{2019}]{Bisbas19}
{Bisbas} T.~G. et~al. (2019) \emph{\mnras}, \emph{485}, 3, 3097.

\bibitem[\protect\astroncite{\emph{{Blair} et~al.}}{1978}]{Blair78a}
{Blair} G.~N. et~al. (1978) \emph{\apj}, \emph{219}, 896.

\bibitem[\protect\astroncite{\emph{{Blitz}}}{1993}]{1993prpl.conf..125B}
{Blitz} L. (1993) in: \emph{Protostars and Planets III}, (edited by E.~H.
  {Levy} and J.~I. {Lunine}), p. 125.

\bibitem[\protect\astroncite{\emph{{Blitz} and {Rosolowsky}}}{2006}]{Blitz2006}
{Blitz} L. and {Rosolowsky} E. (2006) \emph{\apj}, \emph{650}, 2, 933.

\bibitem[\protect\astroncite{\emph{{Blitz} and {Thaddeus}}}{1980}]{Blitz80b}
{Blitz} L. and {Thaddeus} P. (1980) \emph{\apj}, \emph{241}, 676.

\bibitem[\protect\astroncite{\emph{{Blitz} et~al.}}{2007}]{Blitz2007}
{Blitz} L. et~al. (2007) in: \emph{Protostars and Planets V}, (edited by
  B.~{Reipurth}, D.~{Jewitt}, and K.~{Keil}), p.~81.

\bibitem[\protect\astroncite{\emph{{Bolatto} et~al.}}{2008}]{Bolatto08}
{Bolatto} A.~D. et~al. (2008) \emph{\apj}, \emph{686}, 2, 948.

\bibitem[\protect\astroncite{\emph{{Bolatto} et~al.}}{2013}]{bolatto13}
{Bolatto} A.~D. et~al. (2013) \emph{\araa}, \emph{51}, 1, 207.

\bibitem[\protect\astroncite{\emph{{Bonnell} et~al.}}{1992}]{Bonnell1992}
{Bonnell} I. et~al. (1992) \emph{\apj}, \emph{400}, 579.

\bibitem[\protect\astroncite{\emph{{Bonnell} et~al.}}{1997}]{Bonnell97a}
{Bonnell} I.~A. et~al. (1997) \emph{\mnras}, \emph{285}, 201.

\bibitem[\protect\astroncite{\emph{{Bonnell} et~al.}}{2011}]{Bonnell11a}
{Bonnell} I.~A. et~al. (2011) \emph{\mnras}, \emph{410}, 2339.

\bibitem[\protect\astroncite{\emph{{Boyden} et~al.}}{2018}]{boyden18}
{Boyden} R.~D. et~al. (2018) \emph{\apj}, \emph{860}, 2, 157.

\bibitem[\protect\astroncite{\emph{{Braine} et~al.}}{2018}]{braine18}
{Braine} J. et~al. (2018) \emph{\aap}, \emph{612}, A51.

\bibitem[\protect\astroncite{\emph{{Braine} et~al.}}{2020}]{braine20}
{Braine} J. et~al. (2020) \emph{\aap}, \emph{633}, A17.

\bibitem[\protect\astroncite{\emph{{Brunetti} et~al.}}{2021}]{brunetti21}
{Brunetti} N. et~al. (2021) \emph{\mnras}, \emph{500}, 4, 4730.

\bibitem[\protect\astroncite{\emph{{Brunt} et~al.}}{2009}]{brunt09}
{Brunt} C.~M. et~al. (2009) \emph{\aap}, \emph{504}, 3, 883.

\bibitem[\protect\astroncite{\emph{{Burkhart}}}{2018}]{Burkhart18}
{Burkhart} B. (2018) \emph{\apj}, \emph{863}, 2, 118.

\bibitem[\protect\astroncite{\emph{{Burkhart} et~al.}}{2015}]{burkhart2015}
{Burkhart} B. et~al. (2015) \emph{\apj}, \emph{808}, 1, 48.

\bibitem[\protect\astroncite{\emph{{Burkhart} et~al.}}{2017}]{Burkhart17a}
{Burkhart} B. et~al. (2017) \emph{\apjl}, \emph{834}, L1.

\bibitem[\protect\astroncite{\emph{{Caldwell} and
  {Chang}}}{2018}]{CaldwellChang18}
{Caldwell} S. and {Chang} P. (2018) \emph{\mnras}, \emph{474}, 4, 4818.

\bibitem[\protect\astroncite{\emph{{Calzetti} et~al.}}{1989}]{Calzetti89a}
{Calzetti} D. et~al. (1989) \emph{\aap}, \emph{226}, 1, 1.

\bibitem[\protect\astroncite{\emph{{Calzetti} et~al.}}{2015}]{Calzetti15a}
{Calzetti} D. et~al. (2015) \emph{\aj}, \emph{149}, 51.

\bibitem[\protect\astroncite{\emph{{Carilli} and {Walter}}}{2013}]{Carilli2013}
{Carilli} C.~L. and {Walter} F. (2013) \emph{\araa}, \emph{51}, 1, 105.

\bibitem[\protect\astroncite{\emph{{Chabrier}}}{2003}]{Chabrier_2003}
{Chabrier} G. (2003) \emph{\pasp}, \emph{115}, 809, 763.

\bibitem[\protect\astroncite{\emph{{Chen} et~al.}}{2020}]{chen20}
{Chen} B.~Q. et~al. (2020) \emph{\mnras}, \emph{493}, 1, 351.

\bibitem[\protect\astroncite{\emph{{Chen} et~al.}}{2019}]{Chen19}
{Chen} H.-R.~V. et~al. (2019) \emph{\apj}, \emph{875}, 1, 24.

\bibitem[\protect\astroncite{\emph{{Chevance}
  et~al.}}{2020{\natexlab{a}}}]{Chevance2020a}
{Chevance} M. et~al. (2020{\natexlab{a}}) \emph{\mnras}, \emph{493}, 2, 2872.

\bibitem[\protect\astroncite{\emph{{Chevance}
  et~al.}}{2020{\natexlab{b}}}]{Chevance2020b}
{Chevance} M. et~al. (2020{\natexlab{b}}) \emph{\ssr}, \emph{216}, 4, 50.

\bibitem[\protect\astroncite{\emph{{Chevance} et~al.}}{2022}]{chevance2022}
{Chevance} M. et~al. (2022) \emph{\mnras}, \emph{509}, 1, 272.

\bibitem[\protect\astroncite{\emph{{Clark} and {Bonnell}}}{2004}]{Clark04a}
{Clark} P.~C. and {Bonnell} I.~A. (2004) \emph{\mnras}, \emph{347}, L36.

\bibitem[\protect\astroncite{\emph{{Clark} et~al.}}{2008}]{Clark08a}
{Clark} P.~C. et~al. (2008) \emph{\mnras}, \emph{386}, 3.

\bibitem[\protect\astroncite{\emph{{Clarke} et~al.}}{2017}]{Clarke2017}
{Clarke} S.~D. et~al. (2017) \emph{\mnras}, \emph{468}, 2, 2489.

\bibitem[\protect\astroncite{\emph{{Collins} et~al.}}{2012}]{Collins12a}
{Collins} D.~C. et~al. (2012) \emph{\apj}, \emph{750}, 13.

\bibitem[\protect\astroncite{\emph{{Colombo} et~al.}}{2014}]{colombo14}
{Colombo} D. et~al. (2014) \emph{\apj}, \emph{784}, 1, 3.

\bibitem[\protect\astroncite{\emph{{Colombo} et~al.}}{2015}]{colombo15}
{Colombo} D. et~al. (2015) \emph{\mnras}, \emph{454}, 2, 2067.

\bibitem[\protect\astroncite{\emph{{Colombo} et~al.}}{2018}]{Colombo18}
{Colombo} D. et~al. (2018) \emph{\mnras}, \emph{475}, 2, 1791.

\bibitem[\protect\astroncite{\emph{{Colombo} et~al.}}{2019}]{colombo19}
{Colombo} D. et~al. (2019) \emph{\mnras}, \emph{483}, 4, 4291.

\bibitem[\protect\astroncite{\emph{{Combes}}}{2018}]{Combes2018}
{Combes} F. (2018) \emph{\aapr}, \emph{26}, 1, 5.

\bibitem[\protect\astroncite{\emph{{Corbelli} et~al.}}{2017}]{Corbelli17}
{Corbelli} E. et~al. (2017) \emph{\aap}, \emph{601}, A146.

\bibitem[\protect\astroncite{\emph{{Crocker} et~al.}}{2018}]{Crocker_2018}
{Crocker} R.~M. et~al. (2018) \emph{\mnras}, \emph{481}, 4, 4895.

\bibitem[\protect\astroncite{\emph{{Crutcher}}}{2012}]{Crutcher12a}
{Crutcher} R.~M. (2012) \emph{\araa}, \emph{50}, 29.

\bibitem[\protect\astroncite{\emph{{Cunningham} et~al.}}{2018}]{Cunningham18a}
{Cunningham} A.~J. et~al. (2018) \emph{\mnras}, \emph{476}, 771.

\bibitem[\protect\astroncite{\emph{{Da Rio} et~al.}}{2014}]{DaRio14}
{Da Rio} N. et~al. (2014) \emph{\apj}, \emph{795}, 1, 55.

\bibitem[\protect\astroncite{\emph{{Da Rio} et~al.}}{2016}]{DaRio16}
{Da Rio} N. et~al. (2016) \emph{\apj}, \emph{818}, 1, 59.

\bibitem[\protect\astroncite{\emph{{da Silva} et~al.}}{2012}]{SLUG_2012}
{da Silva} R.~L. et~al. (2012) \emph{\apj}, \emph{745}, 2, 145.

\bibitem[\protect\astroncite{\emph{{Dale}}}{2015}]{Dale15}
{Dale} J.~E. (2015) \emph{\nar}, \emph{68}, 1.

\bibitem[\protect\astroncite{\emph{{Dale}}}{2017}]{Dale17a}
{Dale} J.~E. (2017) \emph{\mnras}, \emph{467}, 1067.

\bibitem[\protect\astroncite{\emph{{Dale} and {Bonnell}}}{2008}]{Dale08a}
{Dale} J.~E. and {Bonnell} I.~A. (2008) \emph{\mnras}, \emph{391}, 2.

\bibitem[\protect\astroncite{\emph{{Dale} et~al.}}{2007}]{Dale07a}
{Dale} J.~E. et~al. (2007) \emph{\mnras}, \emph{375}, 1291.

\bibitem[\protect\astroncite{\emph{{Dale} et~al.}}{2012}]{Dale_2012}
{Dale} J.~E. et~al. (2012) \emph{\mnras}, \emph{424}, 1, 377.

\bibitem[\protect\astroncite{\emph{{Dale}
  et~al.}}{2013{\natexlab{a}}}]{Dale_2013a}
{Dale} J.~E. et~al. (2013{\natexlab{a}}) \emph{\mnras}, \emph{430}, 1, 234.

\bibitem[\protect\astroncite{\emph{{Dale}
  et~al.}}{2013{\natexlab{b}}}]{Dale_2013b}
{Dale} J.~E. et~al. (2013{\natexlab{b}}) \emph{\mnras}, \emph{436}, 4, 3430.

\bibitem[\protect\astroncite{\emph{{Dale} et~al.}}{2014}]{Dale_2014}
{Dale} J.~E. et~al. (2014) \emph{\mnras}, \emph{442}, 1, 694.

\bibitem[\protect\astroncite{\emph{{Dalgarno}}}{2006}]{Dalgarno06}
{Dalgarno} A. (2006) \emph{Proceedings of the National Academy of Science},
  \emph{103}, 33, 12269.

\bibitem[\protect\astroncite{\emph{{Dame} et~al.}}{1987}]{Dame87a}
{Dame} T.~M. et~al. (1987) \emph{\apj}, \emph{322}, 706.

\bibitem[\protect\astroncite{\emph{{Dame} et~al.}}{2001}]{Dame01a}
{Dame} T.~M. et~al. (2001) \emph{\apj}, \emph{547}, 792.

\bibitem[\protect\astroncite{\emph{{Davis} et~al.}}{2014}]{Davis_2014}
{Davis} S.~W. et~al. (2014) \emph{\apj}, \emph{796}, 2, 107.

\bibitem[\protect\astroncite{\emph{{Decourchelle}
  et~al.}}{2013}]{decourchelle13}
{Decourchelle} A. et~al. (2013) \emph{arXiv e-prints}, arXiv:1306.2335.

\bibitem[\protect\astroncite{\emph{{Della Bruna}
  et~al.}}{2020}]{dellabruna2020}
{Della Bruna} L. et~al. (2020) \emph{\aap}, \emph{635}, A134.

\bibitem[\protect\astroncite{\emph{{Dempsey} et~al.}}{2013}]{dempsey13}
{Dempsey} J.~T. et~al. (2013) \emph{\apjs}, \emph{209}, 1, 8.

\bibitem[\protect\astroncite{\emph{{Dessauges-Zavadsky}
  et~al.}}{2019}]{dessauges2019}
{Dessauges-Zavadsky} M. et~al. (2019) \emph{Nature Astronomy}, \emph{3}, 1115.

\bibitem[\protect\astroncite{\emph{{Dib} et~al.}}{2007}]{dib07}
{Dib} S. et~al. (2007) \emph{\apj}, \emph{661}, 1, 262.

\bibitem[\protect\astroncite{\emph{{Dib} et~al.}}{2020}]{Dib20a}
{Dib} S. et~al. (2020) \emph{\aap}, \emph{642}, A177.

\bibitem[\protect\astroncite{\emph{{Dobbs}
  et~al.}}{2011{\natexlab{a}}}]{Dobbs11a}
{Dobbs} C.~L. et~al. (2011{\natexlab{a}}) \emph{\mnras}, \emph{417}, 1318.

\bibitem[\protect\astroncite{\emph{{Dobbs}
  et~al.}}{2011{\natexlab{b}}}]{Dobbs11b}
{Dobbs} C.~L. et~al. (2011{\natexlab{b}}) \emph{\mnras}, \emph{413}, 4, 2935.

\bibitem[\protect\astroncite{\emph{{Dobbs} et~al.}}{2014}]{Dobbs14a}
{Dobbs} C.~L. et~al. (2014) \emph{Protostars and Planets VI}, pp. 3--26.

\bibitem[\protect\astroncite{\emph{{Donovan Meyer}
  et~al.}}{2013}]{DonovanMeyer13}
{Donovan Meyer} J. et~al. (2013) \emph{\apj}, \emph{772}, 2, 107.

\bibitem[\protect\astroncite{\emph{{Draine}}}{2011{\natexlab{a}}}]{Draine_2011}
{Draine} B.~T. (2011{\natexlab{a}}) \emph{732}, 2, 100.

\bibitem[\protect\astroncite{\emph{{Draine}}}{2011{\natexlab{b}}}]{Draine_book2011}
{Draine} B.~T. (2011{\natexlab{b}}) \emph{{Physics of the Interstellar and
  Intergalactic Medium}}.

\bibitem[\protect\astroncite{\emph{{Dzib} et~al.}}{2021}]{Dzib21a}
{Dzib} S.~A. et~al. (2021) \emph{\apj}, \emph{906}, 1, 24.

\bibitem[\protect\astroncite{\emph{{El-Badry} et~al.}}{2019}]{El-Badry_2019}
{El-Badry} K. et~al. (2019) \emph{\mnras}, \emph{490}, 2, 1961.

\bibitem[\protect\astroncite{\emph{{Elia} et~al.}}{2018}]{Elia18a}
{Elia} D. et~al. (2018) \emph{\mnras}, \emph{481}, 1, 509.

\bibitem[\protect\astroncite{\emph{Elmegreen}}{2000}]{Elmegreen2000}
Elmegreen B.~G. (2000) \emph{\apj}, \emph{530}, 277.

\bibitem[\protect\astroncite{\emph{{Elmegreen}}}{2018}]{Elmegreen18}
{Elmegreen} B.~G. (2018) \emph{\apj}, \emph{869}, 2, 119.

\bibitem[\protect\astroncite{\emph{{Engargiola} et~al.}}{2003}]{Engargiola03}
{Engargiola} G. et~al. (2003) \emph{\apjs}, \emph{149}, 2, 343.

\bibitem[\protect\astroncite{\emph{{Evans}}}{1978}]{1978prpl.conf..153E}
{Evans} N.~J. I. (1978) in: \emph{IAU Colloq. 52: Protostars and Planets},
  (edited by T.~{Gehrels} and M.~S. {Matthews}), p. 153.

\bibitem[\protect\astroncite{\emph{{Evans} et~al.}}{2014}]{Evans14}
{Evans} Neal~J. I. et~al. (2014) \emph{\apj}, \emph{782}, 2, 114.

\bibitem[\protect\astroncite{\emph{{Evans} et~al.}}{2021}]{evans2021}
{Evans} Neal~J. I. et~al. (2021) \emph{\apj}, \emph{920}, 2, 126.

\bibitem[\protect\astroncite{\emph{{Faesi} et~al.}}{2018}]{faesi18}
{Faesi} C.~M. et~al. (2018) \emph{\apj}, \emph{857}, 1, 19.

\bibitem[\protect\astroncite{\emph{{Falgarone} et~al.}}{1991}]{Falgarone91a}
{Falgarone} E. et~al. (1991) \emph{\apj}, \emph{378}, 186.

\bibitem[\protect\astroncite{\emph{{Fall} et~al.}}{2010}]{Fall10a}
{Fall} S.~M. et~al. (2010) \emph{\apjl}, \emph{710}, L142.

\bibitem[\protect\astroncite{\emph{{Federrath}}}{2015}]{Federrath15b}
{Federrath} C. (2015) \emph{\mnras}, \emph{450}, 4, 4035.

\bibitem[\protect\astroncite{\emph{{Federrath} and
  {Klessen}}}{2012}]{Federrath2012}
{Federrath} C. and {Klessen} R.~S. (2012) \emph{\apj}, \emph{761}, 2, 156.

\bibitem[\protect\astroncite{\emph{{Feldmann} and
  {Gnedin}}}{2011}]{Feldmann11a}
{Feldmann} R. and {Gnedin} N.~Y. (2011) \emph{\apjl}, \emph{727}, L12+.

\bibitem[\protect\astroncite{\emph{{Feldmann} et~al.}}{2011}]{Feldmann11b}
{Feldmann} R. et~al. (2011) \emph{\apj}, \emph{732}, 2, 115.

\bibitem[\protect\astroncite{\emph{{Fielding} et~al.}}{2020}]{Fielding_2020}
{Fielding} D.~B. et~al. (2020) \emph{\apjl}, \emph{894}, 2, L24.

\bibitem[\protect\astroncite{\emph{{Fisher} et~al.}}{2017}]{fisher2017}
{Fisher} D.~B. et~al. (2017) \emph{\mnras}, \emph{464}, 1, 491.

\bibitem[\protect\astroncite{\emph{{Forbrich} et~al.}}{2020}]{forbrich20}
{Forbrich} J. et~al. (2020) \emph{\apj}, \emph{890}, 1, 42.

\bibitem[\protect\astroncite{\emph{{Frank} et~al.}}{2014}]{Frank_2014}
{Frank} A. et~al. (2014) in: \emph{Protostars and Planets VI}, (edited by
  H.~{Beuther}, R.~S. {Klessen}, C.~P. {Dullemond}, and T.~{Henning}), p. 451.

\bibitem[\protect\astroncite{\emph{{Freeman} et~al.}}{2017}]{freeman17}
{Freeman} P. et~al. (2017) \emph{\mnras}, \emph{468}, 2, 1769.

\bibitem[\protect\astroncite{\emph{{Fujii} et~al.}}{2021}]{Fujii21}
{Fujii} K. et~al. (2021) \emph{\mnras}, \emph{505}, 1, 459.

\bibitem[\protect\astroncite{\emph{{Fujimoto} et~al.}}{2016}]{Fujimoto16a}
{Fujimoto} Y. et~al. (2016) \emph{\mnras}, \emph{461}, 2, 1684.

\bibitem[\protect\astroncite{\emph{{Fujimoto} et~al.}}{2019}]{Fujimoto19a}
{Fujimoto} Y. et~al. (2019) \emph{\mnras}, \emph{487}, 2, 1717.

\bibitem[\protect\astroncite{\emph{{Fukui} et~al.}}{1999}]{Fukui99}
{Fukui} Y. et~al. (1999) \emph{\pasj}, \emph{51}, 745.

\bibitem[\protect\astroncite{\emph{{Fukui} et~al.}}{2001}]{Fukui01a}
{Fukui} Y. et~al. (2001) \emph{\pasj}, \emph{53}, L41.

\bibitem[\protect\astroncite{\emph{{Fukui} et~al.}}{2008}]{Fukui08}
{Fukui} Y. et~al. (2008) \emph{\apjs}, \emph{178}, 1, 56.

\bibitem[\protect\astroncite{\emph{{Fukushima} et~al.}}{2020}]{Fukushima_2020}
{Fukushima} H. et~al. (2020) \emph{\mnras}, \emph{497}, 3, 3830.

\bibitem[\protect\astroncite{\emph{{Gallagher} et~al.}}{2018}]{Gallagher18a}
{Gallagher} M.~J. et~al. (2018) \emph{\apj}, \emph{858}, 90.

\bibitem[\protect\astroncite{\emph{{Gammie} and {Ostriker}}}{1996}]{Gammie1996}
{Gammie} C.~F. and {Ostriker} E.~C. (1996) \emph{\apj}, \emph{466}, 814.

\bibitem[\protect\astroncite{\emph{{Gaskin} et~al.}}{2019}]{gaskin19}
{Gaskin} J.~A. et~al. (2019) \emph{Journal of Astronomical Telescopes,
  Instruments, and Systems}, \emph{5}, 021001.

\bibitem[\protect\astroncite{\emph{{Geen} et~al.}}{2015}]{Geen2015}
{Geen} S. et~al. (2015) \emph{\mnras}, \emph{454}, 4, 4484.

\bibitem[\protect\astroncite{\emph{{Geen} et~al.}}{2016}]{Geen_2016}
{Geen} S. et~al. (2016) \emph{\mnras}, \emph{463}, 3, 3129.

\bibitem[\protect\astroncite{\emph{{Geen} et~al.}}{2018}]{Geen_2018}
{Geen} S. et~al. (2018) \emph{\mnras}, \emph{481}, 2, 2548.

\bibitem[\protect\astroncite{\emph{{Geen} et~al.}}{2021}]{Geen_2021}
{Geen} S. et~al. (2021) \emph{\mnras}, \emph{501}, 1, 1352.

\bibitem[\protect\astroncite{\emph{{Gentry} et~al.}}{2017}]{Gentry_2017}
{Gentry} E.~S. et~al. (2017) \emph{\mnras}, \emph{465}, 2, 2471.

\bibitem[\protect\astroncite{\emph{{Gentry} et~al.}}{2019}]{Gentry_2019}
{Gentry} E.~S. et~al. (2019) \emph{\mnras}, \emph{483}, 3, 3647.

\bibitem[\protect\astroncite{\emph{{Ginsburg} et~al.}}{2013}]{ginsburg13}
{Ginsburg} A. et~al. (2013) \emph{\apj}, \emph{779}, 1, 50.

\bibitem[\protect\astroncite{\emph{{Girichidis} et~al.}}{2018}]{Girichidis18}
{Girichidis} P. et~al. (2018) \emph{\mnras}, \emph{480}, 3, 3511.

\bibitem[\protect\astroncite{\emph{{Glover} and {Mac Low}}}{2007}]{Glover07}
{Glover} S. C.~O. and {Mac Low} M.-M. (2007) \emph{\apjs}, \emph{169}, 2, 239.

\bibitem[\protect\astroncite{\emph{{Goldbaum} et~al.}}{2011}]{Goldbaum11a}
{Goldbaum} N.~J. et~al. (2011) \emph{\apj}, \emph{738}, 101.

\bibitem[\protect\astroncite{\emph{{Gong} et~al.}}{2017}]{Gong_2017}
{Gong} M. et~al. (2017) \emph{\apj}, \emph{843}, 1, 38.

\bibitem[\protect\astroncite{\emph{{Gong} et~al.}}{2018}]{Gong_18}
{Gong} M. et~al. (2018) \emph{\apj}, \emph{858}, 1, 16.

\bibitem[\protect\astroncite{\emph{{Gong} et~al.}}{2020}]{Gong20}
{Gong} M. et~al. (2020) \emph{\apj}, \emph{903}, 2, 142.

\bibitem[\protect\astroncite{\emph{{Gonz{\'a}lez} et~al.}}{2021}]{Gonzalez21a}
{Gonz{\'a}lez} M. et~al. (2021) \emph{\aap}, \emph{647}, A14.

\bibitem[\protect\astroncite{\emph{{Gonzalez} and
  {Battaglia}}}{2018}]{gonzalez18}
{Gonzalez} O.~A. and {Battaglia} G. (2018) \emph{arXiv e-prints},
  arXiv:1810.04422.

\bibitem[\protect\astroncite{\emph{{G{\'o}rski} et~al.}}{2005}]{Gorski05}
{G{\'o}rski} K.~M. et~al. (2005) \emph{\apj}, \emph{622}, 2, 759.

\bibitem[\protect\astroncite{\emph{{Gouliermis}}}{2018}]{Gouliermis18a}
{Gouliermis} D.~A. (2018) \emph{\pasp}, \emph{130}, 072001.

\bibitem[\protect\astroncite{\emph{{Gouliermis} et~al.}}{2015}]{Gouliermis15a}
{Gouliermis} D.~A. et~al. (2015) \emph{\mnras}, \emph{452}, 3508.

\bibitem[\protect\astroncite{\emph{{Gouliermis} et~al.}}{2017}]{Gouliermis17a}
{Gouliermis} D.~A. et~al. (2017) \emph{\mnras}, \emph{468}, 509.

\bibitem[\protect\astroncite{\emph{{Grasha} et~al.}}{2017}]{Grasha17a}
{Grasha} K. et~al. (2017) \emph{\apj}, \emph{840}, 113.

\bibitem[\protect\astroncite{\emph{{Grasha} et~al.}}{2018}]{Grasha18}
{Grasha} K. et~al. (2018) \emph{\mnras}, \emph{481}, 1, 1016.

\bibitem[\protect\astroncite{\emph{{Grasha} et~al.}}{2019}]{Grasha19}
{Grasha} K. et~al. (2019) \emph{\mnras}, \emph{483}, 4, 4707.

\bibitem[\protect\astroncite{\emph{{Gratier} et~al.}}{2012}]{gratier12}
{Gratier} P. et~al. (2012) \emph{\aap}, \emph{542}, A108.

\bibitem[\protect\astroncite{\emph{{Grisdale}}}{2021}]{Grisdale21a}
{Grisdale} K. (2021) \emph{\mnras}, \emph{500}, 3, 3552.

\bibitem[\protect\astroncite{\emph{{Grisdale} et~al.}}{2018}]{Grisdale_2018}
{Grisdale} K. et~al. (2018) \emph{\mnras}, \emph{479}, 3, 3167.

\bibitem[\protect\astroncite{\emph{{Grisdale} et~al.}}{2019}]{Grisdale_2019}
{Grisdale} K. et~al. (2019) \emph{\mnras}, \emph{486}, 4, 5482.

\bibitem[\protect\astroncite{\emph{{Gro{\ss}schedl}
  et~al.}}{2018}]{grossschedel18}
{Gro{\ss}schedl} J.~E. et~al. (2018) \emph{\aap}, \emph{619}, A106.

\bibitem[\protect\astroncite{\emph{{Grudi{\'c}} and
  {Hopkins}}}{2019}]{Grudic19b}
{Grudi{\'c}} M.~Y. and {Hopkins} P.~F. (2019) \emph{\mnras}, \emph{488}, 2,
  2970.

\bibitem[\protect\astroncite{\emph{{Grudi{\'c}} et~al.}}{2018}]{Grudic18a}
{Grudi{\'c}} M.~Y. et~al. (2018) \emph{\mnras}, \emph{475}, 3511.

\bibitem[\protect\astroncite{\emph{{Grudi{\'c}} et~al.}}{2019}]{Grudic19a}
{Grudi{\'c}} M.~Y. et~al. (2019) \emph{\mnras}, \emph{488}, 2, 1501.

\bibitem[\protect\astroncite{\emph{{Grudi{\'c}} et~al.}}{2021}]{Grudic21a}
{Grudi{\'c}} M.~Y. et~al. (2021) \emph{\mnras}, \emph{506}, 3, 3239.

\bibitem[\protect\astroncite{\emph{{G{\"u}del} et~al.}}{2008}]{Gudel08}
{G{\"u}del} M. et~al. (2008) \emph{Science}, \emph{319}, 5861, 309.

\bibitem[\protect\astroncite{\emph{{Guszejnov} et~al.}}{2017}]{Guszejnov17a}
{Guszejnov} D. et~al. (2017) \emph{\mnras}, \emph{468}, 4093.

\bibitem[\protect\astroncite{\emph{{Guszejnov} et~al.}}{2018}]{Guszejnov18b}
{Guszejnov} D. et~al. (2018) \emph{\mnras}, \emph{477}, 4, 5139.

\bibitem[\protect\astroncite{\emph{{Hacar} et~al.}}{2013}]{Hacar2013}
{Hacar} A. et~al. (2013) \emph{\aap}, \emph{554}, A55.

\bibitem[\protect\astroncite{\emph{{Haid} et~al.}}{2018}]{Haid18}
{Haid} S. et~al. (2018) \emph{\mnras}, \emph{478}, 4, 4799.

\bibitem[\protect\astroncite{\emph{{Haid} et~al.}}{2019}]{Haid_2019}
{Haid} S. et~al. (2019) \emph{\mnras}, \emph{482}, 3, 4062.

\bibitem[\protect\astroncite{\emph{{Hannon} et~al.}}{2019}]{Hannon19}
{Hannon} S. et~al. (2019) \emph{\mnras}, \emph{490}, 4, 4648.

\bibitem[\protect\astroncite{\emph{{Harper-Clark} and
  {Murray}}}{2009}]{Harper-Clark_2009}
{Harper-Clark} E. and {Murray} N. (2009) \emph{\apj}, \emph{693}, 2, 1696.

\bibitem[\protect\astroncite{\emph{{Hartmann} et~al.}}{2001}]{Hartmann01}
{Hartmann} L. et~al. (2001) \emph{\apj}, \emph{562}, 2, 852.

\bibitem[\protect\astroncite{\emph{{Hartmann} et~al.}}{2012}]{Hartmann12}
{Hartmann} L. et~al. (2012) \emph{\mnras}, \emph{420}, 2, 1457.

\bibitem[\protect\astroncite{\emph{{Haydon}
  et~al.}}{2020{\natexlab{a}}}]{Haydon20a}
{Haydon} D.~T. et~al. (2020{\natexlab{a}}) \emph{\mnras}, \emph{498}, 1, 235.

\bibitem[\protect\astroncite{\emph{{Haydon}
  et~al.}}{2020{\natexlab{b}}}]{Haydon20b}
{Haydon} D.~T. et~al. (2020{\natexlab{b}}) \emph{\mnras}, \emph{497}, 4, 5076.

\bibitem[\protect\astroncite{\emph{{He} et~al.}}{2019}]{He_Ricotti2019}
{He} C.-C. et~al. (2019) \emph{\mnras}, \emph{489}, 2, 1880.

\bibitem[\protect\astroncite{\emph{{Heitsch} et~al.}}{2001}]{Heitsch01}
{Heitsch} F. et~al. (2001) \emph{\apj}, \emph{547}, 1, 280.

\bibitem[\protect\astroncite{\emph{{Heitsch} et~al.}}{2009}]{Heitsch09}
{Heitsch} F. et~al. (2009) \emph{\apj}, \emph{704}, 2, 1735.

\bibitem[\protect\astroncite{\emph{{Hennebelle} and
  {Chabrier}}}{2008}]{Hennebelle2008}
{Hennebelle} P. and {Chabrier} G. (2008) \emph{\apj}, \emph{684}, 1, 395.

\bibitem[\protect\astroncite{\emph{{Hennebelle} and
  {Chabrier}}}{2011}]{Hennebelle11}
{Hennebelle} P. and {Chabrier} G. (2011) \emph{\apjl}, \emph{743}, 2, L29.

\bibitem[\protect\astroncite{\emph{{Hennekemper}
  et~al.}}{2008}]{Hennekemper08a}
{Hennekemper} E. et~al. (2008) \emph{\apj}, \emph{672}, 914.

\bibitem[\protect\astroncite{\emph{{Henshaw} et~al.}}{2014}]{Henshaw14}
{Henshaw} J.~D. et~al. (2014) \emph{\mnras}, \emph{440}, 3, 2860.

\bibitem[\protect\astroncite{\emph{{Henshaw}
  et~al.}}{2016{\natexlab{a}}}]{Henshaw16b}
{Henshaw} J.~D. et~al. (2016{\natexlab{a}}) \emph{\mnras}, \emph{463}, 1, 146.

\bibitem[\protect\astroncite{\emph{{Henshaw}
  et~al.}}{2016{\natexlab{b}}}]{Henshaw16a}
{Henshaw} J.~D. et~al. (2016{\natexlab{b}}) \emph{\mnras}, \emph{457}, 3, 2675.

\bibitem[\protect\astroncite{\emph{{Henshaw} et~al.}}{2020}]{Henshaw20}
{Henshaw} J.~D. et~al. (2020) \emph{Nature Astronomy}, \emph{4}, 1064.

\bibitem[\protect\astroncite{\emph{{Heyer} and {Dame}}}{2015}]{heyerdame15}
{Heyer} M. and {Dame} T.~M. (2015) \emph{\araa}, \emph{53}, 583.

\bibitem[\protect\astroncite{\emph{{Heyer} et~al.}}{2009}]{Heyer09}
{Heyer} M. et~al. (2009) \emph{\apj}, \emph{699}, 2, 1092.

\bibitem[\protect\astroncite{\emph{{Heyer} et~al.}}{2016}]{Heyer16a}
{Heyer} M. et~al. (2016) \emph{\aap}, \emph{588}, A29.

\bibitem[\protect\astroncite{\emph{{Heyer} et~al.}}{2020}]{heyer20}
{Heyer} M. et~al. (2020) \emph{\mnras}, \emph{496}, 4, 4546.

\bibitem[\protect\astroncite{\emph{{Heyer} and {Brunt}}}{2004}]{heyer04}
{Heyer} M.~H. and {Brunt} C.~M. (2004) \emph{\apjl}, \emph{615}, 1, L45.

\bibitem[\protect\astroncite{\emph{{Heyer} et~al.}}{2001}]{heyer01}
{Heyer} M.~H. et~al. (2001) \emph{\apj}, \emph{551}, 2, 852.

\bibitem[\protect\astroncite{\emph{{Hirota} et~al.}}{2018}]{hirota18}
{Hirota} A. et~al. (2018) \emph{\pasj}, \emph{70}, 4, 73.

\bibitem[\protect\astroncite{\emph{{Hollyhead} et~al.}}{2015}]{Hollyhead15}
{Hollyhead} K. et~al. (2015) \emph{\mnras}, \emph{449}, 1, 1106.

\bibitem[\protect\astroncite{\emph{{Hopkins}}}{2013{\natexlab{a}}}]{Hopkins13b}
{Hopkins} P.~F. (2013{\natexlab{a}}) \emph{\mnras}, \emph{430}, 1653.

\bibitem[\protect\astroncite{\emph{{Hopkins}}}{2013{\natexlab{b}}}]{Hopkins13a}
{Hopkins} P.~F. (2013{\natexlab{b}}) \emph{\mnras}, \emph{428}, 1950.

\bibitem[\protect\astroncite{\emph{{Hopkins}}}{2015}]{Hopkins15}
{Hopkins} P.~F. (2015) \emph{\mnras}, \emph{450}, 1, 53.

\bibitem[\protect\astroncite{\emph{{Hopkins} and
  {Grudi{\'c}}}}{2019}]{Hopkins18a}
{Hopkins} P.~F. and {Grudi{\'c}} M.~Y. (2019) \emph{\mnras}, \emph{483}, 3,
  4187.

\bibitem[\protect\astroncite{\emph{{Hosek} et~al.}}{2019}]{Hosek19}
{Hosek} Matthew~W. J. et~al. (2019) \emph{\apj}, \emph{870}, 1, 44.

\bibitem[\protect\astroncite{\emph{{Hu} et~al.}}{2021{\natexlab{a}}}]{Hu21}
{Hu} C.-Y. et~al. (2021{\natexlab{a}}) \emph{\apj}, \emph{920}, 1, 44.

\bibitem[\protect\astroncite{\emph{{Hu} et~al.}}{2019}]{hu19}
{Hu} Y. et~al. (2019) \emph{Nature Astronomy}, \emph{3}, 776.

\bibitem[\protect\astroncite{\emph{{Hu} et~al.}}{2021{\natexlab{b}}}]{Hu21a}
{Hu} Z. et~al. (2021{\natexlab{b}}) \emph{\mnras}, \emph{502}, 4, 5997.

\bibitem[\protect\astroncite{\emph{{Hu} et~al.}}{2022}]{Hu22a}
{Hu} Z. et~al. (2022) \emph{\mnras~in press}, arXiv:2109.04665.

\bibitem[\protect\astroncite{\emph{{Hughes} et~al.}}{2013}]{hughes13}
{Hughes} A. et~al. (2013) \emph{\apj}, \emph{779}, 1, 46.

\bibitem[\protect\astroncite{\emph{{Hunter}}}{1964}]{Hunter_1964}
{Hunter} C. (1964) \emph{\apj}, \emph{139}, 570.

\bibitem[\protect\astroncite{\emph{{Iffrig} and
  {Hennebelle}}}{2015}]{Iffrig_Hennebelle2015}
{Iffrig} O. and {Hennebelle} P. (2015) \emph{\aap}, \emph{576}, A95.

\bibitem[\protect\astroncite{\emph{{Imara} and {Faesi}}}{2019}]{imara19}
{Imara} N. and {Faesi} C.~M. (2019) \emph{\apj}, \emph{876}, 2, 141.

\bibitem[\protect\astroncite{\emph{{Inutsuka} et~al.}}{2015}]{Inutsuka15}
{Inutsuka} S.-i. et~al. (2015) \emph{\aap}, \emph{580}, A49.

\bibitem[\protect\astroncite{\emph{{Iwasaki} et~al.}}{2019}]{Iwasaki19}
{Iwasaki} K. et~al. (2019) \emph{\apj}, \emph{873}, 1, 6.

\bibitem[\protect\astroncite{\emph{{James} et~al.}}{2020}]{James2020}
{James} B.~L. et~al. (2020) \emph{\mnras}, \emph{495}, 3, 2564.

\bibitem[\protect\astroncite{\emph{{Jaupart} and
  {Chabrier}}}{2020}]{Jaupart20a}
{Jaupart} E. and {Chabrier} G. (2020) \emph{\apjl}, \emph{903}, 1, L2.

\bibitem[\protect\astroncite{\emph{{Jeffreson} and
  {Kruijssen}}}{2018}]{Jeffreson18}
{Jeffreson} S. M.~R. and {Kruijssen} J.~M.~D. (2018) \emph{\mnras}, \emph{476},
  3, 3688.

\bibitem[\protect\astroncite{\emph{{Jeffreson} et~al.}}{2020}]{Jeffreson_2020}
{Jeffreson} S. M.~R. et~al. (2020) \emph{\mnras}, \emph{498}, 1, 385.

\bibitem[\protect\astroncite{\emph{{Jeffreson}
  et~al.}}{2021{\natexlab{a}}}]{Jeffreson_2021a}
{Jeffreson} S. M.~R. et~al. (2021{\natexlab{a}}) \emph{\mnras}, \emph{505}, 2,
  1678.

\bibitem[\protect\astroncite{\emph{{Jeffreson}
  et~al.}}{2021{\natexlab{b}}}]{Jeffreson_2021b}
{Jeffreson} S. M.~R. et~al. (2021{\natexlab{b}}) \emph{\mnras}, \emph{505}, 3,
  3470.

\bibitem[\protect\astroncite{\emph{{Jim{\'e}nez-Donaire}
  et~al.}}{2019}]{jimenezdonaire2019}
{Jim{\'e}nez-Donaire} M.~J. et~al. (2019) \emph{\apj}, \emph{880}, 2, 127.

\bibitem[\protect\astroncite{\emph{{Joncour} et~al.}}{2018}]{Joncour18a}
{Joncour} I. et~al. (2018) \emph{\aap}, \emph{620}, A27.

\bibitem[\protect\astroncite{\emph{{Kainulainen} et~al.}}{2009}]{kainulainen09}
{Kainulainen} J. et~al. (2009) \emph{\aap}, \emph{508}, 3, L35.

\bibitem[\protect\astroncite{\emph{{Kainulainen} et~al.}}{2014}]{kainulainen14}
{Kainulainen} J. et~al. (2014) \emph{Science}, \emph{344}, 6180, 183.

\bibitem[\protect\astroncite{\emph{{Kauffmann} et~al.}}{2017}]{Kauffmann17a}
{Kauffmann} J. et~al. (2017) \emph{\aap}, \emph{605}, L5.

\bibitem[\protect\astroncite{\emph{Kawamura et~al.}}{2009}]{Kawamura2009}
Kawamura A. et~al. (2009) \emph{\apjs}, \emph{184}, 1, 1.

\bibitem[\protect\astroncite{\emph{{Keller} et~al.}}{2022}]{Keller2022}
{Keller} B.~W. et~al. (2022) \emph{MNRAS subm.}

\bibitem[\protect\astroncite{\emph{{Khoperskov} et~al.}}{2016}]{khoperskov16}
{Khoperskov} S.~A. et~al. (2016) \emph{\mnras}, \emph{455}, 2, 1782.

\bibitem[\protect\astroncite{\emph{{Khullar} et~al.}}{2021}]{Khullar21a}
{Khullar} S. et~al. (2021) \emph{\mnras}.

\bibitem[\protect\astroncite{\emph{{Kim} and
  {Ostriker}}}{2015}]{Kim_Ostriker2015a}
{Kim} C.-G. and {Ostriker} E.~C. (2015) \emph{\apj}, \emph{802}, 2, 99.

\bibitem[\protect\astroncite{\emph{{Kim} and {Ostriker}}}{2017}]{KimOstriker17}
{Kim} C.-G. and {Ostriker} E.~C. (2017) \emph{\apj}, \emph{846}, 2, 133.

\bibitem[\protect\astroncite{\emph{{Kim}
  et~al.}}{2017}]{Kim_Ostriker_Raileanu2017}
{Kim} C.-G. et~al. (2017) \emph{\apj}, \emph{834}, 1, 25.

\bibitem[\protect\astroncite{\emph{{Kim} et~al.}}{2019{\natexlab{a}}}]{Kim19a}
{Kim} D. et~al. (2019{\natexlab{a}}) \emph{\aj}, \emph{157}, 3, 109.

\bibitem[\protect\astroncite{\emph{{Kim}
  et~al.}}{2021{\natexlab{a}}}]{KimChevance21}
{Kim} J. et~al. (2021{\natexlab{a}}) \emph{\mnras}, \emph{504}, 1, 487.

\bibitem[\protect\astroncite{\emph{{Kim} et~al.}}{2016}]{Kim_JG2016}
{Kim} J.-G. et~al. (2016) \emph{\apj}, \emph{819}, 2, 137.

\bibitem[\protect\astroncite{\emph{{Kim} et~al.}}{2018}]{Kim_JG2018}
{Kim} J.-G. et~al. (2018) \emph{\apj}, \emph{859}, 1, 68.

\bibitem[\protect\astroncite{\emph{{Kim}
  et~al.}}{2019{\natexlab{b}}}]{Kim_JG2019}
{Kim} J.-G. et~al. (2019{\natexlab{b}}) \emph{\apj}, \emph{883}, 1, 102.

\bibitem[\protect\astroncite{\emph{{Kim}
  et~al.}}{2021{\natexlab{b}}}]{Kim_JG2021}
{Kim} J.-G. et~al. (2021{\natexlab{b}}) \emph{\apj}, \emph{911}, 2, 128.

\bibitem[\protect\astroncite{\emph{{Kim} and
  {Ostriker}}}{2002}]{Kim_Ostriker2002}
{Kim} W.-T. and {Ostriker} E.~C. (2002) \emph{\apj}, \emph{570}, 1, 132.

\bibitem[\protect\astroncite{\emph{{Kim} et~al.}}{2020}]{Kim_WT2020}
{Kim} W.-T. et~al. (2020) \emph{\apj}, \emph{898}, 1, 35.

\bibitem[\protect\astroncite{\emph{{Kirk} et~al.}}{2013}]{Kirk13}
{Kirk} H. et~al. (2013) \emph{\apj}, \emph{766}, 2, 115.

\bibitem[\protect\astroncite{\emph{{Klessen} and {Burkert}}}{2000}]{Klessen00a}
{Klessen} R.~S. and {Burkert} A. (2000) \emph{\apjs}, \emph{128}, 287.

\bibitem[\protect\astroncite{\emph{{Klessen} and
  {Hennebelle}}}{2010}]{Klessen10}
{Klessen} R.~S. and {Hennebelle} P. (2010) \emph{\aap}, \emph{520}, A17.

\bibitem[\protect\astroncite{\emph{{Klessen} et~al.}}{1998}]{Klessen98a}
{Klessen} R.~S. et~al. (1998) \emph{\apjl}, \emph{501}, L205+.

\bibitem[\protect\astroncite{\emph{{Kobayashi} et~al.}}{2018}]{Kobayashi18}
{Kobayashi} M. I.~N. et~al. (2018) \emph{\pasj}, \emph{70}, S59.

\bibitem[\protect\astroncite{\emph{{Koch} et~al.}}{2019}]{koch19}
{Koch} E.~W. et~al. (2019) \emph{\aj}, \emph{158}, 1, 1.

\bibitem[\protect\astroncite{\emph{{Koda} et~al.}}{2009}]{Koda09}
{Koda} J. et~al. (2009) \emph{\apjl}, \emph{700}, 2, L132.

\bibitem[\protect\astroncite{\emph{{Kollmeier} et~al.}}{2017}]{kollmeier17}
{Kollmeier} J.~A. et~al. (2017) \emph{arXiv e-prints}, arXiv:1711.03234.

\bibitem[\protect\astroncite{\emph{{Kondo} et~al.}}{2021}]{kondo21}
{Kondo} H. et~al. (2021) \emph{\apj}, \emph{912}, 1, 66.

\bibitem[\protect\astroncite{\emph{{Kong} et~al.}}{2018}]{Kong18}
{Kong} S. et~al. (2018) \emph{\apjs}, \emph{236}, 2, 25.

\bibitem[\protect\astroncite{\emph{{K{\"o}nyves} et~al.}}{2015}]{Konyves15a}
{K{\"o}nyves} V. et~al. (2015) \emph{\aap}, \emph{584}, A91.

\bibitem[\protect\astroncite{\emph{{Koo} et~al.}}{2020}]{Koo_BC2020}
{Koo} B.-C. et~al. (2020) \emph{\apj}, \emph{905}, 1, 35.

\bibitem[\protect\astroncite{\emph{{Kounkel} et~al.}}{2018}]{Kounkel18}
{Kounkel} M. et~al. (2018) \emph{\aj}, \emph{156}, 3, 84.

\bibitem[\protect\astroncite{\emph{{Kraus} and {Hillenbrand}}}{2008}]{Kraus08a}
{Kraus} A.~L. and {Hillenbrand} L.~A. (2008) \emph{\apjl}, \emph{686}, L111.

\bibitem[\protect\astroncite{\emph{{Kreckel} et~al.}}{2018}]{Kreckel2018}
{Kreckel} K. et~al. (2018) \emph{\apjl}, \emph{863}, L21.

\bibitem[\protect\astroncite{\emph{{Kreckel} et~al.}}{2019}]{kreckel2019}
{Kreckel} K. et~al. (2019) \emph{\apj}, \emph{887}, 1, 80.

\bibitem[\protect\astroncite{\emph{{Kritsuk} et~al.}}{2011}]{Kritsuk2011}
{Kritsuk} A.~G. et~al. (2011) \emph{\apjl}, \emph{727}, 1, L20.

\bibitem[\protect\astroncite{\emph{{Kruijssen}}}{2012}]{Kruijssen2012}
{Kruijssen} J.~M.~D. (2012) \emph{\mnras}, \emph{426}, 4, 3008.

\bibitem[\protect\astroncite{\emph{{Kruijssen} and
  {Longmore}}}{2014}]{Kruijssen2014}
{Kruijssen} J.~M.~D. and {Longmore} S.~N. (2014) \emph{\mnras}, \emph{439}, 4,
  3239.

\bibitem[\protect\astroncite{\emph{{Kruijssen} et~al.}}{2015}]{Kruijssen15}
{Kruijssen} J.~M.~D. et~al. (2015) \emph{\mnras}, \emph{447}, 2, 1059.

\bibitem[\protect\astroncite{\emph{{Kruijssen} et~al.}}{2018}]{Kruijssen2018}
{Kruijssen} J.~M.~D. et~al. (2018) \emph{\mnras}, \emph{479}, 1866.

\bibitem[\protect\astroncite{\emph{{Kruijssen} et~al.}}{2019}]{Kruijssen2019}
{Kruijssen} J.~M.~D. et~al. (2019) \emph{\nat}, \emph{569}, 7757, 519.

\bibitem[\protect\astroncite{\emph{{Krumholz}}}{2014}]{Krumholz14a}
{Krumholz} M.~R. (2014) \emph{\physrep}, \emph{539}, 49.

\bibitem[\protect\astroncite{\emph{{Krumholz}}}{2018}]{Krumholz18c}
{Krumholz} M.~R. (2018) \emph{\mnras}, \emph{480}, 3468.

\bibitem[\protect\astroncite{\emph{{Krumholz} and
  {Gnedin}}}{2011}]{Krumholz11b}
{Krumholz} M.~R. and {Gnedin} N.~Y. (2011) \emph{\apj}, \emph{729}, 36.

\bibitem[\protect\astroncite{\emph{{Krumholz} and
  {Matzner}}}{2009}]{Krumholz_Matzner2009}
{Krumholz} M.~R. and {Matzner} C.~D. (2009) \emph{\apj}, \emph{703}, 2, 1352.

\bibitem[\protect\astroncite{\emph{{Krumholz} and {McKee}}}{2005}]{Krumholz05a}
{Krumholz} M.~R. and {McKee} C.~F. (2005) \emph{\apj}, \emph{630}, 250.

\bibitem[\protect\astroncite{\emph{{Krumholz} and {McKee}}}{2020}]{Krumholz20}
{Krumholz} M.~R. and {McKee} C.~F. (2020) \emph{\mnras}, \emph{494}, 1, 624.

\bibitem[\protect\astroncite{\emph{{Krumholz} and {Tan}}}{2007}]{Krumholz07b}
{Krumholz} M.~R. and {Tan} J.~C. (2007) \emph{\apj}, \emph{654}, 304.

\bibitem[\protect\astroncite{\emph{{Krumholz} and
  {Thompson}}}{2013}]{Krumholz_Thompson2013}
{Krumholz} M.~R. and {Thompson} T.~A. (2013) \emph{\mnras}, \emph{434}, 3,
  2329.

\bibitem[\protect\astroncite{\emph{{Krumholz} et~al.}}{2006}]{Krumholz06a}
{Krumholz} M.~R. et~al. (2006) \emph{\apj}, \emph{653}, 361.

\bibitem[\protect\astroncite{\emph{{Krumholz} et~al.}}{2007}]{Krumholz07a}
{Krumholz} M.~R. et~al. (2007) \emph{\apj}, \emph{656}, 959.

\bibitem[\protect\astroncite{\emph{{Krumholz} et~al.}}{2012}]{Krumholz12a}
{Krumholz} M.~R. et~al. (2012) \emph{\apj}, \emph{754}, 71.

\bibitem[\protect\astroncite{\emph{{Krumholz} et~al.}}{2015}]{SLUG_2015}
{Krumholz} M.~R. et~al. (2015) \emph{\mnras}, \emph{452}, 2, 1447.

\bibitem[\protect\astroncite{\emph{{Krumholz} et~al.}}{2017}]{Krumholz17e}
{Krumholz} M.~R. et~al. (2017) \emph{\mnras}, \emph{471}, 4061.

\bibitem[\protect\astroncite{\emph{{Krumholz} et~al.}}{2018}]{Krumholz18}
{Krumholz} M.~R. et~al. (2018) \emph{\mnras}, \emph{477}, 2, 2716.

\bibitem[\protect\astroncite{\emph{{Krumholz} et~al.}}{2019}]{KrumholzARAA19}
{Krumholz} M.~R. et~al. (2019) \emph{\araa}, \emph{57}, 227.

\bibitem[\protect\astroncite{\emph{{Kuhn} et~al.}}{2014}]{Kuhn14a}
{Kuhn} M.~A. et~al. (2014) \emph{\apj}, \emph{787}, 107.

\bibitem[\protect\astroncite{\emph{{Kuhn} et~al.}}{2019}]{Kuhn19}
{Kuhn} M.~A. et~al. (2019) \emph{\apj}, \emph{870}, 1, 32.

\bibitem[\protect\astroncite{\emph{{Lada}}}{1976}]{Lada76a}
{Lada} C.~J. (1976) \emph{\apjs}, \emph{32}, 603.

\bibitem[\protect\astroncite{\emph{{Lada} and {Dame}}}{2020}]{lada20}
{Lada} C.~J. and {Dame} T.~M. (2020) \emph{\apj}, \emph{898}, 1, 3.

\bibitem[\protect\astroncite{\emph{{Lada} et~al.}}{2017}]{Lada17a}
{Lada} C.~J. et~al. (2017) \emph{\aap}, \emph{606}, A100.

\bibitem[\protect\astroncite{\emph{{Lancaster}
  et~al.}}{2021{\natexlab{a}}}]{Lancaster_2021a}
{Lancaster} L. et~al. (2021{\natexlab{a}}) \emph{\apj}, \emph{914}, 2, 89.

\bibitem[\protect\astroncite{\emph{{Lancaster}
  et~al.}}{2021{\natexlab{b}}}]{Lancaster_2021b}
{Lancaster} L. et~al. (2021{\natexlab{b}}) \emph{\apj}, \emph{914}, 2, 90.

\bibitem[\protect\astroncite{\emph{{Lancaster}
  et~al.}}{2021{\natexlab{c}}}]{Lancaster_2021c}
{Lancaster} L. et~al. (2021{\natexlab{c}}) \emph{\apjl}, \emph{922}, 1, L3.

\bibitem[\protect\astroncite{\emph{{Larson}}}{1981}]{Larson81}
{Larson} R.~B. (1981) \emph{\mnras}, \emph{194}, 809.

\bibitem[\protect\astroncite{\emph{{Lee} et~al.}}{2016}]{lee16}
{Lee} E.~J. et~al. (2016) \emph{\apj}, \emph{833}, 2, 229.

\bibitem[\protect\astroncite{\emph{{Lee} and
  {Hennebelle}}}{2016{\natexlab{a}}}]{LeeHennebelle16a}
{Lee} Y.-N. and {Hennebelle} P. (2016{\natexlab{a}}) \emph{\aap}, \emph{591},
  A30.

\bibitem[\protect\astroncite{\emph{{Lee} and
  {Hennebelle}}}{2016{\natexlab{b}}}]{LeeHennebelle16b}
{Lee} Y.-N. and {Hennebelle} P. (2016{\natexlab{b}}) \emph{\aap}, \emph{591},
  A31.

\bibitem[\protect\astroncite{\emph{{Leisawitz} et~al.}}{1989}]{Leisawitz_1989}
{Leisawitz} D. et~al. (1989) \emph{\apjs}, \emph{70}, 731.

\bibitem[\protect\astroncite{\emph{{Leitherer} et~al.}}{1999}]{Leitherer_1999}
{Leitherer} C. et~al. (1999) \emph{\apjs}, \emph{123}, 1, 3.

\bibitem[\protect\astroncite{\emph{{Leroy} et~al.}}{2009}]{Leroy09}
{Leroy} A.~K. et~al. (2009) \emph{\aj}, \emph{137}, 6, 4670.

\bibitem[\protect\astroncite{\emph{{Leroy} et~al.}}{2013}]{Leroy13}
{Leroy} A.~K. et~al. (2013) \emph{\aj}, \emph{146}, 2, 19.

\bibitem[\protect\astroncite{\emph{{Leroy} et~al.}}{2015}]{leroy15}
{Leroy} A.~K. et~al. (2015) \emph{\apj}, \emph{801}, 1, 25.

\bibitem[\protect\astroncite{\emph{{Leroy} et~al.}}{2016}]{leroy16}
{Leroy} A.~K. et~al. (2016) \emph{\apj}, \emph{831}, 1, 16.

\bibitem[\protect\astroncite{\emph{{Leroy} et~al.}}{2017}]{Leroy17}
{Leroy} A.~K. et~al. (2017) \emph{\apj}, \emph{846}, 1, 71.

\bibitem[\protect\astroncite{\emph{{Leroy} et~al.}}{2018}]{leroy2018}
{Leroy} A.~K. et~al. (2018) in: \emph{Science with a Next Generation Very Large
  Array}, vol. 517 of \emph{Astronomical Society of the Pacific Conference
  Series}, (edited by E.~{Murphy}), p. 499.

\bibitem[\protect\astroncite{\emph{{Leroy}
  et~al.}}{2021{\natexlab{a}}}]{leroy21a}
{Leroy} A.~K. et~al. (2021{\natexlab{a}}) \emph{\apjs}, \emph{257}, 2, 43.

\bibitem[\protect\astroncite{\emph{{Leroy}
  et~al.}}{2021{\natexlab{b}}}]{leroy21b}
{Leroy} A.~K. et~al. (2021{\natexlab{b}}) \emph{\apjs}, \emph{255}, 1, 19.

\bibitem[\protect\astroncite{\emph{{Li} et~al.}}{2019}]{Li_Vogelsberger2019}
{Li} H. et~al. (2019) \emph{\mnras}, \emph{487}, 1, 364.

\bibitem[\protect\astroncite{\emph{{Lim} et~al.}}{2020}]{Lim20a}
{Lim} B. et~al. (2020) \emph{\apj}, \emph{899}, 2, 121.

\bibitem[\protect\astroncite{\emph{{Liu} et~al.}}{2021}]{liu21}
{Liu} L. et~al. (2021) \emph{\mnras}, \emph{505}, 3, 4048.

\bibitem[\protect\astroncite{\emph{{Lombardi} et~al.}}{2010}]{Lombardi2010}
{Lombardi} M. et~al. (2010) \emph{\aap}, \emph{512}, A67.

\bibitem[\protect\astroncite{\emph{{Lombardi} et~al.}}{2011}]{Lombardi2011}
{Lombardi} M. et~al. (2011) \emph{\aap}, \emph{535}, A16.

\bibitem[\protect\astroncite{\emph{{Lombardi} et~al.}}{2014}]{lombardi14}
{Lombardi} M. et~al. (2014) \emph{\aap}, \emph{566}, A45.

\bibitem[\protect\astroncite{\emph{{Lombardi} et~al.}}{2015}]{lombardi15}
{Lombardi} M. et~al. (2015) \emph{\aap}, \emph{576}, L1.

\bibitem[\protect\astroncite{\emph{{Longmore} et~al.}}{2013}]{Longmore13}
{Longmore} S.~N. et~al. (2013) \emph{\mnras}, \emph{429}, 2, 987.

\bibitem[\protect\astroncite{\emph{{Longmore} et~al.}}{2014}]{Longmore14}
{Longmore} S.~N. et~al. (2014) in: \emph{Protostars and Planets VI}, (edited by
  H.~{Beuther}, R.~S. {Klessen}, C.~P. {Dullemond}, and T.~{Henning}), p. 291.

\bibitem[\protect\astroncite{\emph{{Lopez} et~al.}}{2014}]{lopez14}
{Lopez} L.~A. et~al. (2014) \emph{\apj}, \emph{795}, 2, 121.

\bibitem[\protect\astroncite{\emph{{Lu} et~al.}}{2013}]{Lu13}
{Lu} J.~R. et~al. (2013) \emph{\apj}, \emph{764}, 2, 155.

\bibitem[\protect\astroncite{\emph{{Lu} et~al.}}{2018}]{Lu18}
{Lu} X. et~al. (2018) \emph{\apj}, \emph{855}, 1, 9.

\bibitem[\protect\astroncite{\emph{{Lucas} et~al.}}{2020}]{lucas20}
{Lucas} W.~E. et~al. (2020) \emph{\mnras}, \emph{493}, 4, 4700.

\bibitem[\protect\astroncite{\emph{{Mac Low}}}{1999}]{maclow1999}
{Mac Low} M.-M. (1999) \emph{\apj}, \emph{524}, 1, 169.

\bibitem[\protect\astroncite{\emph{{Mac Low} and {Klessen}}}{2004}]{Mac-Low04a}
{Mac Low} M.-M. and {Klessen} R.~S. (2004) \emph{Reviews of Modern Physics},
  \emph{76}, 125.

\bibitem[\protect\astroncite{\emph{{Mao} et~al.}}{2020}]{Mao_2020}
{Mao} S.~A. et~al. (2020) \emph{\apj}, \emph{898}, 1, 52.

\bibitem[\protect\astroncite{\emph{{Martizzi} et~al.}}{2015}]{Martizzi_2015}
{Martizzi} D. et~al. (2015) \emph{\mnras}, \emph{450}, 1, 504.

\bibitem[\protect\astroncite{\emph{{Matzner}}}{2002}]{Matzner02a}
{Matzner} C.~D. (2002) \emph{\apj}, \emph{566}, 302.

\bibitem[\protect\astroncite{\emph{{Matzner} and {Jumper}}}{2015}]{Matzner15}
{Matzner} C.~D. and {Jumper} P.~H. (2015) \emph{\apj}, \emph{815}, 1, 68.

\bibitem[\protect\astroncite{\emph{{Matzner} and
  {McKee}}}{2000}]{Matzner_McKee2000}
{Matzner} C.~D. and {McKee} C.~F. (2000) \emph{\apj}, \emph{545}, 1, 364.

\bibitem[\protect\astroncite{\emph{{McCrady} et~al.}}{2005}]{McCrady05}
{McCrady} N. et~al. (2005) \emph{\apj}, \emph{621}, 1, 278.

\bibitem[\protect\astroncite{\emph{{McCray} and
  {Kafatos}}}{1987}]{McCray_Kafatos1987}
{McCray} R. and {Kafatos} M. (1987) \emph{\apj}, \emph{317}, 190.

\bibitem[\protect\astroncite{\emph{{McDermid} et~al.}}{2020}]{mcdermid20}
{McDermid} R.~M. et~al. (2020) \emph{arXiv e-prints}, arXiv:2009.09242.

\bibitem[\protect\astroncite{\emph{{McKee} and
  {Ostriker}}}{2007}]{McKee_Ostriker07}
{McKee} C.~F. and {Ostriker} E.~C. (2007) \emph{\araa}, \emph{45}, 565.

\bibitem[\protect\astroncite{\emph{{McKee} and {Williams}}}{1997}]{McKee97a}
{McKee} C.~F. and {Williams} J.~P. (1997) \emph{\apj}, \emph{476}, 144.

\bibitem[\protect\astroncite{\emph{{McKee} and
  {Zweibel}}}{1992}]{mckee_zweibel1992}
{McKee} C.~F. and {Zweibel} E.~G. (1992) \emph{\apj}, \emph{399}, 551.

\bibitem[\protect\astroncite{\emph{{McKellar}}}{1940}]{McKellar40a}
{McKellar} A. (1940) \emph{\pasp}, \emph{52}, 307, 187.

\bibitem[\protect\astroncite{\emph{{McKinnon} et~al.}}{2019}]{mckinnon19}
{McKinnon} M. et~al. (2019) in: \emph{Bulletin of the American Astronomical
  Society}, vol.~51, p.~81.

\bibitem[\protect\astroncite{\emph{{McLeod} et~al.}}{2019}]{mcleod19}
{McLeod} A.~F. et~al. (2019) \emph{\mnras}, \emph{486}, 4, 5263.

\bibitem[\protect\astroncite{\emph{{McLeod} et~al.}}{2020}]{mcleod2020}
{McLeod} A.~F. et~al. (2020) \emph{\apj}, \emph{891}, 1, 25.

\bibitem[\protect\astroncite{\emph{{McLeod} et~al.}}{2021}]{mcleod21}
{McLeod} A.~F. et~al. (2021) \emph{\mnras}, \emph{508}, 4, 5425.

\bibitem[\protect\astroncite{\emph{{Meidt} et~al.}}{2015}]{Meidt15}
{Meidt} S.~E. et~al. (2015) \emph{\apj}, \emph{806}, 1, 72.

\bibitem[\protect\astroncite{\emph{{Meidt} et~al.}}{2018}]{meidt18}
{Meidt} S.~E. et~al. (2018) \emph{\apj}, \emph{854}, 2, 100.

\bibitem[\protect\astroncite{\emph{{Meidt} et~al.}}{2020}]{meidt20}
{Meidt} S.~E. et~al. (2020) \emph{\apj}, \emph{892}, 2, 73.

\bibitem[\protect\astroncite{\emph{{Menon} et~al.}}{2021}]{Menon21a}
{Menon} S.~H. et~al. (2021) \emph{\mnras}, \emph{507}, 4, 5542.

\bibitem[\protect\astroncite{\emph{{Messa} et~al.}}{2021}]{Messa21}
{Messa} M. et~al. (2021) \emph{\apj}, \emph{909}, 2, 121.

\bibitem[\protect\astroncite{\emph{{Miura} et~al.}}{2012}]{miura12}
{Miura} R.~E. et~al. (2012) \emph{\apj}, \emph{761}, 1, 37.

\bibitem[\protect\astroncite{\emph{{Miura} et~al.}}{2018}]{miura18}
{Miura} R.~E. et~al. (2018) \emph{\apj}, \emph{864}, 2, 120.

\bibitem[\protect\astroncite{\emph{{Miura} et~al.}}{2021}]{miura21}
{Miura} R.~E. et~al. (2021) \emph{\mnras}, \emph{504}, 4, 6198.

\bibitem[\protect\astroncite{\emph{{Miville-Desch{\^e}nes}
  et~al.}}{2017}]{MivilleDeschenes17}
{Miville-Desch{\^e}nes} M.-A. et~al. (2017) \emph{\apj}, \emph{834}, 1, 57.

\bibitem[\protect\astroncite{\emph{{Miyamoto} et~al.}}{2014}]{Miyamoto_2014}
{Miyamoto} Y. et~al. (2014) \emph{\pasj}, \emph{66}, 2, 36.

\bibitem[\protect\astroncite{\emph{{Mok} et~al.}}{2020}]{mok20}
{Mok} A. et~al. (2020) \emph{\apj}, \emph{893}, 2, 135.

\bibitem[\protect\astroncite{\emph{{Molinari} et~al.}}{2016}]{molinari16}
{Molinari} S. et~al. (2016) \emph{\aap}, \emph{591}, A149.

\bibitem[\protect\astroncite{\emph{{Mooney} and {Solomon}}}{1988}]{Mooney88a}
{Mooney} T.~J. and {Solomon} P.~M. (1988) \emph{\apjl}, \emph{334}, L51.

\bibitem[\protect\astroncite{\emph{{Motte} et~al.}}{2010}]{motte10}
{Motte} F. et~al. (2010) \emph{\aap}, \emph{518}, L77.

\bibitem[\protect\astroncite{\emph{{Motte} et~al.}}{2018}]{Motte18}
{Motte} F. et~al. (2018) \emph{\araa}, \emph{56}, 41.

\bibitem[\protect\astroncite{\emph{{Mouschovias} and
  {Ciolek}}}{1999}]{Mouschovias99a}
{Mouschovias} T.~C. and {Ciolek} G.~E. (1999) in: \emph{NATO Advanced Science
  Institutes (ASI) Series C}, vol. 540, p. 305.

\bibitem[\protect\astroncite{\emph{{Mouschovias} and
  {Spitzer}}}{1976}]{Mouschovias76a}
{Mouschovias} T.~C. and {Spitzer} Jr. L. (1976) \emph{\apj}, \emph{210}, 326.

\bibitem[\protect\astroncite{\emph{{Muraoka} et~al.}}{2020}]{muraoka20}
{Muraoka} K. et~al. (2020) \emph{\apj}, \emph{903}, 2, 94.

\bibitem[\protect\astroncite{\emph{{Murray}}}{2011}]{Murray11}
{Murray} N. (2011) \emph{\apj}, \emph{729}, 133.

\bibitem[\protect\astroncite{\emph{{Murray} et~al.}}{2010}]{Murray_2010}
{Murray} N. et~al. (2010) \emph{\apj}, \emph{709}, 1, 191.

\bibitem[\protect\astroncite{\emph{{Myers} et~al.}}{2014}]{Myers14a}
{Myers} A.~T. et~al. (2014) \emph{\mnras}, \emph{439}, 3420.

\bibitem[\protect\astroncite{\emph{{Myers} et~al.}}{1986}]{Myers86a}
{Myers} P.~C. et~al. (1986) \emph{\apj}, \emph{301}, 398.

\bibitem[\protect\astroncite{\emph{{Nakamura} and {Li}}}{2007}]{Nakamura07a}
{Nakamura} F. and {Li} Z.-Y. (2007) \emph{\apj}, \emph{662}, 395.

\bibitem[\protect\astroncite{\emph{{Nandra} et~al.}}{2013}]{nandra13}
{Nandra} K. et~al. (2013) \emph{arXiv e-prints}, arXiv:1306.2307.

\bibitem[\protect\astroncite{\emph{{Ochsendorf} et~al.}}{2017}]{Ochsendorf17}
{Ochsendorf} B.~B. et~al. (2017) \emph{\apj}, \emph{841}, 2, 109.

\bibitem[\protect\astroncite{\emph{{Ohlin} et~al.}}{2019}]{Ohlin_2019}
{Ohlin} L. et~al. (2019) \emph{\mnras}, \emph{485}, 3, 3887.

\bibitem[\protect\astroncite{\emph{{Oka} et~al.}}{2001}]{oka01}
{Oka} T. et~al. (2001) \emph{\apj}, \emph{562}, 1, 348.

\bibitem[\protect\astroncite{\emph{{Olivier} et~al.}}{2021}]{olivier21}
{Olivier} G.~M. et~al. (2021) \emph{\apj}, \emph{908}, 1, 68.

\bibitem[\protect\astroncite{\emph{{Onodera} et~al.}}{2010}]{Onodera10}
{Onodera} S. et~al. (2010) \emph{\apjl}, \emph{722}, 2, L127.

\bibitem[\protect\astroncite{\emph{{Onus} et~al.}}{2018}]{Onus18a}
{Onus} A. et~al. (2018) \emph{\mnras}, \emph{479}, 2, 1702.

\bibitem[\protect\astroncite{\emph{{Oort} and
  {Spitzer}}}{1955}]{Oort_Spitzer1955}
{Oort} J.~H. and {Spitzer} Lyman J. (1955) \emph{\apj}, \emph{121}, 6.

\bibitem[\protect\astroncite{\emph{{Ostriker} and {Shetty}}}{2011}]{Ostriker11}
{Ostriker} E.~C. and {Shetty} R. (2011) \emph{\apj}, \emph{731}, 1, 41.

\bibitem[\protect\astroncite{\emph{{Ostriker} et~al.}}{1999}]{Ostriker99a}
{Ostriker} E.~C. et~al. (1999) \emph{\apj}, \emph{513}, 259.

\bibitem[\protect\astroncite{\emph{{Ostriker} et~al.}}{2001}]{Ostriker2001}
{Ostriker} E.~C. et~al. (2001) \emph{\apj}, \emph{546}, 2, 980.

\bibitem[\protect\astroncite{\emph{{Ostriker} et~al.}}{2010}]{Ostriker10}
{Ostriker} E.~C. et~al. (2010) \emph{\apj}, \emph{721}, 2, 975.

\bibitem[\protect\astroncite{\emph{{Pabst} et~al.}}{2019}]{Pabst_2019}
{Pabst} C. et~al. (2019) \emph{\nat}, \emph{565}, 7741, 618.

\bibitem[\protect\astroncite{\emph{{Pabst} et~al.}}{2020}]{Pabst_2020}
{Pabst} C.~H.~M. et~al. (2020) \emph{\aap}, \emph{639}, A2.

\bibitem[\protect\astroncite{\emph{{Padoan} and {Nordlund}}}{2011}]{Padoan11}
{Padoan} P. and {Nordlund} {\r{A}}. (2011) \emph{\apj}, \emph{730}, 1, 40.

\bibitem[\protect\astroncite{\emph{{Padoan} et~al.}}{2006}]{padoan06}
{Padoan} P. et~al. (2006) \emph{\apjl}, \emph{653}, 2, L125.

\bibitem[\protect\astroncite{\emph{{Padoan} et~al.}}{2007}]{Padoan07a}
{Padoan} P. et~al. (2007) \emph{\apj}, \emph{661}, 972.

\bibitem[\protect\astroncite{\emph{{Padoan} et~al.}}{2012}]{Padoan12}
{Padoan} P. et~al. (2012) \emph{\apjl}, \emph{759}, 2, L27.

\bibitem[\protect\astroncite{\emph{{Padoan} et~al.}}{2016}]{Padoan16}
{Padoan} P. et~al. (2016) \emph{\apj}, \emph{822}, 1, 11.

\bibitem[\protect\astroncite{\emph{{Palla} and
  {Stahler}}}{1999}]{PallaStahler99}
{Palla} F. and {Stahler} S.~W. (1999) \emph{\apj}, \emph{525}, 2, 772.

\bibitem[\protect\astroncite{\emph{{Palla} and
  {Stahler}}}{2000}]{PallaStahler00}
{Palla} F. and {Stahler} S.~W. (2000) \emph{\apj}, \emph{540}, 1, 255.

\bibitem[\protect\astroncite{\emph{{Pan} et~al.}}{2015}]{pan15}
{Pan} H.-A. et~al. (2015) \emph{\mnras}, \emph{453}, 3, 3082.

\bibitem[\protect\astroncite{\emph{{Pereira-Santaella}
  et~al.}}{2016}]{pereira16}
{Pereira-Santaella} M. et~al. (2016) \emph{\aap}, \emph{587}, A44.

\bibitem[\protect\astroncite{\emph{{Peretto} et~al.}}{2014}]{Peretto14}
{Peretto} N. et~al. (2014) \emph{\aap}, \emph{561}, A83.

\bibitem[\protect\astroncite{\emph{{Peters} et~al.}}{2010}]{Peters10a}
{Peters} T. et~al. (2010) \emph{\apj}, \emph{711}, 1017.

\bibitem[\protect\astroncite{\emph{{Pettitt} et~al.}}{2020}]{Pettit2020}
{Pettitt} A.~R. et~al. (2020) \emph{\mnras}, \emph{498}, 1, 1159.

\bibitem[\protect\astroncite{\emph{{Pety} et~al.}}{2013}]{Pety13}
{Pety} J. et~al. (2013) \emph{\apj}, \emph{779}, 1, 43.

\bibitem[\protect\astroncite{\emph{{Pety} et~al.}}{2017}]{pety17}
{Pety} J. et~al. (2017) \emph{\aap}, \emph{599}, A98.

\bibitem[\protect\astroncite{\emph{{Pineda} et~al.}}{2008}]{Pineda2008}
{Pineda} J.~E. et~al. (2008) \emph{\apj}, \emph{679}, 1, 481.

\bibitem[\protect\astroncite{\emph{{Pineda} et~al.}}{2009}]{pineda09}
{Pineda} J.~E. et~al. (2009) \emph{\apjl}, \emph{699}, 2, L134.

\bibitem[\protect\astroncite{\emph{{Pingel} et~al.}}{2013}]{pingel13}
{Pingel} N.~M. et~al. (2013) \emph{\apj}, \emph{779}, 1, 36.

\bibitem[\protect\astroncite{\emph{{Planck Collaboration}
  et~al.}}{2016}]{planckpaper35}
{Planck Collaboration} et~al. (2016) \emph{\aap}, \emph{586}, A138.

\bibitem[\protect\astroncite{\emph{{Pokhrel} et~al.}}{2020}]{Pokhrel20a}
{Pokhrel} R. et~al. (2020) \emph{\apj}, \emph{896}, 1, 60.

\bibitem[\protect\astroncite{\emph{{Pokhrel} et~al.}}{2021}]{Pokhrel21}
{Pokhrel} R. et~al. (2021) \emph{\apjl}, \emph{912}, 1, L19.

\bibitem[\protect\astroncite{\emph{{Pon} et~al.}}{2012}]{Pon12}
{Pon} A. et~al. (2012) \emph{\apj}, \emph{756}, 2, 145.

\bibitem[\protect\astroncite{\emph{{Rahner} et~al.}}{2017}]{Rahner_2017}
{Rahner} D. et~al. (2017) \emph{\mnras}, \emph{470}, 4, 4453.

\bibitem[\protect\astroncite{\emph{{Raskutti} et~al.}}{2016}]{Raskutti2016}
{Raskutti} S. et~al. (2016) \emph{\apj}, \emph{829}, 2, 130.

\bibitem[\protect\astroncite{\emph{{Raskutti} et~al.}}{2017}]{Raskutti17a}
{Raskutti} S. et~al. (2017) \emph{\apj}, \emph{850}, 112.

\bibitem[\protect\astroncite{\emph{{Rebolledo} et~al.}}{2015}]{rebolledo15}
{Rebolledo} D. et~al. (2015) \emph{\apj}, \emph{808}, 1, 99.

\bibitem[\protect\astroncite{\emph{{Retter} et~al.}}{2021}]{Retter21a}
{Retter} B. et~al. (2021) \emph{\mnras}, \emph{507}, 2, 1904.

\bibitem[\protect\astroncite{\emph{{Rezaei Kh.} et~al.}}{2020}]{rezaeikhan20}
{Rezaei Kh.} S. et~al. (2020) \emph{\aap}, \emph{643}, A151.

\bibitem[\protect\astroncite{\emph{{Rice} et~al.}}{2016}]{rice16}
{Rice} T.~S. et~al. (2016) \emph{\apj}, \emph{822}, 1, 52.

\bibitem[\protect\astroncite{\emph{{Richard} et~al.}}{2019}]{richard19}
{Richard} J. et~al. (2019) \emph{arXiv e-prints}, arXiv:1906.01657.

\bibitem[\protect\astroncite{\emph{{Richings} and {Schaye}}}{2016}]{Richings16}
{Richings} A.~J. and {Schaye} J. (2016) \emph{\mnras}, \emph{460}, 3, 2297.

\bibitem[\protect\astroncite{\emph{{Rieder} et~al.}}{2022}]{Rieder2022}
{Rieder} S. et~al. (2022) \emph{\mnras}, \emph{509}, 4, 6155.

\bibitem[\protect\astroncite{\emph{{Robitaille} et~al.}}{2019}]{Robitaille19a}
{Robitaille} J.~F. et~al. (2019) \emph{\aap}, \emph{628}, A33.

\bibitem[\protect\astroncite{\emph{{Rogers} and {Pittard}}}{2013}]{Rogers13a}
{Rogers} H. and {Pittard} J.~M. (2013) \emph{\mnras}, \emph{431}, 1337.

\bibitem[\protect\astroncite{\emph{{Roman-Duval} et~al.}}{2010}]{RomanDuval10}
{Roman-Duval} J. et~al. (2010) \emph{\apj}, \emph{723}, 1, 492.

\bibitem[\protect\astroncite{\emph{{Rosen} et~al.}}{2014}]{Rosen_2014}
{Rosen} A.~L. et~al. (2014) \emph{\mnras}, \emph{442}, 3, 2701.

\bibitem[\protect\astroncite{\emph{{Rosolowsky}}}{2005}]{rosolowsky05}
{Rosolowsky} E. (2005) \emph{\pasp}, \emph{117}, 838, 1403.

\bibitem[\protect\astroncite{\emph{{Rosolowsky}}}{2007}]{Rosolowsky07}
{Rosolowsky} E. (2007) \emph{\apj}, \emph{654}, 1, 240.

\bibitem[\protect\astroncite{\emph{{Rosolowsky} and {Leroy}}}{2006}]{cprops}
{Rosolowsky} E. and {Leroy} A. (2006) \emph{\pasp}, \emph{118}, 842, 590.

\bibitem[\protect\astroncite{\emph{{Rosolowsky} et~al.}}{2003}]{Rosolowsky03a}
{Rosolowsky} E. et~al. (2003) \emph{\apj}, \emph{599}, 1, 258.

\bibitem[\protect\astroncite{\emph{{Rosolowsky} et~al.}}{2021}]{rosolowsky21}
{Rosolowsky} E. et~al. (2021) \emph{\mnras}, \emph{502}, 1, 1218.

\bibitem[\protect\astroncite{\emph{{Safranek-Shrader}
  et~al.}}{2017}]{Safranek-Shrader17}
{Safranek-Shrader} C. et~al. (2017) \emph{\mnras}, \emph{465}, 1, 885.

\bibitem[\protect\astroncite{\emph{{Saintonge} et~al.}}{2011}]{Saintonge11}
{Saintonge} A. et~al. (2011) \emph{\mnras}, \emph{415}, 1, 32.

\bibitem[\protect\astroncite{\emph{{Sandstrom} et~al.}}{2013}]{Sandstrom13}
{Sandstrom} K.~M. et~al. (2013) \emph{\apj}, \emph{777}, 1, 5.

\bibitem[\protect\astroncite{\emph{{Scheepmaker}
  et~al.}}{2009}]{Scheepmaker09a}
{Scheepmaker} R.~A. et~al. (2009) \emph{\aap}, \emph{494}, 81.

\bibitem[\protect\astroncite{\emph{{Schinnerer} et~al.}}{2013}]{Schinnerer13}
{Schinnerer} E. et~al. (2013) \emph{\apj}, \emph{779}, 1, 42.

\bibitem[\protect\astroncite{\emph{{Schinnerer} et~al.}}{2019}]{Schinnerer2019}
{Schinnerer} E. et~al. (2019) \emph{\apj}, \emph{887}, 1, 49.

\bibitem[\protect\astroncite{\emph{{Schneider} et~al.}}{2011}]{Schneider11a}
{Schneider} N. et~al. (2011) \emph{\aap}, \emph{529}, A1.

\bibitem[\protect\astroncite{\emph{{Schneider}
  et~al.}}{2015{\natexlab{a}}}]{Schneider_2015a}
{Schneider} N. et~al. (2015{\natexlab{a}}) \emph{\aap}, \emph{575}, A79.

\bibitem[\protect\astroncite{\emph{{Schneider}
  et~al.}}{2015{\natexlab{b}}}]{Schneider_2015b}
{Schneider} N. et~al. (2015{\natexlab{b}}) \emph{\aap}, \emph{578}, A29.

\bibitem[\protect\astroncite{\emph{{Schruba} et~al.}}{2010}]{Schruba10}
{Schruba} A. et~al. (2010) \emph{\apj}, \emph{722}, 2, 1699.

\bibitem[\protect\astroncite{\emph{{Schruba} et~al.}}{2017}]{schruba17}
{Schruba} A. et~al. (2017) \emph{\apj}, \emph{835}, 2, 278.

\bibitem[\protect\astroncite{\emph{{Schruba} et~al.}}{2018}]{Schruba18}
{Schruba} A. et~al. (2018) \emph{\apj}, \emph{862}, 2, 110.

\bibitem[\protect\astroncite{\emph{{Schruba} et~al.}}{2019}]{Schruba19}
{Schruba} A. et~al. (2019) \emph{\apj}, \emph{883}, 1, 2.

\bibitem[\protect\astroncite{\emph{{Schuller} et~al.}}{2021}]{schuller21}
{Schuller} F. et~al. (2021) \emph{\mnras}, \emph{500}, 3, 3064.

\bibitem[\protect\astroncite{\emph{{Sciortino} et~al.}}{2013}]{sciortino13}
{Sciortino} S. et~al. (2013) \emph{arXiv e-prints}, arXiv:1306.2333.

\bibitem[\protect\astroncite{\emph{Scoville and Hersh}}{1979}]{Scoville1979}
Scoville N.~Z. and Hersh K. (1979) \emph{\apj}, \emph{229}, 578.

\bibitem[\protect\astroncite{\emph{{Scoville} et~al.}}{1987}]{Scoville87a}
{Scoville} N.~Z. et~al. (1987) \emph{\apjs}, \emph{63}, 821.

\bibitem[\protect\astroncite{\emph{{Seifried} et~al.}}{2020}]{Seifried20}
{Seifried} D. et~al. (2020) \emph{\mnras}, \emph{492}, 1, 1465.

\bibitem[\protect\astroncite{\emph{{Semenov} et~al.}}{2003}]{Semenov_2003}
{Semenov} D. et~al. (2003) \emph{\aap}, \emph{410}, 611.

\bibitem[\protect\astroncite{\emph{{Semenov} et~al.}}{2016}]{Semenov16a}
{Semenov} V.~A. et~al. (2016) \emph{\apj}, \emph{826}, 200.

\bibitem[\protect\astroncite{\emph{{Semenov} et~al.}}{2017}]{Semenov17}
{Semenov} V.~A. et~al. (2017) \emph{\apj}, \emph{845}, 2, 133.

\bibitem[\protect\astroncite{\emph{{Semenov} et~al.}}{2018}]{Semenov18}
{Semenov} V.~A. et~al. (2018) \emph{\apj}, \emph{861}, 1, 4.

\bibitem[\protect\astroncite{\emph{{Semenov} et~al.}}{2021}]{Semenov_2021}
{Semenov} V.~A. et~al. (2021) \emph{\apj}, \emph{918}, 1, 13.

\bibitem[\protect\astroncite{\emph{{Shetty} et~al.}}{2007}]{shetty_2007}
{Shetty} R. et~al. (2007) \emph{\apj}, \emph{665}, 2, 1138.

\bibitem[\protect\astroncite{\emph{{Shetty} et~al.}}{2011}]{Shetty2011}
{Shetty} R. et~al. (2011) \emph{\mnras}, \emph{415}, 4, 3253.

\bibitem[\protect\astroncite{\emph{{Shetty} et~al.}}{2012}]{shetty12}
{Shetty} R. et~al. (2012) \emph{\mnras}, \emph{425}, 1, 720.

\bibitem[\protect\astroncite{\emph{{Shimajiri} et~al.}}{2019}]{Shimajiri19}
{Shimajiri} Y. et~al. (2019) \emph{\aap}, \emph{623}, A16.

\bibitem[\protect\astroncite{\emph{{Shu} et~al.}}{1987}]{Shu87a}
{Shu} F.~H. et~al. (1987) \emph{\araa}, \emph{25}, 23.

\bibitem[\protect\astroncite{\emph{Simon et~al.}}{2019}]{simon2019astro2020}
Simon R. et~al. (2019) Astro2020: The cycling of matter from the interstellar
  medium to stars and back.

\bibitem[\protect\astroncite{\emph{{Simon} et~al.}}{2019}]{GEcoDecadal}
{Simon} R. et~al. (2019) \emph{\baas}, \emph{51}, 3, 367.

\bibitem[\protect\astroncite{\emph{{Sivanandam} et~al.}}{2018}]{sivanandam18}
{Sivanandam} S. et~al. (2018) in: \emph{Ground-based and Airborne
  Instrumentation for Astronomy VII}, vol. 10702 of \emph{Society of
  Photo-Optical Instrumentation Engineers (SPIE) Conference Series}, (edited by
  C.~J. {Evans}, L.~{Simard}, and H.~{Takami}), p. 107021J.

\bibitem[\protect\astroncite{\emph{{Skinner} and
  {Ostriker}}}{2015}]{Skinner_Ostriker2015}
{Skinner} M.~A. and {Ostriker} E.~C. (2015) \emph{\apj}, \emph{809}, 2, 187.

\bibitem[\protect\astroncite{\emph{{Smilgys} and
  {Bonnell}}}{2017}]{Smilgys2017}
{Smilgys} R. and {Bonnell} I.~A. (2017) \emph{\mnras}, \emph{472}, 4, 4982.

\bibitem[\protect\astroncite{\emph{{Smith} et~al.}}{2021}]{Smith_MC2021}
{Smith} M.~C. et~al. (2021) \emph{\mnras}, \emph{506}, 3, 3882.

\bibitem[\protect\astroncite{\emph{{Smith} et~al.}}{2020}]{Smith_R2020}
{Smith} R.~J. et~al. (2020) \emph{\mnras}, \emph{492}, 2, 1594.

\bibitem[\protect\astroncite{\emph{{Soler}}}{2019}]{soler19}
{Soler} J.~D. (2019) \emph{\aap}, \emph{629}, A96.

\bibitem[\protect\astroncite{\emph{{Solomon} et~al.}}{1987}]{Solomon87a}
{Solomon} P.~M. et~al. (1987) \emph{\apj}, \emph{319}, 730.

\bibitem[\protect\astroncite{\emph{{Spilker} et~al.}}{2021}]{spilker21}
{Spilker} A. et~al. (2021) \emph{\aap}, \emph{653}, A63.

\bibitem[\protect\astroncite{\emph{{Spitzer}}}{1978}]{Spitzer1978}
{Spitzer} L. (1978) \emph{{Physical processes in the interstellar medium}}.

\bibitem[\protect\astroncite{\emph{{Stacey} et~al.}}{2018}]{stacey2018}
{Stacey} G.~J. et~al. (2018) in: \emph{Ground-based and Airborne Telescopes
  VII}, vol. 10700 of \emph{Society of Photo-Optical Instrumentation Engineers
  (SPIE) Conference Series}, (edited by H.~K. {Marshall} and J.~{Spyromilio}),
  p. 107001M.

\bibitem[\protect\astroncite{\emph{{Stephens} et~al.}}{2016}]{Stephens16a}
{Stephens} I.~W. et~al. (2016) \emph{\apj}, \emph{824}, 29.

\bibitem[\protect\astroncite{\emph{{Sternberg} and
  {Dalgarno}}}{1995}]{Sternberg95}
{Sternberg} A. and {Dalgarno} A. (1995) \emph{\apjs}, \emph{99}, 565.

\bibitem[\protect\astroncite{\emph{{Sternberg} et~al.}}{2014}]{Sternberg14}
{Sternberg} A. et~al. (2014) \emph{\apj}, \emph{790}, 1, 10.

\bibitem[\protect\astroncite{\emph{{Stone} et~al.}}{1998}]{Stone1998}
{Stone} J.~M. et~al. (1998) \emph{\apjl}, \emph{508}, 1, L99.

\bibitem[\protect\astroncite{\emph{{Su} et~al.}}{2019}]{su19}
{Su} Y. et~al. (2019) \emph{\apjs}, \emph{240}, 1, 9.

\bibitem[\protect\astroncite{\emph{{Sun} et~al.}}{2018}]{Sun18}
{Sun} J. et~al. (2018) \emph{\apj}, \emph{860}, 2, 172.

\bibitem[\protect\astroncite{\emph{{Sun} et~al.}}{2020{\natexlab{a}}}]{Sun20b}
{Sun} J. et~al. (2020{\natexlab{a}}) \emph{\apj}, \emph{892}, 2, 148.

\bibitem[\protect\astroncite{\emph{{Sun} et~al.}}{2020{\natexlab{b}}}]{sun20a}
{Sun} J. et~al. (2020{\natexlab{b}}) \emph{\apjl}, \emph{901}, 1, L8.

\bibitem[\protect\astroncite{\emph{{Sun} et~al.}}{2017}]{Sun17a}
{Sun} N.-C. et~al. (2017) \emph{\apj}, \emph{849}, 2, 149.

\bibitem[\protect\astroncite{\emph{{Swiggum} et~al.}}{2021}]{Swiggum21a}
{Swiggum} C. et~al. (2021) \emph{\apj}, \emph{917}, 1, 21.

\bibitem[\protect\astroncite{\emph{{Swings} and {Rosenfeld}}}{1937}]{Swings37a}
{Swings} P. and {Rosenfeld} L. (1937) \emph{\apj}, \emph{86}, 483.

\bibitem[\protect\astroncite{\emph{{Tacconi} et~al.}}{2013}]{Tacconi13}
{Tacconi} L.~J. et~al. (2013) \emph{\apj}, \emph{768}, 1, 74.

\bibitem[\protect\astroncite{\emph{{Tacconi} et~al.}}{2020}]{Tacconi20}
{Tacconi} L.~J. et~al. (2020) \emph{\araa}, \emph{58}, 157.

\bibitem[\protect\astroncite{\emph{{Tenorio-Tagle}}}{1979}]{Tenorio-Tagle1979}
{Tenorio-Tagle} G. (1979) \emph{\aap}, \emph{71}, 59.

\bibitem[\protect\astroncite{\emph{{Theissen} et~al.}}{2021}]{Theissen21a}
{Theissen} C.~A. et~al. (2021) \emph{arXiv e-prints}, arXiv:2105.05871.

\bibitem[\protect\astroncite{\emph{{Thompson} and
  {Krumholz}}}{2016}]{Thompson_2016}
{Thompson} T.~A. and {Krumholz} M.~R. (2016) \emph{\mnras}, \emph{455}, 1, 334.

\bibitem[\protect\astroncite{\emph{{Toal{\'a}} et~al.}}{2012}]{Toala12}
{Toal{\'a}} J.~A. et~al. (2012) \emph{\apj}, \emph{744}, 2, 190.

\bibitem[\protect\astroncite{\emph{{Tosaki} et~al.}}{2017}]{tosaki17}
{Tosaki} T. et~al. (2017) \emph{\pasj}, \emph{69}, 2, 18.

\bibitem[\protect\astroncite{\emph{{Traficante} et~al.}}{2018}]{Traficante18}
{Traficante} A. et~al. (2018) \emph{\mnras}, \emph{477}, 2, 2220.

\bibitem[\protect\astroncite{\emph{{Tress} et~al.}}{2020}]{Tress_2020}
{Tress} R.~G. et~al. (2020) \emph{\mnras}, \emph{492}, 2, 2973.

\bibitem[\protect\astroncite{\emph{{Tsang} and
  {Milosavljevi{\'c}}}}{2018}]{Tsang_Milos2018}
{Tsang} B. T.~H. and {Milosavljevi{\'c}} M. (2018) \emph{\mnras}, \emph{478},
  3, 4142.

\bibitem[\protect\astroncite{\emph{{Umemoto} et~al.}}{2017}]{umemoto17}
{Umemoto} T. et~al. (2017) \emph{\pasj}, \emph{69}, 5, 78.

\bibitem[\protect\astroncite{\emph{{Urquhart} et~al.}}{2018}]{Urquhart18a}
{Urquhart} J.~S. et~al. (2018) \emph{\mnras}, \emph{473}, 1059.

\bibitem[\protect\astroncite{\emph{{Usero} et~al.}}{2015}]{Usero15a}
{Usero} A. et~al. (2015) \emph{\aj}, \emph{150}, 115.

\bibitem[\protect\astroncite{\emph{{Utomo} et~al.}}{2015}]{utomo15}
{Utomo} D. et~al. (2015) \emph{\apj}, \emph{803}, 1, 16.

\bibitem[\protect\astroncite{\emph{{Utomo} et~al.}}{2018}]{Utomo18}
{Utomo} D. et~al. (2018) \emph{\apjl}, \emph{861}, 2, L18.

\bibitem[\protect\astroncite{\emph{{van Dishoeck} and
  {Black}}}{1988}]{vanDishoeck88}
{van Dishoeck} E.~F. and {Black} J.~H. (1988) \emph{\apj}, \emph{334}, 771.

\bibitem[\protect\astroncite{\emph{{Vazquez-Semadeni}}}{1994}]{Vazquez-Semadeni94a}
{Vazquez-Semadeni} E. (1994) \emph{\apj}, \emph{423}, 681.

\bibitem[\protect\astroncite{\emph{{Vazquez-Semadeni}
  et~al.}}{1995}]{Vazquez-Semadeni95a}
{Vazquez-Semadeni} E. et~al. (1995) \emph{\apj}, \emph{441}, 702.

\bibitem[\protect\astroncite{\emph{{Vazquez-Semadeni}
  et~al.}}{2000}]{PPIV_2000}
{Vazquez-Semadeni} E. et~al. (2000) in: \emph{Protostars and Planets IV},
  (edited by V.~{Mannings}, A.~P. {Boss}, and S.~S. {Russell}), p.~3.

\bibitem[\protect\astroncite{\emph{{V{\'a}zquez-Semadeni}
  et~al.}}{2017}]{Vazquez-Semadeni17}
{V{\'a}zquez-Semadeni} E. et~al. (2017) \emph{\mnras}, \emph{467}, 2, 1313.

\bibitem[\protect\astroncite{\emph{{V{\'a}zquez-Semadeni}
  et~al.}}{2019}]{Vazquez-Semadeni19}
{V{\'a}zquez-Semadeni} E. et~al. (2019) \emph{\mnras}, \emph{490}, 3, 3061.

\bibitem[\protect\astroncite{\emph{{Vogel} et~al.}}{1987}]{Vogel87}
{Vogel} S.~N. et~al. (1987) \emph{\apjl}, \emph{321}, L145.

\bibitem[\protect\astroncite{\emph{{Vutisalchavakul}
  et~al.}}{2016}]{Vutisalchavakul16}
{Vutisalchavakul} N. et~al. (2016) \emph{\apj}, \emph{831}, 1, 73.

\bibitem[\protect\astroncite{\emph{{Wakelam} et~al.}}{2017}]{Wakelam17}
{Wakelam} V. et~al. (2017) \emph{Molecular Astrophysics}, \emph{9}, 1.

\bibitem[\protect\astroncite{\emph{{Walch} and {Naab}}}{2015}]{Walch_Naab2015}
{Walch} S. and {Naab} T. (2015) \emph{\mnras}, \emph{451}, 3, 2757.

\bibitem[\protect\astroncite{\emph{{Walch} et~al.}}{2015}]{Walch15}
{Walch} S. et~al. (2015) \emph{\mnras}, \emph{454}, 1, 238.

\bibitem[\protect\astroncite{\emph{{Walch} et~al.}}{2012}]{Walch2012}
{Walch} S.~K. et~al. (2012) \emph{\mnras}, \emph{427}, 1, 625.

\bibitem[\protect\astroncite{\emph{{Ward} and {Kruijssen}}}{2018}]{Ward18}
{Ward} J.~L. and {Kruijssen} J.~M.~D. (2018) \emph{\mnras}, \emph{475}, 4,
  5659.

\bibitem[\protect\astroncite{\emph{{Ward}
  et~al.}}{2020{\natexlab{a}}}]{Ward20_gaia}
{Ward} J.~L. et~al. (2020{\natexlab{a}}) \emph{\mnras}, \emph{495}, 1, 663.

\bibitem[\protect\astroncite{\emph{{Ward}
  et~al.}}{2020{\natexlab{b}}}]{ward2020}
{Ward} J.~L. et~al. (2020{\natexlab{b}}) \emph{\mnras}, \emph{497}, 2, 2286.

\bibitem[\protect\astroncite{\emph{{Weaver} et~al.}}{1977}]{Weaver_1977}
{Weaver} R. et~al. (1977) \emph{\apj}, \emph{218}, 377.

\bibitem[\protect\astroncite{\emph{{Westpfahl} et~al.}}{1999}]{Westpfahl99a}
{Westpfahl} D.~J. et~al. (1999) \emph{\aj}, \emph{117}, 2, 868.

\bibitem[\protect\astroncite{\emph{{Whitmore} et~al.}}{2014}]{Whitmore14}
{Whitmore} B.~C. et~al. (2014) \emph{\apj}, \emph{795}, 2, 156.

\bibitem[\protect\astroncite{\emph{{Whitworth}}}{1979}]{Whitworth79a}
{Whitworth} A. (1979) \emph{\mnras}, \emph{186}, 59.

\bibitem[\protect\astroncite{\emph{{Wilcock} et~al.}}{2012}]{Wilcock12a}
{Wilcock} L.~A. et~al. (2012) \emph{\mnras}, \emph{422}, 2, 1071.

\bibitem[\protect\astroncite{\emph{{Williams} and {McKee}}}{1997}]{Williams97a}
{Williams} J.~P. and {McKee} C.~F. (1997) \emph{\apj}, \emph{476}, 166.

\bibitem[\protect\astroncite{\emph{{Williams} et~al.}}{1994}]{williams94}
{Williams} J.~P. et~al. (1994) \emph{\apj}, \emph{428}, 693.

\bibitem[\protect\astroncite{\emph{{Williams} et~al.}}{2018}]{williams18}
{Williams} T.~G. et~al. (2018) \emph{\mnras}, \emph{479}, 1, 297.

\bibitem[\protect\astroncite{\emph{{Wilson} et~al.}}{2005}]{wilson05}
{Wilson} B.~A. et~al. (2005) \emph{\aap}, \emph{430}, 523.

\bibitem[\protect\astroncite{\emph{{Wilson} and {Scoville}}}{1990}]{Wilson90a}
{Wilson} C.~D. and {Scoville} N. (1990) \emph{\apj}, \emph{363}, 435.

\bibitem[\protect\astroncite{\emph{{Wilson} et~al.}}{2012}]{Wilson12}
{Wilson} C.~D. et~al. (2012) \emph{\mnras}, \emph{424}, 4, 3050.

\bibitem[\protect\astroncite{\emph{{Wilson} et~al.}}{1970}]{Wilson70a}
{Wilson} R.~W. et~al. (1970) \emph{\apjl}, \emph{161}, L43.

\bibitem[\protect\astroncite{\emph{{Wong} et~al.}}{2011}]{Wong11}
{Wong} T. et~al. (2011) \emph{\apjs}, \emph{197}, 2, 16.

\bibitem[\protect\astroncite{\emph{{Wong} et~al.}}{2013}]{Wong2013}
{Wong} T. et~al. (2013) \emph{\apjl}, \emph{777}, 1, L4.

\bibitem[\protect\astroncite{\emph{{Wong} et~al.}}{2019}]{Wong19}
{Wong} T. et~al. (2019) \emph{\apj}, \emph{885}, 1, 50.

\bibitem[\protect\astroncite{\emph{{Wu} et~al.}}{2005}]{Wu05}
{Wu} J. et~al. (2005) \emph{\apjl}, \emph{635}, 2, L173.

\bibitem[\protect\astroncite{\emph{{Yahia} et~al.}}{2021}]{Yahia21a}
{Yahia} H. et~al. (2021) \emph{\aap}, \emph{649}, A33.

\bibitem[\protect\astroncite{\emph{{Yorke}}}{1986}]{Yorke1986}
{Yorke} H.~W. (1986) \emph{\araa}, \emph{24}, 49.

\bibitem[\protect\astroncite{\emph{{Zabel} et~al.}}{2020}]{zabel2020}
{Zabel} N. et~al. (2020) \emph{\mnras}, \emph{496}, 2, 2155.

\bibitem[\protect\astroncite{\emph{{Zamora-Avil{\'e}s} and
  {V{\'a}zquez-Semadeni}}}{2014}]{Zamora-Aviles14a}
{Zamora-Avil{\'e}s} M. and {V{\'a}zquez-Semadeni} E. (2014) \emph{\apj},
  \emph{793}, 84.

\bibitem[\protect\astroncite{\emph{{Zamora-Avil{\'e}s}
  et~al.}}{2012}]{Zamora-Aviles12a}
{Zamora-Avil{\'e}s} M. et~al. (2012) \emph{\apj}, \emph{751}, 77.

\bibitem[\protect\astroncite{\emph{{Zanella} et~al.}}{2015}]{zanella2015}
{Zanella} A. et~al. (2015) \emph{\nat}, \emph{521}, 7550, 54.

\bibitem[\protect\astroncite{\emph{{Zhang} and
  {Chevalier}}}{2019}]{Zhang_Chevalier2019}
{Zhang} D. and {Chevalier} R.~A. (2019) \emph{\mnras}, \emph{482}, 2, 1602.

\bibitem[\protect\astroncite{\emph{{Zinnecker}}}{1984}]{Zinnecker84a}
{Zinnecker} H. (1984) \emph{\mnras}, \emph{210}, 43.

\end{thebibliography}
\end{document}